\documentclass[longauth]{aa} 
\usepackage{amsmath,inputenx,nicefrac,mathrsfs}
\usepackage[dvipsnames]{xcolor}
\usepackage{rotating}
\usepackage{graphicx,subcaption}
\usepackage{multicol,multirow,blindtext}
\usepackage{wasysym,longtable}
\usepackage{pstricks}
\usepackage{orcidlink}
\usepackage{txfonts,url,hyperref}
\usepackage{nicefrac,cancel,dashbox}
\usepackage{tikz,tkz-euclide,longtable}

\begin{document} 

\title{The First Look of {\it Gaia}: Daily data quality and instrument health assessment with automated early warnings}

\titlerunning{The {\it Gaia} First Look}
\authorrunning{M.~Altmann et al.}

\author{
M. Altmann $^{\orcidlink{0000-0002-0530-0913}}$\inst{1,2},
Z. Balog $^{\orcidlink{0000-0003-1748-2926}}$\inst{1},
W. L\"offler $^{\orcidlink{0009-0003-1319-5601}}$\inst{1},
U. Bastian $^{\orcidlink{0000-0002-8667-1715}}$\inst{1},
M. Biermann $^{\orcidlink{0000-0002-5791-9056}}$\inst{1},
A. Sagrista Selles $^{\orcidlink{0000-0001-6191-2028}}$\inst{1},
M. Davidson $^{\orcidlink{0000-0001-9271-4411}}$\inst{3},
N. Rowell $^{\orcidlink{0000-0003-3809-1895}}$\inst{3},
E. Serpell $^{\orcidlink{0009-0005-2537-1793}}$\inst{4},
A. Abreu Aramburu $^{\orcidlink{0000-0003-3959-0856}}$\inst{5,6},
T. Br\"usemeister\inst{1},
C. Crowley $^{\orcidlink{0000-0002-9391-9630}}$\inst{7},
M. Hauser\inst{1},
S. Jordan $^{\orcidlink{0000-0001-6316-6831}}$\inst{1},
J. Mart\'in-Fleitas $^{\orcidlink{0000-0002-8594-569X}}$\inst{8},
A. Mora $^{\orcidlink{0000-0002-1922-8529}}$\inst{9},
E. Fernandez del Peloso $^{\orcidlink{0009-0008-1357-8895}}$\inst{1},
U. Stampa$^{\orcidlink{0009-0001-0723-7137}}$\inst{1}
}

\institute{
 Zentrum f\"ur Astronomie der Universit\"at Heidelberg, Astronomisches Recheninstitut, M\"onchhofstr. 12-14, 69120 Heidelberg, Germany
 \and
LTE, Observatoire de Paris, PSL Research University, CNRS, Sorbonne Universit\'e, UPMC Univ. Paris 06, LNE, 61 avenue de l’Observatoire, 75014 Paris, France
\and
Institute for Astronomy, School of Physics and Astronomy, University of Edinburgh, Royal Observatory, Blackford Hill, Edinburgh,
EH9 3HJ, United Kingdom
\and
Telespazio Germany GmbH, Europaplatz~5, Darmstadt, Germany 
\and
European Space Agency (ESA), European Space Astronomy Centre
(ESAC), Camino bajo del Castillo, s/n, Urbanizacion Villafranca del
Castillo, Villanueva de la Cañada, 28692 Madrid, Spain
\and
Universidad Complutense de Madrid, Av. Complutense, s/n, Moncloa - Aravaca, 28040 (Madrid), Spain
\and
HE Space Operations BV for European Space Agency (ESA),
Camino bajo del Castillo, s/n, Urbanizacion Villafranca del Castillo,
Villanueva de la Cañada, 28692 Madrid, Spain
\and
Aurora Technology B.V. for European Space Agency (ESA), Camino bajo del Castillo, s/n, Urbanizacion Villafranca del Castillo, Villanueva de la Ca\~nada, 28692 Madrid, Spain.
\and
Telespazio UK S.L. for European Space Agency (ESA), Camino bajo del Castillo, s/n, Urbanizacion Villafranca del Castillo, Villanueva de la Ca\~nada, 28692 Madrid, Spain
}

\date{Received 19 January 2026 / Accepted 27 March 2026 }

\abstract{
The ESA {\it Gaia} mission is a 10+ year astrometric whole-sky scan, demanding
 consistent data quality over the whole timespan of operations}{The {\it Gaia} First Look (FL) is a system whose aim is 
monitoring the data quality to identify problems, which includes early warning capabilities
 for potential upcoming issues.}{In 
order to achieve its goals, the {\it Gaia}  FL  implemented its own limited astrometric solution, 
and used the daily calibrations from other segments of the Data Processing and Analysis Consortium (DPAC),
 as well as the diagnostic data from the satellite itself, in order to obtain a complete
picture of the situation of the {\it Gaia} satellite on a daily basis. This led to a short-term health and data quality check,
but also to a broader overview of the longer-term trends and evolutions within the payload. 
Potential issues that were encountered were
 reported to other groups within DPAC for further analysis purposes. 
When required, ways to mitigate the problems were discussed, and implemented.}{We show a number of findings by the 
{\it Gaia} FL concerning longer-term evolution, individual but common effects, as well as
 detrimental impacts, all of which occurred  over the operational phase of the {\it Gaia} mission.}{} 
 
      \keywords{methods: observational - methods: data analysis - astrometry - Space vehicles - techniques: photometric 
}
   \maketitle  

\section{Introduction\label{sect:introduction}}

{\it Gaia}, ESA's ground-breaking astrometric satellite mission, has produced a database of the highest-quality astrometry of
almost two billion objects (see e.g. \citealt{2016A&A...595A...1G}) 
In its currently three data releases, 
{\it Gaia} has helped to create the most accurate 3D map of the Milky Way Galaxy, and the precise parallaxes and proper motions
have provided an unprecedented insight into the kinematics and dynamics of our galaxy, 
enabling us to revolutionise our understanding of its formation and evolution. Apart from the Milky Way 
and its satellites (e.g. the two Magellanic clouds), it has provided the platform for increasing our knowledge of stellar evolution, as well as our own Solar System, including asteroids and minor planets. 
At present, all of this has been secured by just a fraction of the total data, procured 
in its 10.5 year operational lifespan. Therefore, when the complete dataset has been processed and published, 
much more is to be expected. 

However, this promise depends on a mostly consistent data quality, 
obtained by a delicate space probe over this long
timespan in a harsh environment, and a good understanding of issues and general trends in the inevitable evolution of
the {\it Gaia} spacecraft.  
The main
aim of {\it Gaia} is the assembly of a global high-precision astrometric database of all objects down to 20.7~mag
 in the {\it Gaia} $G$-band \citep{2016A&A...595A...1G}.
This highly ambitious undertaking relies on the consistent quality of the input data, as bad portions
of data have the potential of compromising the complete result. Moreover this project mandates the knowledge of 
long-term evolutionary processes within the payload. This encompasses all relevant parts of the instrumentation,
the two telescopes, also known as fields of view  (FoVs), the focal plane, with its array of detectors, called
the focal plane array (FPA; see Fig.~\ref{fig:FPA.fig}), and many more. Since any spacecraft is exposed to the harsh environmental conditions
in space, ageing of the components of the instrument set-up is expected, and must be closely monitored. 

The {\it Gaia} First Look (FL)  task  monitored on a daily basis whether {\it Gaia} had reached the targeted level
of precision, as this could not be verified by the main data reduction processes until
many months of observational data had been incorporated in a global, coherent, and interleaved
data reduction. Neither the instrument state nor the data quality could be checked at the desired
level of precision by standard procedures applied to typical space missions. It was desirable to
know the measurement precision and instrument stability as soon as possible since unperceived
subtle effects could arise during the mission that could affect all data and potentially result in a
loss of many months of data if not detected in a timely manner.

On the other hand, the resources allocated to the data quality monitoring components within any project are usually 
very limited. For obvious reasons, the actual procurement of the data and subsequent scientific processing are highly prioritised.
Moreover, in order to be able to achieve timely conclusions in a regular fashion, the time intervals covered by 
the data monitoring system need to be short (e.g. daily). This leads to compromises as well. 
Therefore, a method needed to be found, which was able to
\begin{itemize}
	\item cope with limited computational and human resources;
	\item produce meaningful diagnostic outputs covering a small time interval of data ($\sim$24~hours) 
		and be processed in a similar timespan in order to produce diagnostic results in a timely fashion;
	\item assess all critical components of the main measurements of the {\it Gaia} satellite and their output, such as the precision of the stellar, attitude, and instrument parameters against the mission target level.
\end{itemize}

The resulting infrastructure for the {\it Gaia} mission is called the First Look (FL). 
This served to process and analyse the incoming mission data 
and to present the diagnostic output for a given timespan, 
so that it could be evaluated by a dedicated group of First Look scientists (FLSs).
The findings made by the FLSs were then reported to other relevant parts of {\it Gaia}'s Data Processing and Analysis Consortium (DPAC) for further 
action, if required.  

Section~\ref{sect:FL} gives a detailed description 
of the First Look, its components, and inner structure. Section~\ref{sect:results} highlights some of the results of the FL, such
as the investigation of long-term trends (Sect.~\ref{sect:results:ambient}); the analysis of the {\it Gaia} attitude,
 and its disturbances 
(Sect.~\ref{sect:results:RE}); and the
effects of some exemplary disturbances, such as the impact of micro-meteoroids (Sect.~\ref{sect:results:MM}). Presenting all findings involving the {\it Gaia}
First Look is far beyond the scope of this publication. Therefore we  restrict ourselves to a few
 instructive and important examples.

Certain key parameters, mostly the output of the one-day calibrations (ODC) have been qualified by the  FLSs on a 
daily basis, with the verdicts stored in the main database of {\it Gaia}, for later usage. If modifications to the satellite operations appeared necessary, 
this was communicated to the Science Operations Centre (SOC) for assessment and action. 
The same applied to subtle trends, and degradations seen in the 
diagnostic output. Through this feedback loop, the FL  played a vital role in maintaining {\it Gaia}'s scientific performance.

\section{Structure and components of the {\it Gaia} First Look}\label{sect:FL}

\begin{figure}[ht]
       \centering
        \includegraphics[width=0.67\hsize]{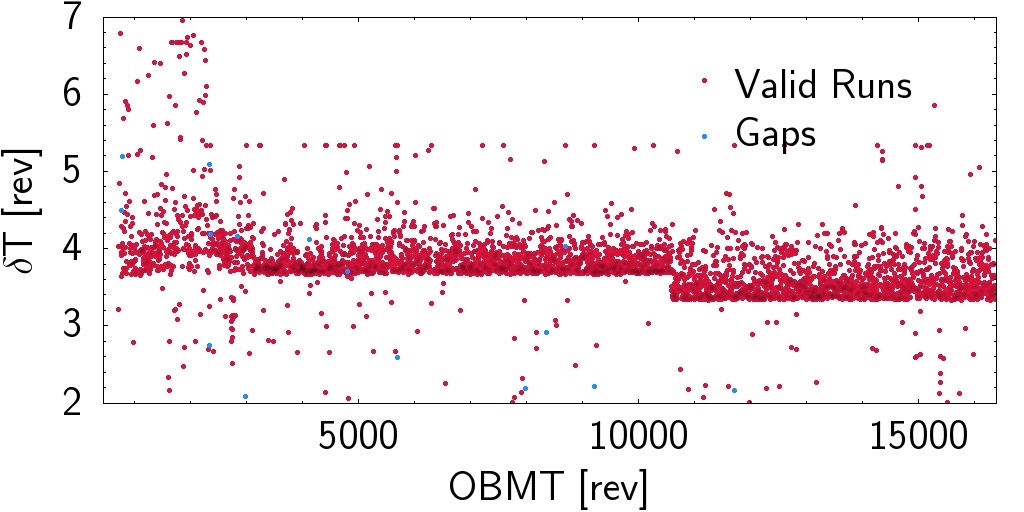}
        \includegraphics[width=0.32\hsize]{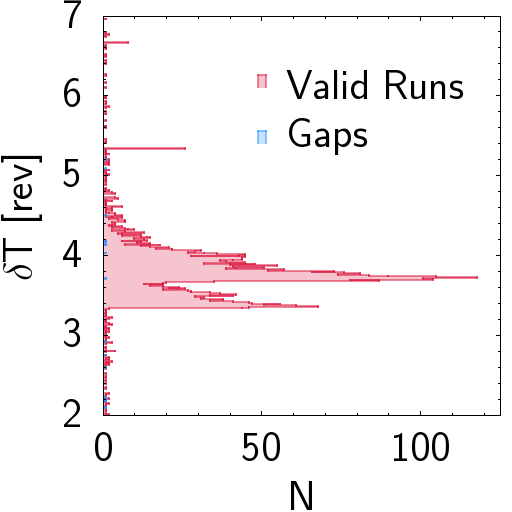}\\
        \caption{\small Length of the First Look days over the entire mission. 
$\delta T$ is the length of each FL day, i.e. the timespan covered by each FL run, 
given in units of revolution. The left panel shows the evolution over time, 
while the right panel shows the corresponding histogram, which highlights the overall distribution of the FL day
lengths. Valid FL runs, i.e. those that have real data in them, are shown in red, 
while gaps are shown in blue.}
       \label{fig:dflerunstat.fig}
\end{figure}

\begin{figure}[b]
       \centering
	\includegraphics[width=\hsize]{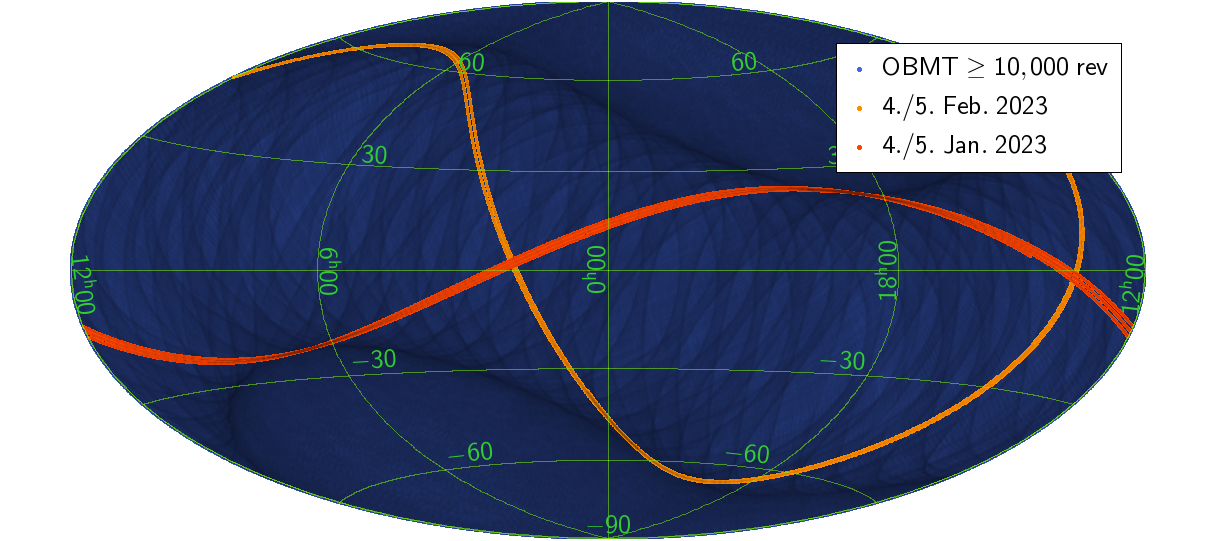}
	\caption{\small Sky coverage (given in equatorial coordinates) of two FL days
 spaced about one month apart. The blue pattern is the complete {\it Gaia} sky coverage starting at 
	OBMT=10,000~rev to the end of operations (OBMT=16,385.671~rev), the red track represents a FL day that lasted from January 4 to 5, 2023, and the orange track another day, between February 4 and 5, 2023.
	In this plot the trace for just the first telescope (FoV 1) is shown}
       \label{fig:skycov.fig}
\end{figure}

In this section we present the {\it Gaia} First Look in detail. As outlined in the previous section, the FL needed to digest and produce 
a very heterogeneous set of data. Some of these were available continuously, such as the data from the on board counters (see Sect.~\ref{sect:FL:ASD}), 
housekeeping data (see Sect.~\ref{sect:FL:HK}), the astrometric observations of the sources. Others, such as the source cross matching
(see Sect.~\ref{sect:FL:IDT}), and the one-day calibrations (see Sects.~\ref{sect:FL:ODAS} to \ref{sect:FL:RODC}), were produced once in a given time period, i.e. 
about once per roughly 24~hours.

In order to fulfil its role, the FL required several kinds of input data. These are
\begin{itemize}
\item the astrometric, photometric, and spectroscopic source data and their cross matches (see Sect.~\ref{sect:FL:IDT});
\item the results of the One-Day Astrometric Solution (ODAS; see Sect.~\ref{sect:FL:ODAS});
\item CCD health data, i.e. output of the CCD One-Day Calibration (CODC; see Sect.~\ref{sect:FL:CODC});
\item Line spread function (LSF) data, i.e. output of the LSF/PSF One-Day Calibration (LODC; see Sect.~\ref{sect:FL:LODC});
\item the results of the RVS One-Day Calibration (RODC; see Sect.~\ref{sect:FL:RODC});	
\item data from the Basic Angle Monitor (BAM;  see Sect.~\ref{sect:FL:BAM});
\item the on board auxiliary science data (ASD) counter values, which show how many objects have been observed and processed on board {\it Gaia} (see Sect.~\ref{sect:FL:ASD});
\item housekeeping data, such as focal plane assembly, mirror, and tank temperatures, attitude thruster activation, the filling level of the Payload Data Handling Unit (PDHU), 
	i.e. the amount of data in the on board storage system, the compression, and deletion rate (see Sect.~\ref{sect:FL:HK}).  
\end{itemize}
The data inserted into the FL system to create one FL report
nominally covered about 24 hours, or four satellite revolutions, which was known as a First Look day (FL day). 
One revolution as a unit of time corresponded to one actual revolution of the spacecraft, and lasted exactly
6~hours. The On-Board Mission Timeline (OBMT), based on the on board atomic clock time is the time coordinate used by DPAC for tagging the observational data with time stamps \citep{2018A&A...616A...1G}. We note that this is not 
strictly a physical time scale, as it is not completely contiguous. 
The time coordinates used by {\it Gaia} are described in more detail in \citet{2015jsrs.conf...55K} and \citet{2017SSRv..212.1423K}.
In reality,
most undisturbed FL days were a bit shorter, typically between three and four revolutions, with only a few being slightly more than four revolutions long (see Fig.~\ref{fig:dflerunstat.fig}). 
Examples of areas of the sky covered during typical FL days are shown in Fig.~\ref{fig:skycov.fig}. 

\subsection{The First Look toolkit}\label{sect:FL:tools}
\begin{figure}[ht]
       \centering
        \includegraphics[width=\hsize]{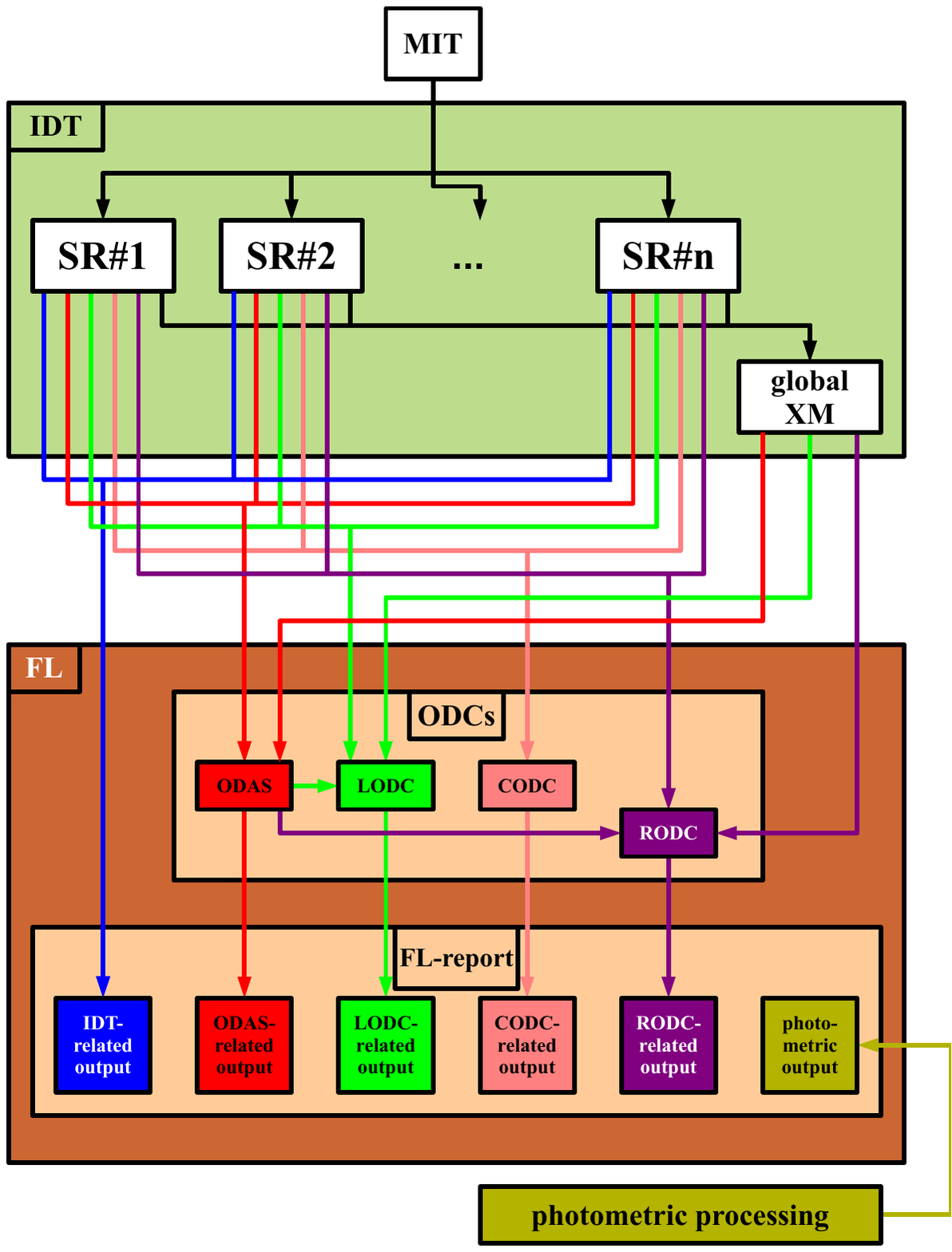}
        \caption{\small Data flow diagram of the FL-system. SR\#1 to SR\#n indicate the IDT subruns (see Sect.~\ref{sect:FL:IDT})}
       \label{fig:dataflow.fig}
\end{figure}
In order to prepare the various types of data for the inspection by the FLS, two sets of functionalities have been created. The first set, called 
'diagnostics', were rather basic statistics procedures, which counted the number of objects, 
binned the measurements into time or magnitude intervals, created simple histograms, and computed
simple statistical quantities. Typical
values for the length of sampling time intervals were 5~min and 30~min. 
In many cases these were then presented as plots and tables. 
Many results of these diagnostics were then piped into tasks belonging to the second set,
 i.e. the 'analyses'. These were more sophisticated and elaborate procedures,
presenting a more physical understanding of the results. Again, typical outputs were tables, graphs, and histogram plots. 
In order to give access to longer trends in the performance of
the spacecraft, many diagnostics and analyses additionally stored their results in trend files. 
These cover significantly longer stretches of time than one day, usually spanning several 100 revolutions. 

The majority of diagnostics and analyses contained guide values which aid the FLS to assess the quality of the output data. There are different types of such
parameters:
\begin{itemize}
	\item expectation values, which indicated the expected value of a certain parameter in question at a given time 
	\item alarm values, which gave the limits of the allowed range of variation from the expectation value
	\item warning values, which were set closer to the expectation values than the alarm values, served to indicate that the measured values are approaching the allowed limits. 
\end{itemize}
Since it was expected from the beginning that during the time of spacecraft operations the overall system would
 change, the FL implementation allowed the expectation
values, alarm and warning limits can be adapted, to reflect this evolution. This was done by the FLS, 
either by accepting a change automatically suggested by the FL system (based on a trend analysis of previous results),
 or by setting the values manually. 

\subsection{MOC (Mission Operations Centre) interface tasks (MIT), Initial Data Treatment (IDT), 
and FL}\label{sect:FL:IDT}
The data, as coming from the {\it Gaia} satellite, needed to be arranged and pre-processed, before it could be made available to downstream components of the
DPAC, such as FL. 
The pre-processing  has been described in \citet{2016A&A...595A...3F}, therefore we focus on the parts
most relevant to the FL. Telemetry data encompass several types, which are arranged into two main categories. The first are science rpckets (SP), 
which comprise the scientific
data from the star mapper (SM), astrometric field (AF), red and blue photometer (RP, BP), and radial velocity spectrometer (RVS) detectors, as well as from the 
Basic Angle Monitor (BAM) and wavefront sensor (WFS) (see Fig.~\ref{fig:FPA.fig}). 
Additional SP types contain data from suspected moving objects (SMO), and for data observed in various 
special observing modes. The second main category of data are auxiliary science data (ASD) packets, 
which do not contain actual science data, but data required to
process and categorise it. The seven types of ASD telemetry give information on, amongst others, the 
pre-scan pixels
(essential for the determination of the read-out noise and bias level estimation), 
the charge injections log, object count rates. These data were then
inserted into the Initial Data Treatment (IDT) and First Look (FL) database (IDTFLDB) via the MOC interface task (MIT), 
which also provided the OBMT times. Further data required by IDT and FL were housekeeping data (HK), which
includes a large number of system temperatures, thruster activation information, 
but also the spacecraft attitude as determined on board. 
The input data were completed by the Video Processing Unit (VPU) parameters and
reference data. The latter consists of the predicted {\it Gaia} orbit, the calibration of each CCD detector and the IDT source catalogue.

IDT worked as follows: 
Once a sufficient amount of data, i.e. about 30~minutes worth of high-priority data, 
(data, which was absolutely required to perform the daily calibration and analysis tasks) had been inserted into the  
IDTFLDB, IDT would start a subrun (see Fig.~\ref{fig:dataflow.fig}) which performed all the necessary tasks. 
These produced the information needed for the source match, the astro- and photometric elementaries 
(extracted basic image parameters, i.e. photometric flux and astrometric centroid, for each detected source image), the first On-Ground Attitude (OGA1), which is a 
refinement of the input Initial On-Ground Attitude (IOGA) (for more details see \citet{2016A&A...595A...3F}).
FL could then start with the first diagnostics immediately afterwards.
After a certain number of these subruns had been completed, no further subruns were started, 
but a global cross match was performed, which assigned the observations to sources. Once all this was done, 
the next IDT run started. Lower-priority data were not organised this way, but processed continuously as they 
became available.
 After an IDT run was complete, the output was then passed on to the one-day calibrations, which further processed the data products (see Sect.~\ref{sect:FL:ODAS} to \ref{sect:FL:RODC} and also Fig.~\ref{fig:dataflow.fig}). 
\subsection{The One-Day Astrometric Solution}\label{sect:FL:ODAS}
The One-rDy Astrometric Solution (ODAS), covering 
the data accumulated in one FL day, 
serves to assess the astrometric quality of the data
measured by {\it Gaia} on a daily basis and to identify any problems that might occur. The global astrometric solution (AGIS), 
which provides the astrometry of the {\it Gaia} 
catalogues, is much more elaborate and relies on several years worth of data (see \citet{2021A&A...649A...2L}). In addition, the main aim of these two approaches is somewhat different. While the
full astrometry aims at minimising any systematic effects, these are among the main results of ODAS, because
 their magnitude and their dependence on other parameters needs to be determined as quickly as possible in order to allow the global astrometric processing to properly calibrate these effects.
The ODAS method in its entirety is described in \citet{ODASpaper}, here we restrict ourselves to the input and
output parameters, and essentially treat the ODAS itself as a black box, which performs a limited astrometric solution 
based on only one FL day of data. Its main results are the magnitude and dependence of systematic effects, and 
an improved attitude determination (OGA2). 

As input, ODAS required the astrometric measurements of about 100,000-300,000 stars, of which about 10\% 
 were two-dimensional, i.e. of Window Class 0 (WC0, i.e. observations brighter than $G$=13.0~mag).\footnote{Also included
in this sample are the 
calibration faint stars (CFS). These were a randomly selected small subset of observations of stars fainter than G=13.0~mag,
i.e. the lower magnitude limit of WC0, which are treated as WC0 stars. Thus, unlike their non-CFS counterparts,
they have the same 2D cutout-windows as the WC0 stars, and were used to monitor a variety of parameters, 
and in our case for the Across Scan (AC) calibrations of the fainter stars.} 
About 90\% were one-dimensional, i.e. of Window Class 1 (WC1), i.e. $G$=13.0~mag to 16.0~mag (for more information on window classes and gating
of observations of bright stars, see Sect. 2.2 of \citet{2021A&A...649A..11R}).
Additionally there was a small subsample of about 1,000 to 3,000~stars per day, which had astrometric observations of both WC0 and WC1 
due to uncertainties in the on board magnitude determination (or intrinsic stellar variability). These are required to connect the calibrations of 
both Window Classes. Since {\it Gaia} essentially scanned great circles in the sky, it necessarily scanned both regions near and far away from the Galactic plane -- except in those cases, when the scan direction lay in the Galactic plane itself. Therefore the object density within a scan was highly variable. This would present problems for the 
integrity of the ODAS procedure, since the densely populated parts of the scan would dominate the results. In order to mitigate this problem, a homogenisation routine
had been implemented, which smoothed out the numbers of objects from various parts of the sky as well as possible.
 Furthermore, the input stars for ODAS were selected such as to fulfil
additional constraints. They were selected to be unblemished and ungated WC0 and WC1 observations,
 which ruled out the brighter part of WC0 ($G\le$11.98~mag).
These objects were called ODAS 
primaries, and they formed the base of the ODAS processing. 
All remaining sources were called secondaries, and the ODAS calibration was applied to these. Additional 
input into ODAS was the IDT-generated OGA1 (see Sect.~\ref{sect:FL:IDT}), the {\it Gaia} orbit and Solar System ephemerides. 

The outputs from ODAS are:
\begin{itemize}
	\item updates of the large-scale astrometric calibration parameters (LSCP)
for all SM and AF detectors, both FoV, WC0 and WC1, as well as WC2 (the window class encompassing the faintest detections, i.e. those fainter than $G$=16.0~mag) based on those stars 
		near the WC1/WC2 magnitude border, i.e. which in some {\it Gaia} scans are allocated as WC1 and 
in other instances as WC2. There are three  degrees
                of these astrometric calibration parameters.
Altogether this leads to
		1,764 parameters (see also \citet{2012A&A...538A..78L}). 
	\item source updates (i.e. the update of the parameters of existing sources, and the creation of new sources, if applicable, overall similar to the source update 
              for AGIS; see \citet{2021A&A...649A...2L}, but limited to positions only).
	\item astrometric residuals
	\item OGA2, a much more refined On-Ground Attitude determination
\end{itemize}
\subsection{The CCD One-Day Calibration (CODC)}\label{sect:FL:CODC}
The CODC mainly served to monitor and provide an initial calibration of the SM, AF, and XP detectors on {\it Gaia}'s Focal Plane Array (FPA). At a most basic level, changes in the CCDs during the mission, such as 
new hot or dead columns, were reported. The sub-system also had a role as a cosmic-radiation monitor via its tracking of charge transfer inefficiency (CTI) levels. In space, CTI is a significant issue since cosmic and solar 
particles cause traps in the silicon matrix, which hamper the transfer of the accumulated charge from one pixel 
to the next. The effect is a complex distortion of the observed image profiles, affecting the 
astrometric performance of the mission. This was mitigated on board by periodic insertion of charge via a dedicated injection structure on each detector \citep{2016A&A...595A...6C}. These injections occurred periodically 
every 2 or 5 seconds for AF and XP devices respectively. The injected charge filled some of the traps,
 making them unavailable to damage science charge packets during their transfer. The CODC calibrated these 
injections and watched the subsequent trail of released charge. Both calibrations could be analysed to indicate the radiation damage state. As {\it Gaia} scanned the sky incessantly and had no shutters, conventional means to access 
these parameters, such as obtaining a flat-field or dark exposures, were not an option. The measurements had to be made from the routine data, either using the available science observations or with dedicated windows 
inserted into the resource allocation process, known as virtual objects. 

While the gross electronic prescan bias levels were determined by IDT, the CODC performed a calibration of the Proximity
 Electronics Module bias non-uniformity -- a coupling of the read-out chain causing complex variation of 
the bias level during each TDI line. To do this, each VPU was briefly taken out of service to acquire a special sequence of virtual objects. Although these on board activities happened only with a cadence of a few 
months, the CODC process would check daily for the corresponding observations, making the creation of bias non-uniformity calibrations a data-driven product. Further information is available in \citet{2016A&A...595A...6C}.  

Another product of the CODC included statistics and monitoring of the background levels over the focal plane. Although this is clearly an optical rather than CCD calibration the CODC offered a convenient framework for 
monitoring the low-resolution background calibrations produced in the preceding IDT runs.

We defer from providing a detailed description of the various algorithms used in CODC as they were naturally superseded by alternative methods in the cyclic processing (IDU; see \citet{2016A&A...595A...3F}) and will be extensively documented in future work. 
The cyclic processing offered many advantages such as opportunities for global modelling and more computing resources.
 In FL CODC the focus was on alerting the team in case of significant on board changes in near-real time.  

\subsection{The LSF/PSF One-Day Calibration (LODC)}\label{sect:FL:LODC}
A fundamental property of any optical system is the point spread function (PSF), which in the context 
of the Gaia data processing describes the 2D distribution of photoelectrons in the digitised and 
pixelated image of a point source. For Gaia observations we additionally have the line spread function (LSF), 
which is the equivalent quantity for 1D images. The great majority of Gaia observations are 1D and span the 
AL direction only (see \citet{2021A&A...649A..11R}, Sect.~2.2). We refer to both of these quantities together as the PLSF.

Accurate knowledge of the PLSF is vital for the determination of the astrometric and photometric properties of each observation, 
and also for monitoring the evolution of the image quality, which occurs due to a combination of optical effects 
(e.g. mirror contamination and focus changes) and electronic effects (e.g. charge transfer inefficiency induced by radiation damage). 
The LODC performs a continuous near-real-time calibration of the PLSF model with the primary aim of tracking the evolving image quality. 
The calibrations it produced were not fed back to IDT for use in the estimation of locations and fluxes of observations, 
which instead used a fixed non-evolving PLSF calibration largely for stability reasons.

The PLSF modelling is complex due to the presence of numerous secondary dependences on time, source colour, position, 
stellar image AC drift rate, and others. The model actually implemented in LODC is simpler than that used in the ongoing 
cyclic data processing. Specifically, the running solution (see \citet{2021A&A...649A..11R}, Sect.~3.4.5) 
that tracks the evolution in time is (by necessity) computed in the forward direction only, leading to some time lag 
in the calibration solution. No calibration breakpoints are included, and the solution therefore takes some time to recover 
following discontinuous changes in the true PLSF. Such events can occur without warning (e.g. safe modes, micro-meteoroid impacts), 
whereas more predictable events (e.g. decontaminations, refocuses) face the practical difficulty of coordinating the breakpoint 
insertion with the satellite operations.
Finally, the PSF model is factored into 
two 1D functions that model the marginal AL and AC LSFs; this model was used in the production of DR1 and DR2 
(see \citet{2016A&A...595A...6C}, Sect.~5.1.4; also  \citet{2018gdr2.reptE...2H}, Sect.~2.3.2) but has since been abandoned 
in the cyclic data processing in favour of a complete 2D model (see \citet{2021A&A...649A..11R}).
Despite these limitations, the LODC provides important insight on the image quality in near-real time and gives vital input to decisions to refocus the telescopes.

\subsection{The RVS one-day calibration (RODC)}\label{sect:FL:RODC}
The Radial Velocity Spectrometer (RVS) was a spectrograph with $R=11,500$ resolution, 
covering the wavelength range between 8470 and 8740~{\AA}. 
The primary objective for this instrument was to measure 
radial velocities of a bright subset of stars, but also abundances, rotation velocities and astrophysical parameters,
 such as $\log g$  and $T_{\rm eff}$. 
Like the previous ODC, the RODC
characterised the calibration of the detectors and the instrument. 
The information available to the FL encompassed dispersion and resolution as well as centroiding and limiting wavelength diagnostics.

\subsection{The Basic Angle Monitor (BAM)}\label{sect:FL:BAM}

\begin{figure}[ht]
       \centering
        \includegraphics[width=\hsize]{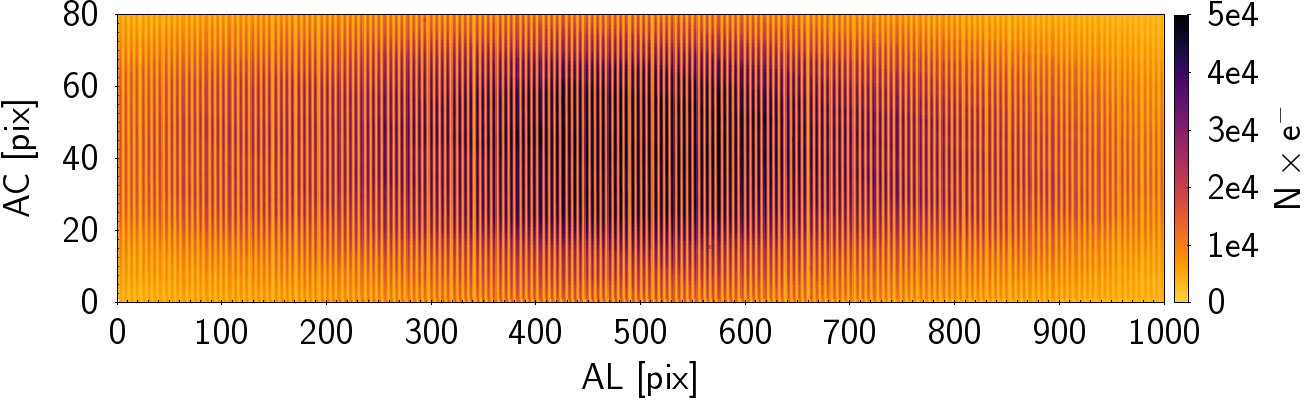}
        \includegraphics[width=\hsize]{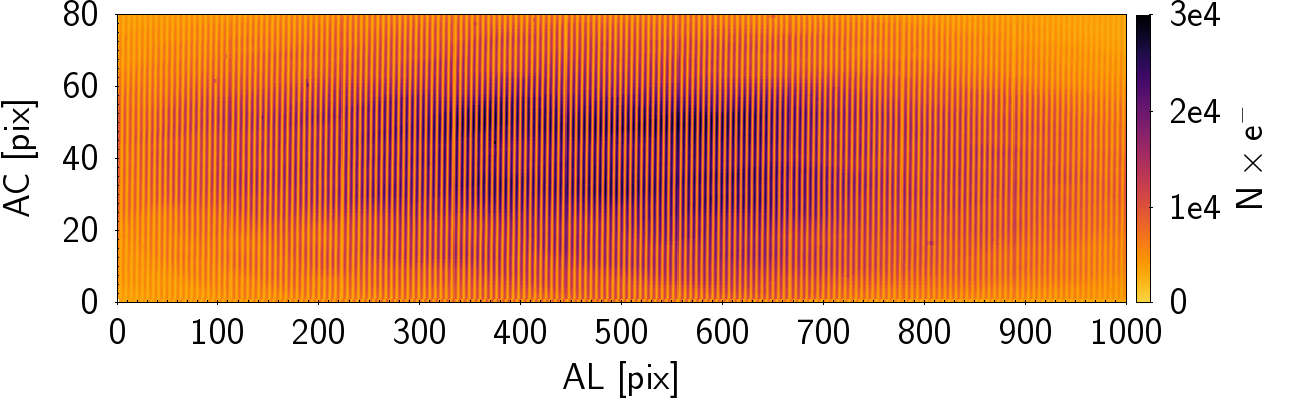}
        \caption{\small Samples of the BAM fringe pattern. The upper panel shows FoV1, the lower panel FoV2.}
       \label{fig:BAM_raw.fig}
\end{figure}

\begin{figure}[ht]
       \centering
        \includegraphics[width=\hsize]{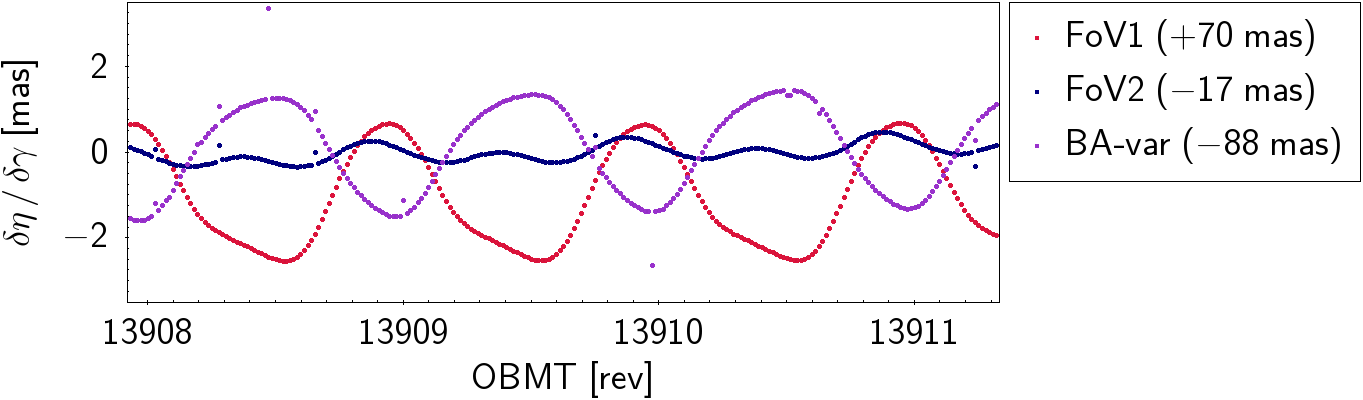}
        \caption{\small Six-hour periodic oscillations of the Basic Angle. 
Shown are the data for one arbitrarily chosen FL day, which does not show any anomalies. 
The red points denote the AL fringe phase
of FoV1 and the blue points FoV2. The combined values, i.e. the total BA variations are shown in violet. 
We note that the absolute values are shifted by the amount indicated in the 
legend in order to fit them in the same plot}
       \label{fig:BAvar.fig}
\end{figure}

The Basic Angle (BA) is the (supposedly) fixed angle of 106.5$^\circ$ between the line of sight of the first and second field of view (FoV1 and FoV2). 
According to the fundamental {\it Gaia} measurement principle, the astrometric
detectors in columns AF1-9 record stellar images from both FoV. 
The positions of all objects in the detector data stream were measured regardless of their FoV of origin. 
All of these relative coordinates
need to be known with the same precision and accuracy. For intra-FoV objects, i.e. those in the same FoV, 
the level of accuracy is determined by the detector pixel structure. However, the
angular distance on the sky between objects from different FoV additionally requires the angle between the two FoV, i.e. the BA to be 
known to the same level of accuracy as that of the distance within the detector.
This put extremely tight constraints on the tolerances of variation of the BA, which were an order of magnitude below the total accuracy aimed at. 

The BAM was an interferometric device, designed to monitor the BA to such levels of precision. Its design and function is described in \citet{10.1117/12.2309016}. 
The instrument had two branches, one for each FoV. These produced the fringe patterns shown in Fig.~\ref{fig:BAM_raw.fig}. From these fringe patterns,
a set of parameters can be derived via Fourier transforms (see \citet{2014SPIE.9143E..0XM}), 
the most important of which being the Along Scan (AL) fringe phase. This is the foundation of the Basic Angle measurements, with
the resulting BA being the difference between the fringe phases for FoV1 and FoV2.

Soon after the launch of {\it Gaia}, it was discovered with the help of the FL that the Basic Angle did not behave as anticipated, but oscillates with the heliotropic spin phase $\Omega$, as shown in Fig.~\ref{fig:BAvar.fig}.
This pattern is seen in both FoV, with a higher amplitude in FoV1. The peak-to-peak amplitude of the BA 
caused by this effect is of the order of 3~mas. This is about three orders of magnitude higher than
than the limit specified for the allowed variations of the BA, and thus has to be accounted for 
in the {\it Gaia} data reduction. For the ODAS it did not play a significant role, 
but monitoring the BA within the scope of the FL became more important than originally planned. 

In some cases, usually caused by some disturbance of the payload, for example by a micro-meteoroid impact or a planned intervention into the system, there can be a jump
in one or both branches of the AL-fringe phase as shown in Fig.~\ref{fig:BAvar_jumps.fig}. Some of these instances are discussed in Sect.~\ref{sect:results:MM} 
\subsection{The auxiliary science data}\label{sect:FL:ASD}
\begin{figure*}[ht]
       \centering
        \includegraphics[width=0.48\hsize]{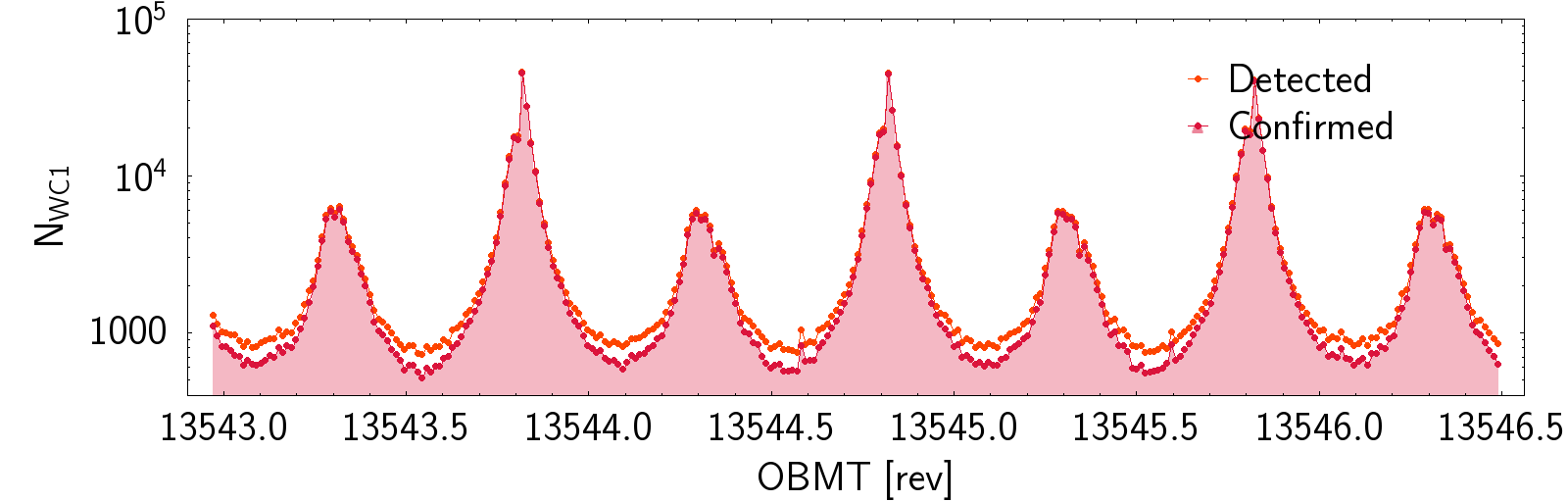}
        \includegraphics[width=0.48\hsize]{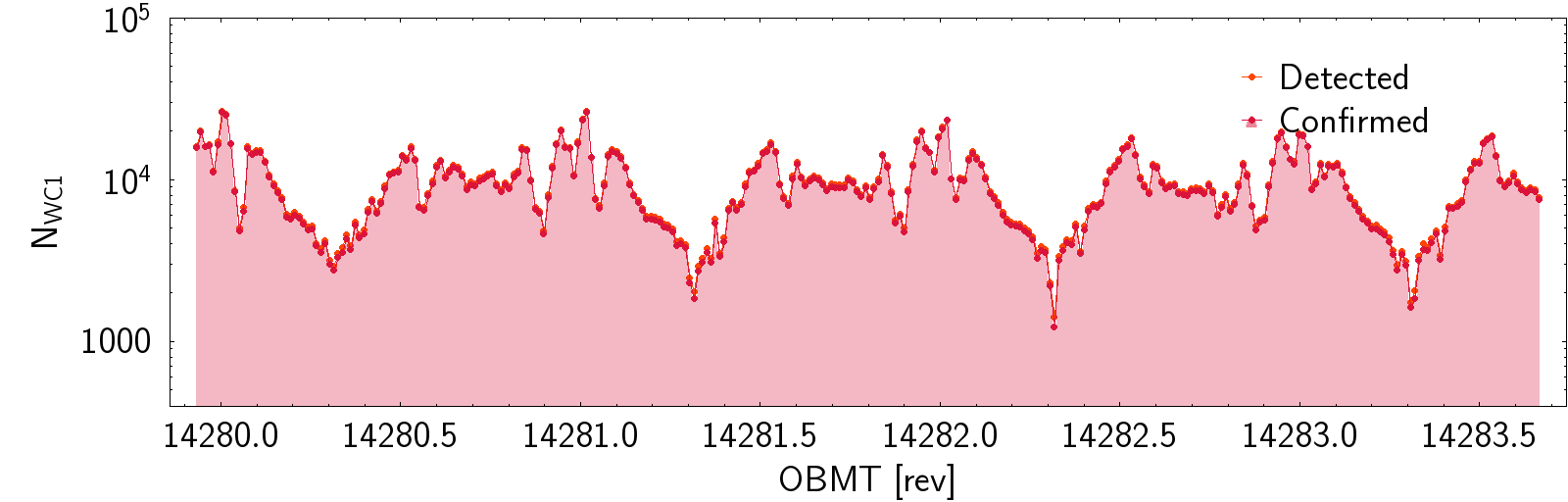}
        \includegraphics[width=0.48\hsize]{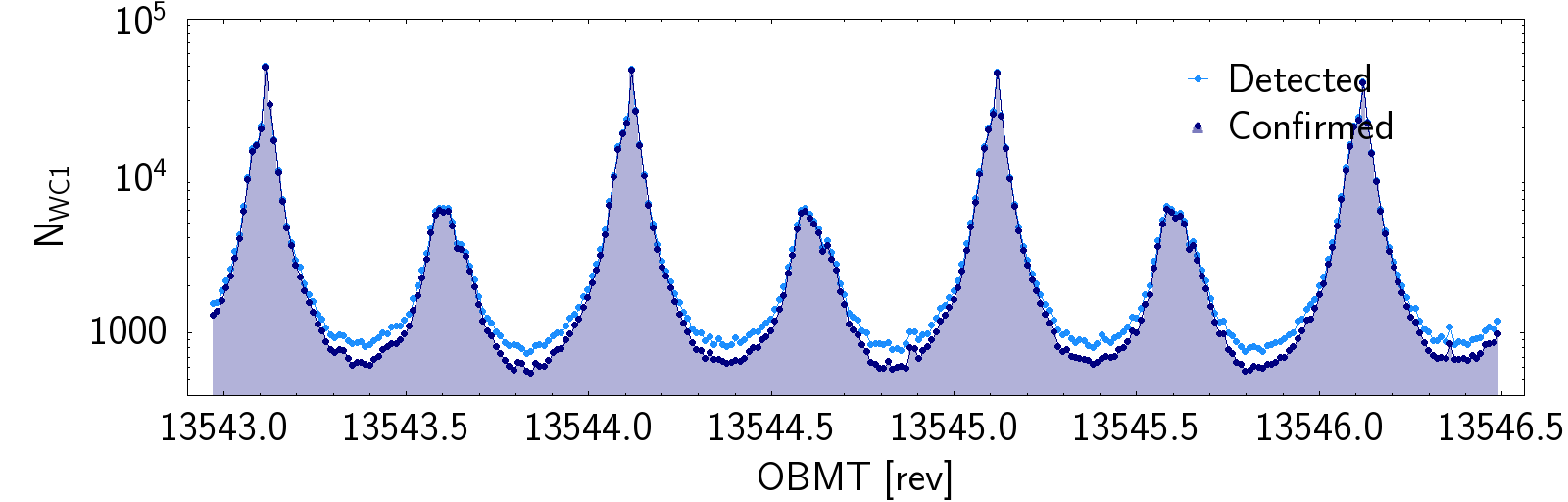}
        \includegraphics[width=0.48\hsize]{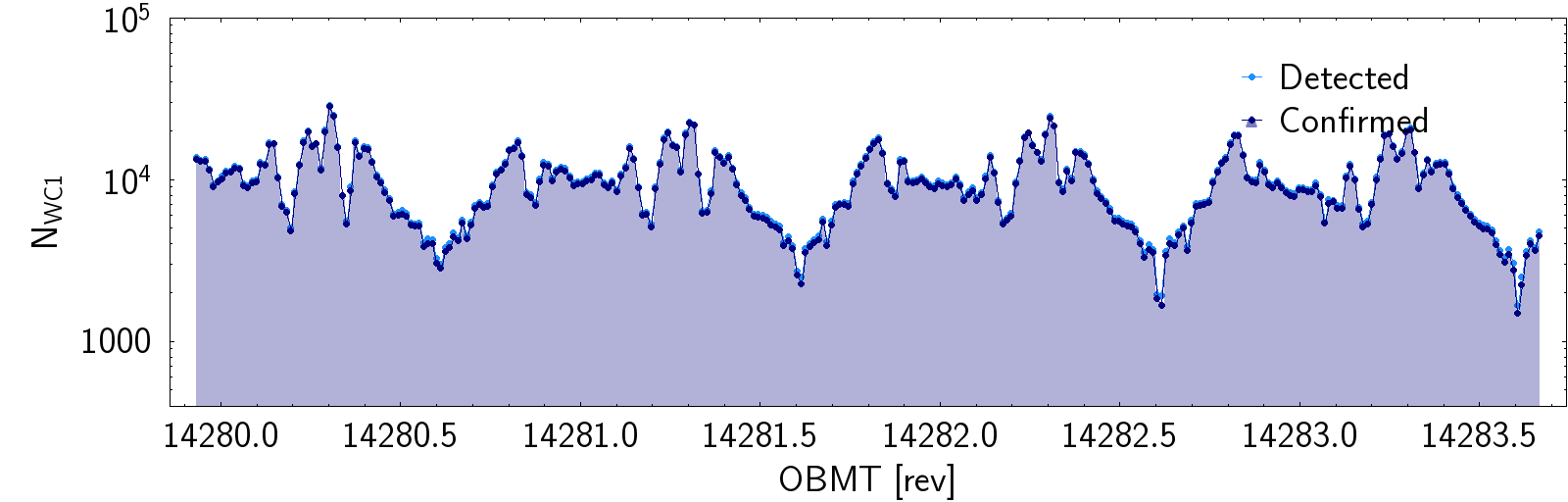}
        \includegraphics[width=0.48\hsize]{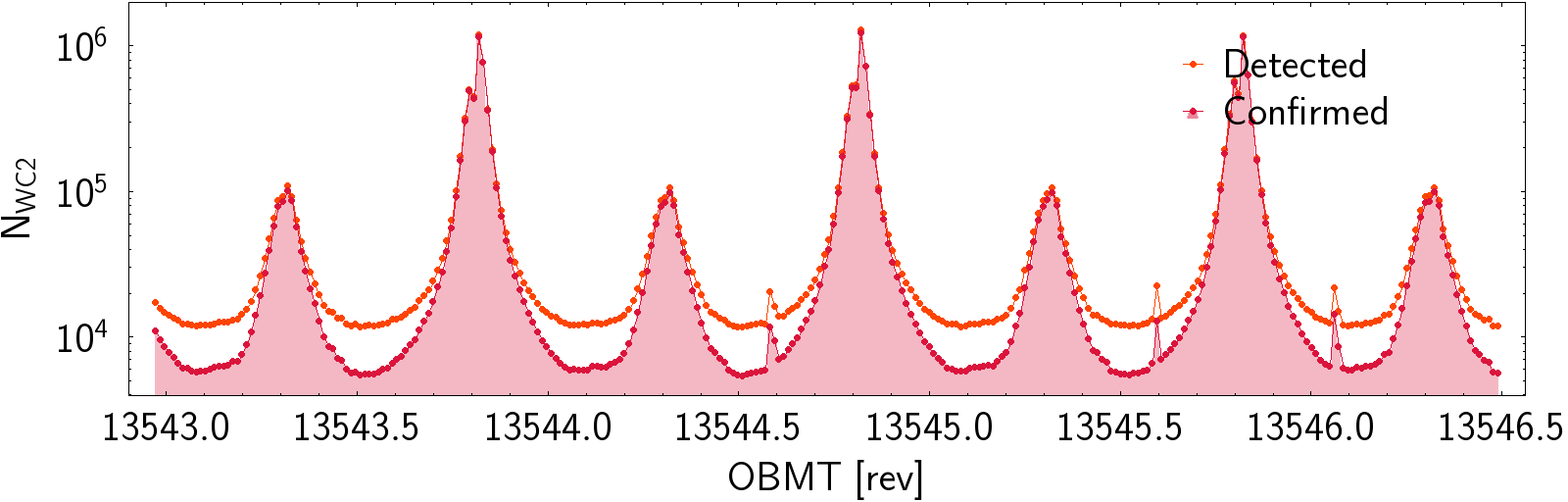}
        \includegraphics[width=0.48\hsize]{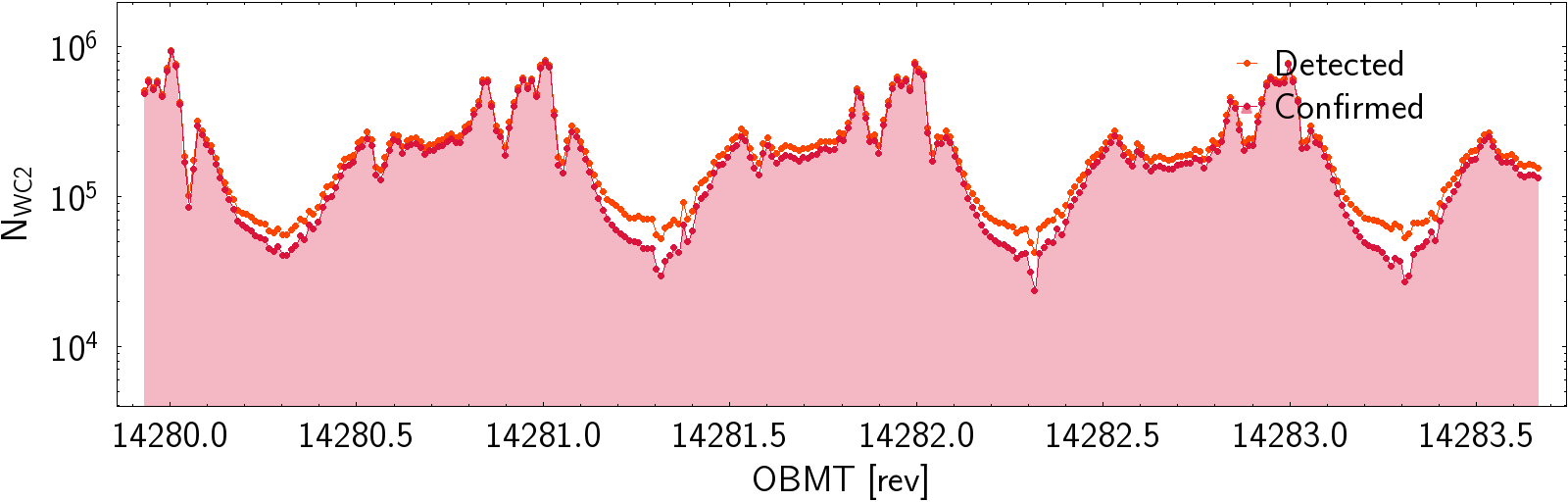}
        \includegraphics[width=0.48\hsize]{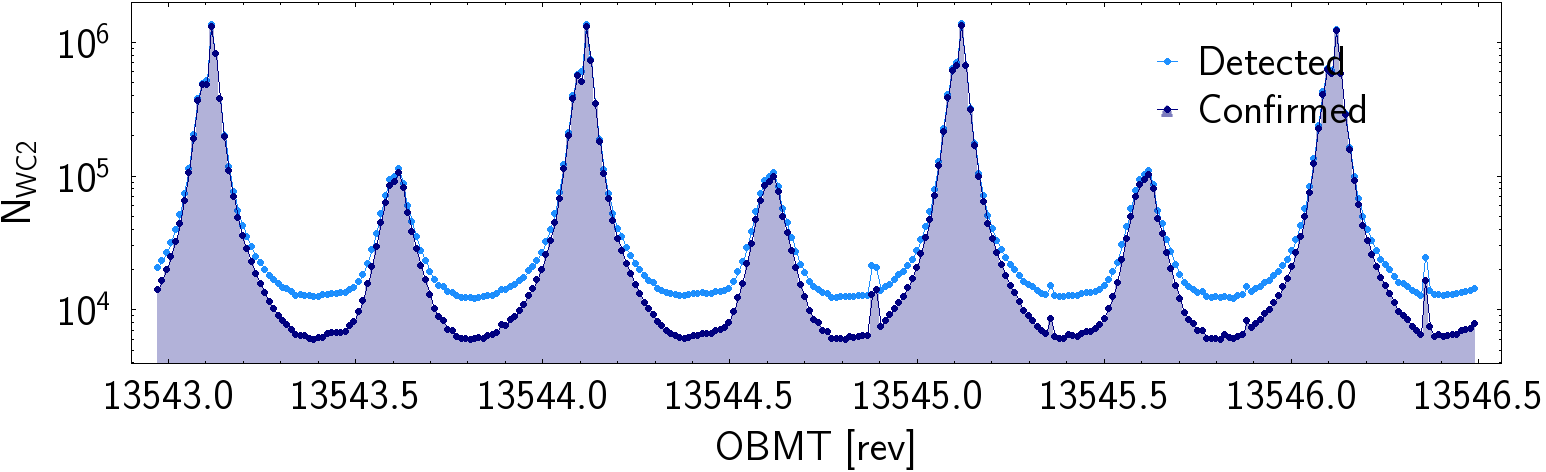}
        \includegraphics[width=0.48\hsize]{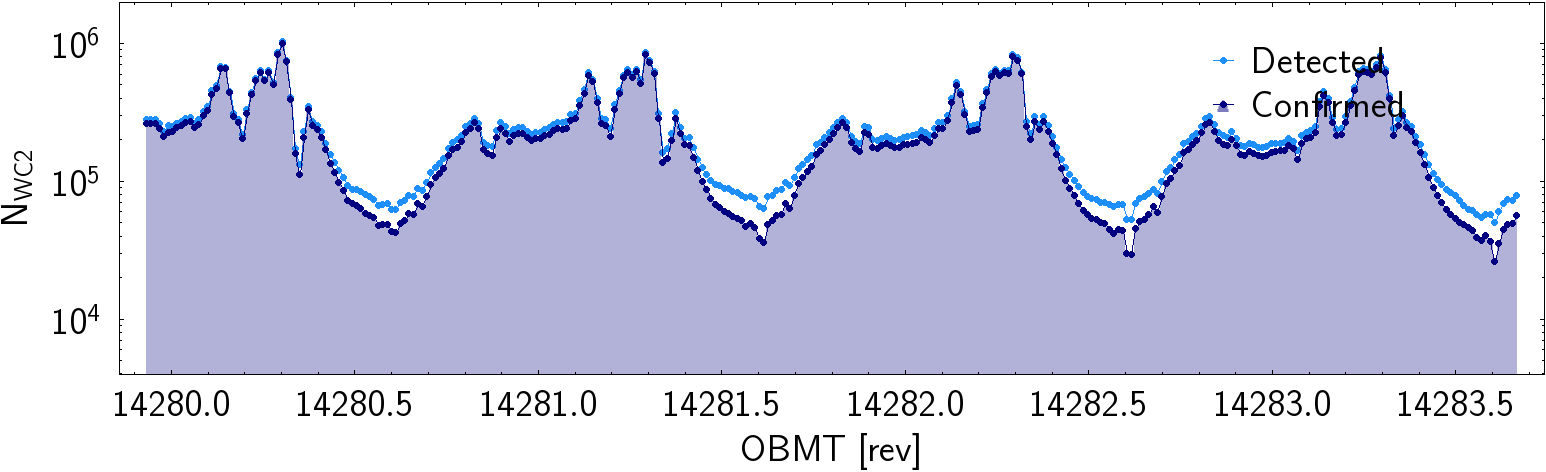}
        \caption{\small Count rates for detected objects and objects confirmed for astrometric observations of the two FL
        days shown in Fig.~\ref{fig:skycov.fig}. FoV1 is shown in red  in the first and third row of plots, FoV2 in blue  in the
        second  and fourth rows. The upper two rows show the rates for WC1, i.e. the objects in the magnitude range between $G$=13.00 and 16.00~mag,
        and the lower rows those of the fainter WC2, i.e. stars fainter than $G$=16.00~mag. The left column shows the time between 
        February 4, 2023, 11:01 UTC and February 5, 2023, 08:06 UTC, while the right column shows the time between Aug. 7, 2023, 19:42 UTC and Aug. 8, 2023, 18:17 UTC. 
        The count rates are given in logarithmic form.} 
       \label{fig:ASD.fig}
\end{figure*}

\begin{figure}[ht]
       \centering
       \includegraphics[width=\hsize]{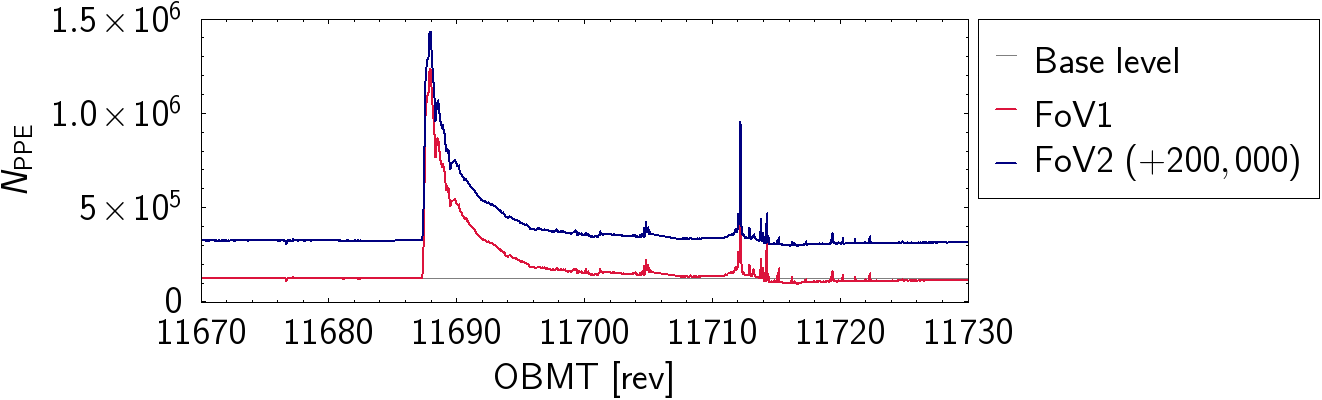}
       \caption{\small {\it Gaia} Prompt particle event   count rates during a larger solar outburst, which took place on October 30 and 31, 2021. The 
combined FoV1 count rates are shown in red and those for FoV2 in  blue. We note that in order to avoid overlap the FoV2 values have been
shifted by 200,000 counts. The grey horizontal line indicates the PPE count rate level without the disturbance (with respect to FoV1 in this plot), i.e. 
defining the base line}
       \label{fig:ASD_PPE.fig}
\end{figure}  
As mentioned in Sect.~\ref{sect:FL:IDT}, {\it Gaia} did not only transmit the science data, which contains the actual observations, 
but also a suite of metadata, called
the Auxiliary Science Data (ASD). Most relevant for the FL were the ASD2 data, which contain non-illuminated pre-scan data from the CCDs, 
the ASD4 counter data, which supply statistical information about the source detection process, and the ASD6 data, which recorded gate changes for bright
stars (see Sect.~\ref{sect:FL:ODAS} and \citet{2021A&A...649A..11R}).

In this section, we  focus on the ASD4 counters, which have provided a wealth of information about the observing process. They have presented
 for all types of
observations, i.e. astrometric, photometric, and spectroscopic, the number of objects, which were detected, confirmed, and allocated to readout windows.
The ASD4 rates also recorded the number of rejected detections, and the reason why they were discarded, 
as well as information about the inserted virtual objects.
Virtual objects are empty  windows (which have the same dimensions as the actual  windows of the various configurations)
allocated in regular intervals to yield information which could not be obtained through a regular window due to
the presence of the object and its signal. Examples
are measurements of the background, noise, or other CCD characteristics. 
The ASD4 statistics were given out separately for each of the three window classes. 
They counted incoming observations, and their values are stored and transmitted to Earth in time intervals 
of 257.6~secs, or slightly more than 4 minutes.  

Further count rates monitored by the ASD4 counters are
\begin{itemize}
\item  suspected moving objects, i.e. objects, with a large offset between the SM position and the expected propagated position in the AF1 strip, which may indicate that such
objects are nearby Solar System objects.
\item detections confirmed or invalidated for the use for the Attitude and Orbit Control Sub-system (AOCS), i.e. the system which controls {\it Gaia}'s attitude -- the direction it points to and the rotation. 
\item peculiar or malformed detections, such as saturated or faint detections, or windows without maxima.
\item the number of ASD packets in the Payload Data Handling Unit (PDHU), 
i.e. the on board storage and transmitting system.
\item the number of prompt particle events (PPE) and ripples. The former are spurious detections with a source profile sharper than that of a star, the latter
detections with a light profile shallower than that of a point source. PPE events are often caused by impact of energetic particles, ripples are mainly caused by stellar diffraction spikes, but also by non-stellar real objects, for example
light reflections. 
\end{itemize}

The ASD4 counter information provides insight into the source density distribution with time,
and into external events, such as solar eruptions, 
and allows to monitor the overall detection efficiency.  Fig.~\ref{fig:ASD.fig} shows the number of sources detected and confirmed as 
valid objects for two different FL days. The left set of plots in the figure shows the situation when {\it Gaia}
was scanning the sky at a moderate inclination with respect to the Galactic plane, 
resulting in a mostly low source density with 
passages through the Galactic Plane (GP) as two peaks of high source density, with the higher peak being the one closer to the 
Galactic Centre (GC). The right column of plots shows the source densities during a  Galactic Plane Scan (GPS), during which
{\it Gaia} scanned the sky at a very low inclination angle with respect to the GP. 
Here, no obvious peaks are seen and the source densities are 
generally high. Due to the much higher number of observations during a GPS, the strain on the source allocation and data downlink was much 
higher.

The 106.5$^\circ$ Basic Angle between the two apertures is also clearly seen in the plots, with the features seen in FoV2 being 0.29~rev later 
than the same features observed with FoV1.  Fig.~\ref{fig:ASD.fig} shows the detection counts
 for window classes 1 and 2 (for WC0 there are no 
counts of confirmed objects, WC0 objects, i.e. the brightest, are guaranteed to be allocated). Overall the source densities are very similar, 
apart from the absolute number, which is much higher for the fainter WC2 category. There are, however, some subtle differences, notably there are more
extra peaks in WC2, which are caused by the passage of small densely populated patches in the sky (e.g. globular star clusters), which mostly 
contain fainter stars. The number of detected objects, which were not confirmed to be valid sources is higher in WC2, but relatively independent 
of the source density within each window class.  

Fig.~\ref{fig:ASD_PPE.fig} shows the output of the PPE counters during a large solar storm. Here the rapid rise and slower decline after the maximum, 
typical of such events, can be clearly seen.  It is also evident that there is no time delay between the signal of the two apertures, since these 
particles do not follow the optical path of the mirror assembly but hit both SM strips simultaneously.

\subsection{The housekeeping data (HK)}\label{sect:FL:HK}
The final source of diagnostic data for the FL was housekeeping data. This mostly consists of temperature and thrust measurements as well as information on the filling
status of the Payload Data Handling Unit (PDHU).
This information was especially important during Galactic Plane Scans, i.e. when the strain on the PDHU was largest, 
due to the high number of
observations.

The temperature data allowed the monitoring of the thermal stability of various components of the spacecraft. 
Given the strong stability requirements of an astrometric
mission like {\it Gaia}, it was necessary that the temperatures were stable too. 
The {\it Gaia} FL had access to the temperatures for the FPA, the mirrors, the propellant, 
and since 2023 of the Deployable
Sunshield Array (DSA). Temperatures were measured using thermistors, 
which have a high measurement cadence, but only a very low thermal resolution of 0.1~K. A higher resolution was 
provided by averaging the large number of individual measurements. 
The caveat of this approach is that aliasing effects can occur in some cases in various components of the spacecraft. 
Usually the measurements were binned into 5-30 min time bins. 

The thrust output of the cold gas thrusters, which had ensured the rotation of {\it Gaia} and are controlled by 
the {\it Gaia} AOCS, was sampled in a similar way. Excess use of thrust output, in order
to correct for rate excursions, shows up as peaks in the thrust output patterns. 

\subsection{Trends}\label{sect:FL:Trends}

For many diagnostics and analyses described in the previous sections, the output was also given out over longer time periods, allowing the investigation of longer-term trends. 
In some cases this was done at a lower temporal resolution (e.g. one value per FL day) than in the main output. 
This allowed the quick evaluation of longer-term trends, and some examples  are shown in Section~\ref{sect:results}. 
\subsection{Other data}\label{sect:FL:Other_data}
As described above, the data generally made available to the FL had been pre-processed to a certain degree,
 for example organised into bins covering a certain amount of time, arranged into histograms, fitted, or smoothed. 
This was optimised for the daily inspection. However, certain findings, which required a more detailed study, also
needed the use of additional data from other DPAC systems.
 Therefore, the FL-team had access to all {\it Gaia} data, usually upon request, 
given the large volumes of such datasets. In the following sections, some examples for the usage of non-FL data are presented, 
especially in Sect.~\ref{sect:results:RE}.
\section{Findings}\label{sect:results}
As the FL was a diagnostic tool, there are actually no scientific results in a conventional sense (e.g. observational evidence for
a hypothesis).  
What we   discuss in this section are our findings in terms of the 
overall data quality, the long-term evolution of this overall data quality, and the impact of disturbances on the payload 
and on the quality of the scientific data as well as the post-disturbance recovery process.
In other words, the fewer abnormalities FL had found, the better, as this means that everything was nominal and stable.
FL looked for any aberrations from the norm, whether previously anticipated or completely unexpected. 
All peculiarities found and identified were
investigated further. Additionally the longer-term trends have also been monitored. All of these findings were reported to other relevant parts of 
DPAC or ESA, and formed part of the basis for decisions on any actions needed to keep up the consistent data quality. These reports were the daily 
qualifications, the semi-regular FLS-reports, or when something urgent had occurred, alarming the relevant groups directly. Furthermore,
the insights FL had obtained formed vital input for discussions of issues, and aided decisions about required actions,
 if needed.  
In the following, as typical output of the FL, we describe and discuss long-term trends 
(Sect.~\ref{sect:results:ambient}), and examples of 
disturbances, both in general terms (Sect.~\ref{sect:results:RE}) by the example of OGA2$-$NSL rate excursions, and particular noteworthy events (Sect.~\ref{sect:results:MM}) caused by micro-meteoroid impacts. 
\subsection{Long-term and seasonal evolution of the satellite and dependences on the ambient conditions}\label{sect:results:ambient}

{\it Gaia} was on a 3D-Lissajous orbit around the Lagrange Point~2 (L2) of the Sun-Earth system. 
One of the most important reasons for this choice is the thermal stability of this region,
which co-rotates with Earth on an orbital radius about 1.5~million km larger than that of the Earth. For an astrometric mission, such as {\it Gaia}, a stable ambience, especially
in terms of temperature is paramount. Only this way, the targeted high precision could be guaranteed. 
Every correction of systematic effects acting on the 
data introduces a source of inaccuracy, which needs to be accounted for by modelling the effect. Given the imperfection of
this type of correction, the fewer and smaller effects to be dealt with, the better.
 Thus the L2 is almost ideal for such missions.

However, even in the L2 region, ambient conditions are not completely stable. 
As the L2 point belongs to the unstable Lagrange points, 
any spacecraft must be nudged regularly to stay near the L2. These
manoeuvres, called station-keeping manoeuvres, exert a disturbance on the stability of the payload. Moreover, the L2 itself lies in a zone where the Sun is perpetually obscured by Earth. As this would exclude the use of solar panels 
for power generation, and cause frequent thermal disruptions, this presents the main reason for the choice of the Lissajous-type orbit. 
This implies a variation of the distance between {\it Gaia} and the Sun, 
depending on the position of the satellite on its orbit. As Earth itself is on
an elliptic rather than a circular orbit, with the distance Earth-Sun changing by approximately 5 million km annually, seasonal changes have an even larger impact 
on the thermal environment of {\it Gaia}. The variation in distance due to the radial boundaries of the Lissajous orbit
in the direction radial to the Sun was about 300,000~km, i.e. about 6\% of the seasonal changes. 

Further systematic influences on {\it Gaia} were solar radiation, the stellar source density, stray light by bright objects close to the fields of view, and 
time dependent changes in the instrument itself, such as ageing of components, for example the mirror throughput and the detector response, and degrading focus. Individual, 
major interventions into the satellite system could also have drastic effects on the performance, unwanted or wanted. In this section, we  
show and discuss how these sources of instability impact the measurements and how the FL helped to improve the scientific quality of the {\it Gaia} catalogues.

Before we start off, 
a cautionary note is required. Firstly, as 
the mission progressed, and with it our understanding of the instrument and its measurements, the need for additional
diagnostic output arose. Therefore for some of the parameters discussed in this section, we do not have the corresponding data from the mission start. 
Secondly, our diagnostic tools have been adapted and improved over time, so that in some cases the results are not always comparable in a straightforward way, 
i.e. suitable for a general science 
publication. For both reasons the coverage of the earlier parts of the mission is more limited than that of the latter 
parts. We do, however, strive to give an as coherent analysis of the long-term effects on {\it Gaia} as possible.

\subsubsection{Significant events}\label{sect:FL:Trends:sign_events}

\begin{table*}
       \caption{\small \label{tab:sign_events.tab} List of significant events}
       \begin{center}
       \begin{tabular}{rrrp{9.5cm}}
 \hline
               Date   & OBMT &  Type & remarks \\
               \multicolumn{1}{c}{yyyy-mm-dd} & \multicolumn{1}{c}{rev} & & \\
               \hline
               2014-09-23 &  1317.30  & De-contamination & --- \\
               2014-10-14 &  1443.96 & Re-focus & --- \\
               2015-06-04 &  2330.80 & De-contamination & --- \\
               2015-08-03 &  2574.66 & Re-focus & --- \\
               2016-08-24 &  4112.80 & De-contamination & --- \\
               2017-06-05 &  5263.26 & Safe-mode & anomaly in the control electronics of the micro propulsion system \\
               2017-07-11 &  5406.21 & Re-focus & --- \\
               2018-02-18 &  6293.37 & Safe-mode & permanent failure of primary transponder \\
               2018-12-03 &  7447.70 & Re-focus & --- \\
               2019-07-16 &  8345.76 & large manoeuvre & Whitehead Eclipse Avoidance Manoeuvre (WEAM) \\
               2024-04-02 & 15235.23 & micro-meteoroid-impact & caused persisting massive stray-light peaks, mostly in FoV1, lasting 40~minutes during each 6-hour revolution\\
               2024-05-15 & 15406.55 & electronic failure & The AF1\_3 detector remained permanently inoperable\\
               2024-06-13 & 15522.88 & Re-focus & --- \\
               2024-10-07 & 15987.80 & Radiation? & aggravation of magnitude and CCD strip coverage of the stray-light peaks, starting on 2024-04-02, possibly caused 
               by solar activity \\
               \hline
\end{tabular}
\tablefoot{These are  events that left a sizeable imprint on the long-term output of the quantities discussed in Sect.~\ref{sect:results:ambient}.
These events are also shown in Figs.~\ref{fig:Long_term_Temps.fig} to \ref{fig:BAM_BAV.fig}, where applicable.
The dates and the OBMT times given refer to the start of the events.}
\end{center}
\end{table*}

Before we start off looking at the long-term evolution of various parameters in Sect.~\ref{sect:FL:Trends:temps} to 
\ref{FL:Trends:BAV}, we need to discuss and define events, which have had a profound longer-term influence onto the payload. These
  manifested themselves in the data, which are discussed in this section. The times of their occurrence are marked as lines in the plots in this section.

Natural influences result from either micro-meteoroid impacts or enhanced solar activity such as 
coronal mass ejections. 
In most cases, the long-term impact of such events is insignificant. 
However, some more noteworthy micro-meteoroid impact occurrences are discussed in Sect.~\ref{sect:results:MM}. The event of April 2, 2024, did have the most drastic
consequences for the operations in the entire mission, thus this is a significant event. This also applies for the severe aggravation of the 
periodic stray-light peaks, introduced by this impact, occurring on October 7, 2024. The cause of this is unknown, but could possibly be linked to 
a coronal mass ejection around this time. 

Most planned events were station-keeping manoeuvres, which were regularly scheduled to maintain the orbit. 
While these did occasionally
have a certain influence onto operations, imprinting themselves in the trend plots in this section, 
they are too common and too regular to be called significant events. Larger Manoeuvres,
such as the Whitehead Eclipse Avoidance Manoeuvre (WEAM), conducted on July 16, 2019, did
 have a larger and longer-term influence, 
thus being a significant event in this context. The same applies for decontamination campaigns,
 which severely changed the thermal stability, and for
refocusings which changed the geometry of the optics.

Finally another issue, which could have had substantial
 consequences for the payload were internal failures. Among these, safe mode events were 
the most dramatic emergencies, as the spacecraft triggered a safe mode when it perceived itself to be in danger. 
There were several levels of safe modes, depending on the gravity of the situation, 
as assessed by the spacecraft software.
In two instances, one occurring in June~2017 and the other in February~2018 (see Table~\ref{tab:sign_events.tab}), 
the spacecraft responded by automatically slewing from the nominal Solar Aspect Angle 
(i.e. the angle between the line Sun-{\it Gaia} and {\it Gaia}'s rotational axis) of 45$^\circ$ to 0$^\circ$. 
This means that the DSA was facing directly to the Sun. 
This is the deepest and most momentous level of safe mode possible.  
Additionally, all electronics not absolutely vital for low level operations were switched off. Both of these measures
severely disturbed the thermal equilibrium. 
Most other temporary outages of an electronic component on board have had some short-term influence, mostly thermal, but not enough to
show a long-term imprint on the parameters discussed in this section. The electronic failure, 
leading to the permanent in-operability
of the AF1 detector in Row~3 (AF1$\_3$; see Fig.~\ref{fig:FPA.fig}), led to a persistent change of operations.

\subsubsection{Temperatures}\label{sect:FL:Trends:temps}

\begin{figure*}[ht]
       \centering
        \includegraphics[width=\hsize]{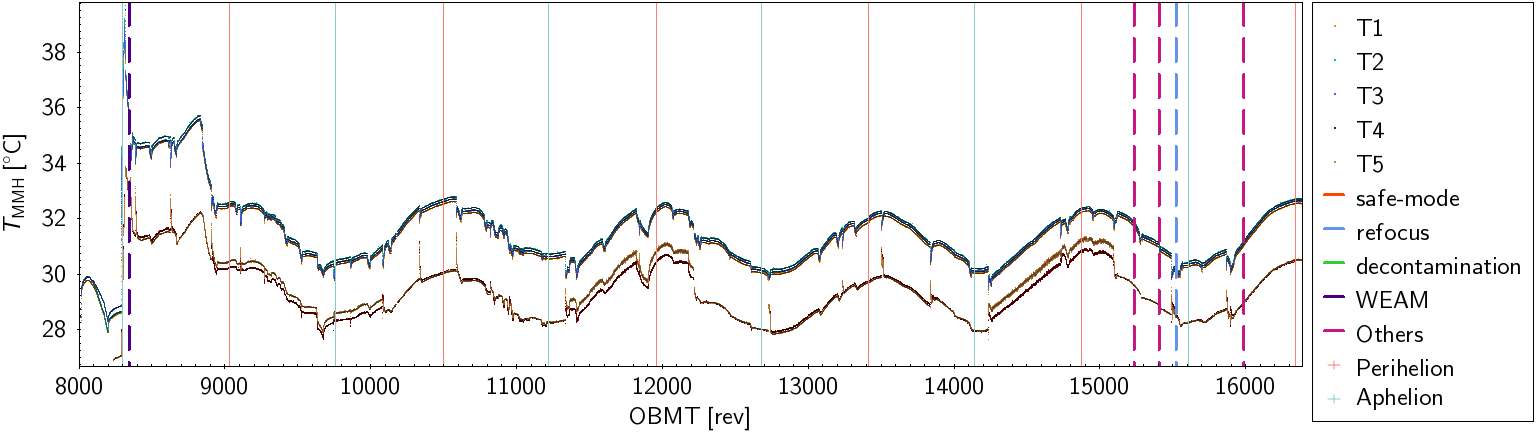}
        \includegraphics[width=\hsize]{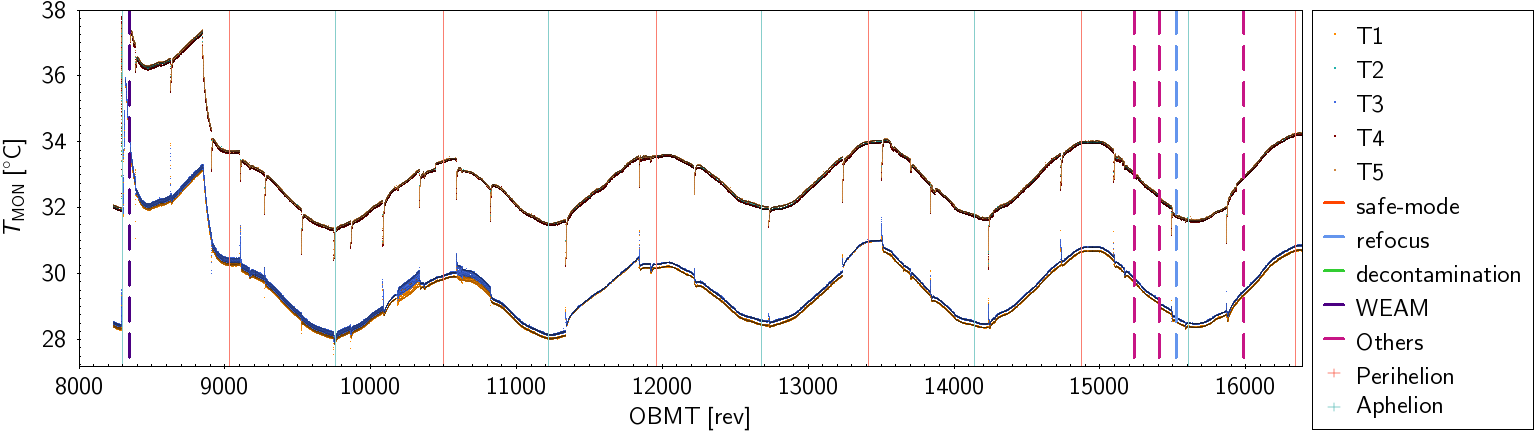}
        \caption{\small Temperatures of the propellant after OBMT=8000~rev. The upper panel shows the fuel temperatures and the lower one the
         oxidiser values. For each propellant component there are five thermistor type sensors, three in the upper half and two in the lower half of each tank.
         The significant events listed in Table~\ref{tab:sign_events.tab} as well
        as the times of the extrema of the distance between Sun and Earth are also  indicated (see legend).}
       \label{fig:Long_term_Temps.fig}
\end{figure*}

\begin{figure*}[ht]
       \centering
        \includegraphics[width=\hsize]{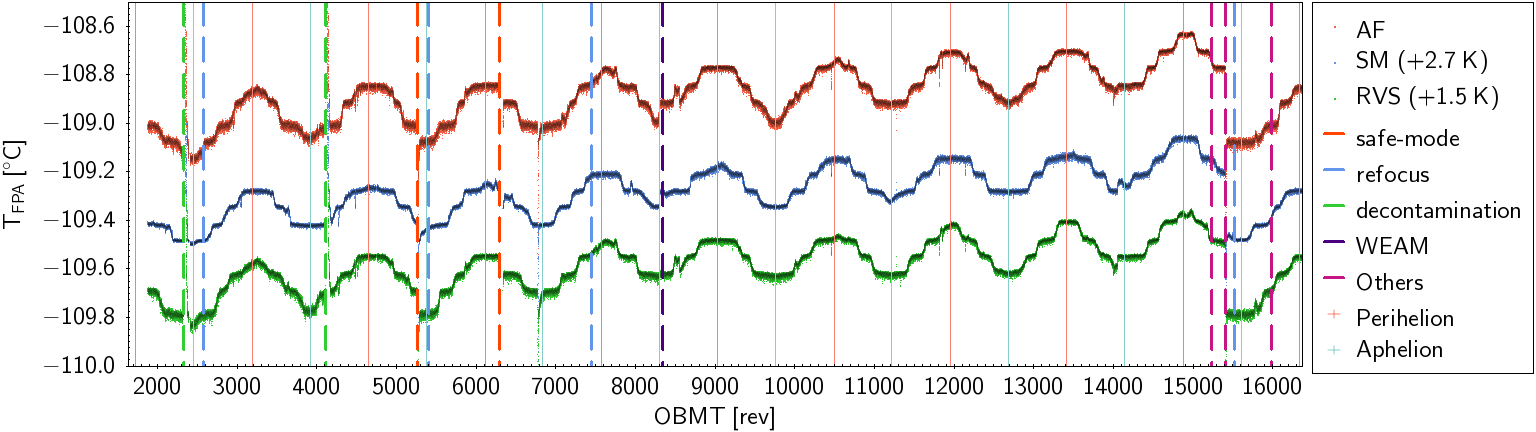}
        \caption{\small Temperatures of the focal plane array. Shown are the readings of the thermistors 
         closest to the AF (red points), SM (blue), and RVS (green) parts of the FPA. 
As in the previous figures, the significant events listed in Table~\ref{tab:sign_events.tab} as well
        as the times of the extrema of the distance between Sun and Earth are also indicated (see legend).
The SM and RVS 
         temperatures have been shifted by +2.7 and +1.5~K respectively in order to fit into the same plot.}
       \label{fig:T_FPA.fig}
\end{figure*}
\begin{figure}[ht]
       \centering
        \includegraphics[width=\hsize]{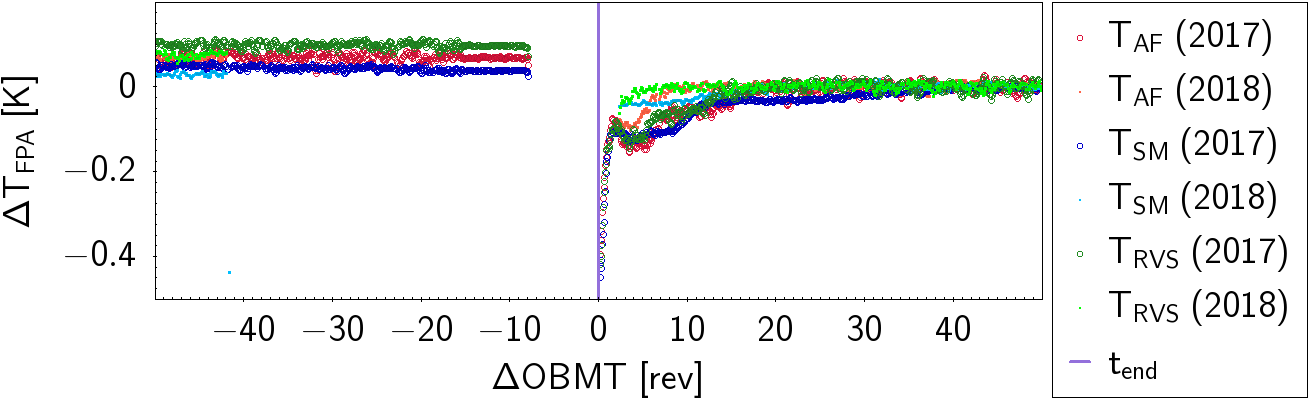}
        \includegraphics[width=\hsize]{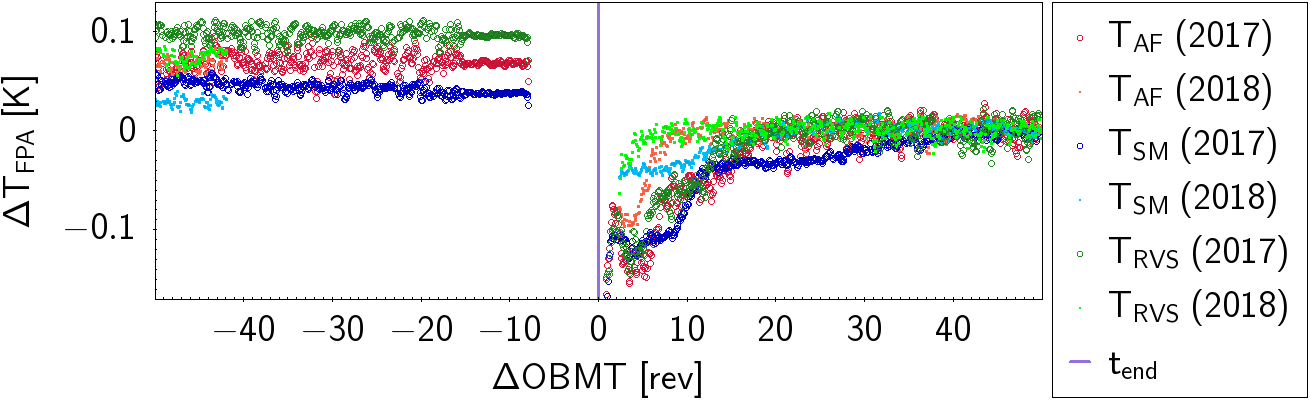}
        \caption{\small Temperatures of the focal plane array during the safe-mode events of 2017 and 2018.
         Shown are the readings of the thermistors
         closest to the AF (in red), SM (in blue), and RVS (in green) parts of the FPA. The values
         corresponding to the 2017 safe-mode are depicted as open circles in darker colours, while the 2018 data are 
         shown as smaller dots  in lighter colours. The OBMT had been shifted for both events so that the 
         official end of the event is at $\Delta$OBMT=0~rev (also marked by the vertical line). 
         The temperatures have been shifted, so that they match 
         at $\Delta T$=0~K at the stability level after the event. The lower panel is a zoom-in of the upper panel, 
         giving better access to the more subtle temperature recovery after the initial warming.}
       \label{fig:T_FPA_safemode.fig}
\end{figure}

The principal factor affecting {\it Gaia} was its temperature. In the case of {\it Gaia} the temperature was
mainly influenced by the influx of heat from the Sun and by internally induced heating (e.g. of 
the propellant or the focal plane assembly). This  was caused intentionally by changing the output of the on board heaters, for example in preparation of a manoeuvre, 
or unintended by the amount of heat produced by the
electronics which changes in relation to the source density on the sky. The absolute 
exposure of {\it Gaia} to sunlight was ruled by the distance
to the Sun, which varies on a seasonal scale, and by variations in radial distance of the Lissajous orbit of {\it Gaia}
around the L2 point. Secondary solar heating effects arose due to the scanning law, with different parts of {\it Gaia} being 
more exposed to the Sun during a rotation. {\it Gaia} was designed with this in mind, as the actual scientific
payload was shielded from direct sunlight by the deployable shield array (DSA), 
the circular Kapton shield attached to the 
base of the satellite, which has a radius of 11~m. While this kept away the vast majority of solar flux from the
sensitive parts of {\it Gaia}, the shielding was not perfect. On-board heating systems were turned on to heat the propellant tanks
prior to station-keeping -- and other manoeuvres, or the mirror assembly during the initial decontamination
of the optical surfaces after launch on further occasions as the contamination of
the mirrors by water ice built up again. In contrast to the seasonal solar distance variations, these instances are short, and less
regular.

The propellant used for the manoeuvering thrusters 
is stored in spherical tanks near the centre of the base of the spacecraft. Each of these was equipped with
five thermistor-type temperature measuring devices, three of which are located in the upper half and two in the
lower half of the tanks.
Figure~\ref{fig:Long_term_Temps.fig} shows the temperature trends of the propellant,\footnote{the propellant consists of
two components, the fuel and the oxidiser} the upper panel those of the 
fuel component, and the lower panel the oxidiser temperatures. The seasonal trend can clearly be seen,
with the highest values reached in early January of each year, i.e. at perihelion, and the lowest values in
early July, when Earth, the L2 and thus {\it Gaia} are near the aphelion. 
The amplitude is quite substantial, $\sim1$~K over the year. The figure also shows shorter-term variations 
in the temperatures of one or both components of the propellant. Those, which were seen in one of the tanks only, were 
usually caused by fuel movement, when bubbles of fuel float around in the tank. By contrast, those which occur
in both the fuel and oxidiser temperatures were  induced on purpose, for example by heating up the tanks prior to one of 
the 62 station-keeping manoeuvres,  by different uses of the thrusters, or by another event such as a safe mode. 
Some of the values tend to oscillate with a heliotropic period,\footnote{Heliotropic means following the Sun, and in the 
context of Gaia, this reflects the fact that due to the orbit of Gaia around the Sun, the orientation of the 
spacecraft with respect to the Sun was not exactly the same after one complete Gaia revolution, 
but a short time thereafter. This is the same reason that the duration of a solar day is different to a sidereal day}
 which in some instances led to
noticeable attitude rate excursions, which are discussed in Sect.~\ref{sect:results:RE}.

Similarly, the temperatures measured at the focal plane assembly, shown in Fig.~\ref{fig:T_FPA.fig}, and at the mirrors
were subject to the seasonal variations, albeit at a much smaller scale. They were also affected by outages in the CCD 
control electronics (visible in the figure as narrow peaks pointing mostly into negative direction). In such instances,
components were partly switched off, which implies less consumption of electrical power on the short-term, 
correlated with less heat production, until everything was cycled up to nominal operation again. In contrast to 
the propellant temperatures, with their much larger amplitude, the FPA and mirror temperature curves show pronounced 
steps. This is an averaging effect caused by the low temperature resolution of the thermistor devices of only 0.1~K,
which at the low level of variations for these temperatures could only partially be compensated by the high readout
frequency. As seen in Fig.~\ref{fig:T_FPA.fig}, the annual peak-to-valley amplitude is only $\sim$0.1~K, i.e. very
similar to the thermistor's thermal resolution. 

Fig.~\ref{fig:T_FPA_safemode.fig} shows the impact of two safe-mode events, which occurred in June 2017 and
February 2018, onto the FPA. As can be seen in the upper panel, the temperature drop at the end of each event was quite
significant, of the order of 0.4~K. While most of this was recovered quite quickly, there was  a residual relaxation
process, which generally takes more than 10, in some cases up to 50 revolutions to stabilise. The lower panel
of Fig.~\ref{fig:T_FPA_safemode.fig} zooms in on this latter more gradual stabilisation. It is to be noted that the 
trends of the 2018 safe-mode stabilised much sooner in all three thermistors than in the case of the 
earlier one in 2017, despite the later one having been of much longer duration. 
One reason is that  
the recovery procedures had been adapted to minimise the thermal impact on the payload based on the analysis
 of the temperature trends after the 2017 event. 
This is an example, of how FL-investigations have contributed to minimise scientific losses in such an unplanned major disruptive
event.

To summarise, the on board temperatures were influenced by several effects. The most pronounced and ubiquitous is
 the seasonal trend by the yearly changing distance of {\it Gaia} from the Sun. Others were only seen in the temperatures
of certain components, and may be caused by several phenomena, as explained in this section. It becomes clear that even
in such a thermally stable environment as the Sun-Earth L2 region, thermal influences on a spacecraft are an
important factor. As we show in the following, these temperature variations do have effects on other parameters.    

\subsubsection{The Cram\'er-Rao lower bound}\label{sect:FL:Trends:CRLB}

\begin{figure*}[ht]
       \centering
        \includegraphics[width=\hsize]{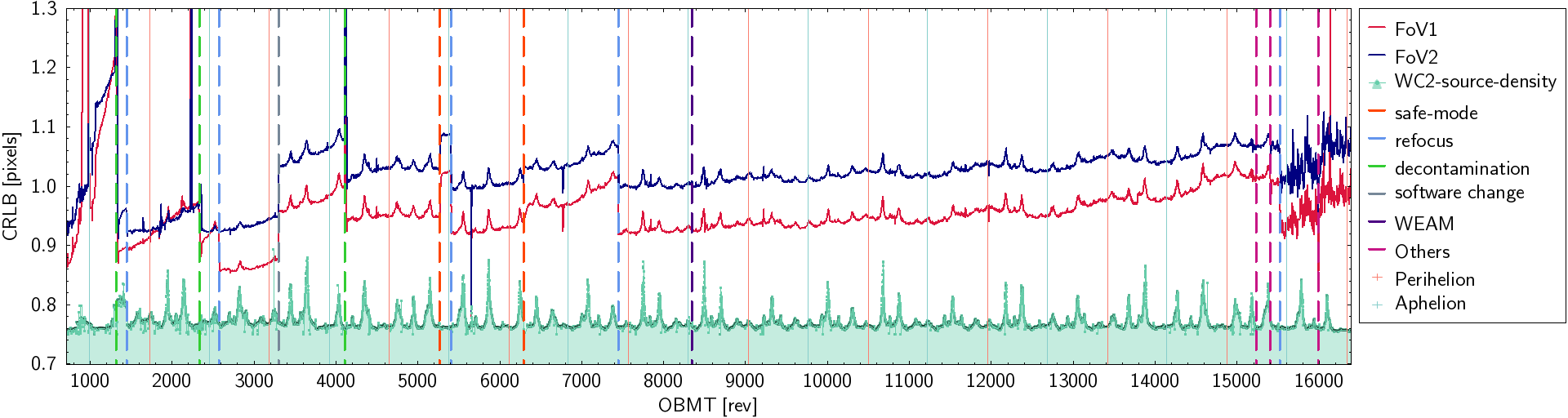}
        \includegraphics[width=\hsize]{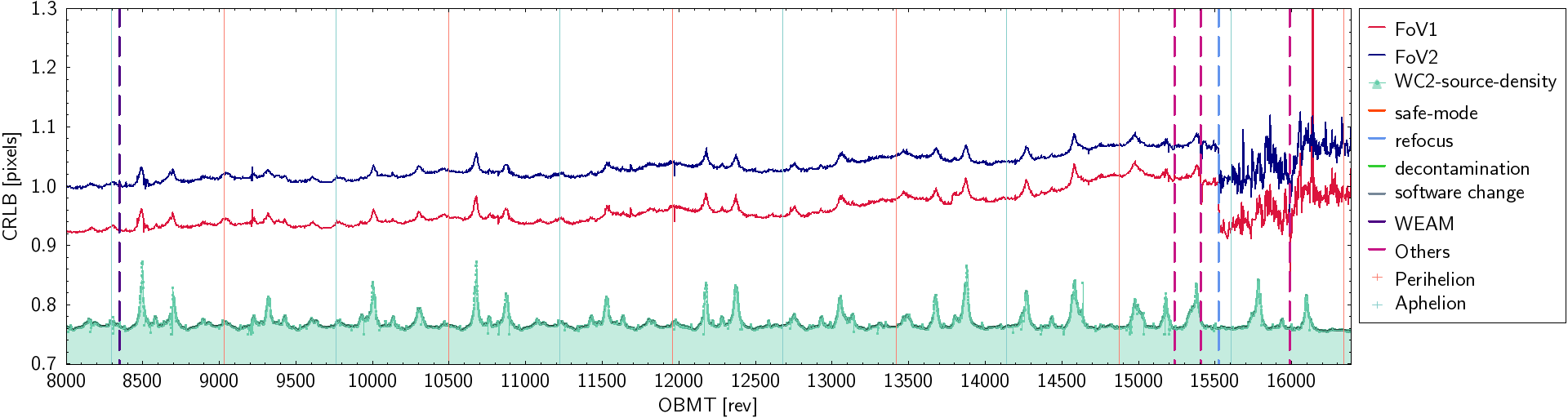}
        \caption{\small Evolution of the Cram\'er-Rao lower bound (CRLB) during the mission. The upper panel shows the complete mission from
         2014 to early 2025. 
As in the previous figures, the significant events listed in Table~\ref{tab:sign_events.tab} as well
        as the times of the extrema of the distance between Sun and Earth are also  indicated (see legend). A change in the way the CRLB is calculated is also indicated by a grey vertical
line.
The lower plot is a zoom-in on both
coordinates, highlighting the time range after OBMT=8000~rev, where there are no large disturbances and interferences into the payload. Here, to indicate the
seasons, both the aphelion and perihelion times of Earth are indicated as vertical lines. In both plots the number of observed sources of Window Class 2 (binned to
1~rev) are also shown.}
       \label{fig:Long_term_CRLB.fig}
\end{figure*}

\begin{figure}[ht]
       \centering
        \includegraphics[width=\hsize]{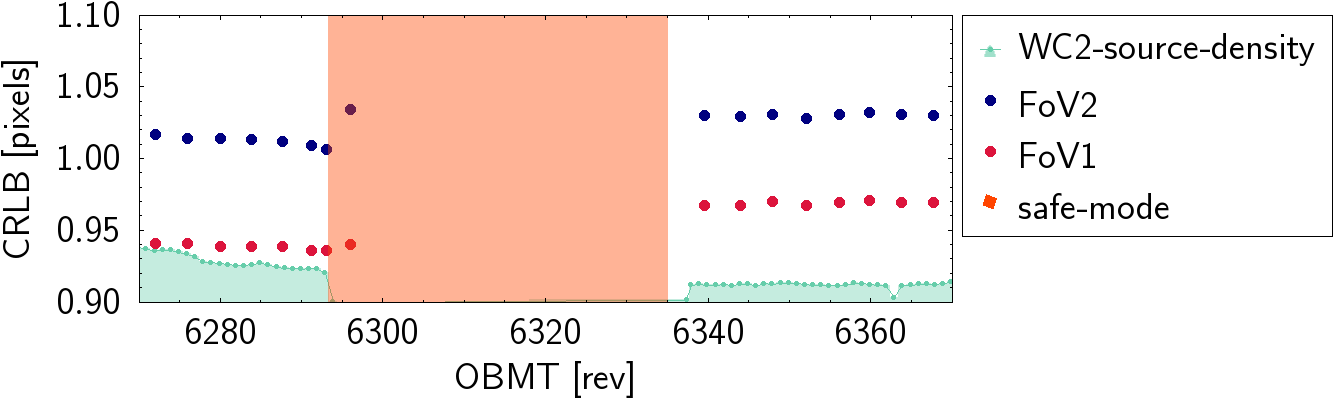}
        \includegraphics[width=\hsize]{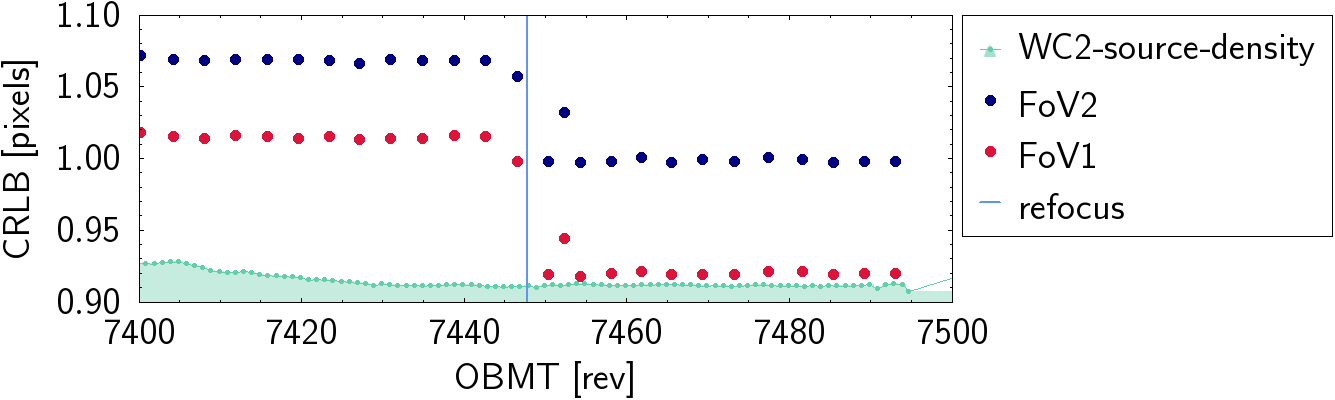}
        \caption{\small Behaviour of the Cram\'er-Rao lower bound (CRLB) during a safe mode (upper panel) and a refocusing (lower panel). The time and duration of the
        are indicated respectively by the shaded area and the vertical line.
In both plots the number of observed sources of Window Class 2 (binned to
1~rev) are also shown} 
       \label{fig:Long_term_CRLB_event.fig}
\end{figure}

One important criterion concerning the quality of individual measurements with a given instrumental set-up is the Cram\'er-Rao lower bound (CRLB; see \citet{Cramer1945} and \citet{MR15748}). While this concept had been developed
in general information science, it has since been adapted for the more specific use for astronomical observations, 
i.e. for optical point-like or near point-like sources (see e.g. 
\citealt{2014PASP..126..798M} and \citealt{2017A&A...606A..27B}). In our context, the CRLB is the minimum-achievable variance of
the AL-location estimated from a Gaia observation that can be achieved by any unbiased maximum-likelihood estimator.
A higher value for the CRLB would indicate a degraded 
signal to noise, which in our context could be caused, for example, by  
a worse focus, higher background level, lower optical throughput, or more detector noise. As the CRLB is highly dependent on the signal strength, i.e. the object magnitude,
and measurements come from sources of a wide range of brightness, the individual CRLB values have been normalised to 
a common flux.  Fig.~\ref{fig:Long_term_CRLB.fig} shows the evolution of the CRLB during the complete mission.
Additionally the times of decisive interventions into the spacecraft and the WC2 source detection counts are shown in this figure. The lower plot shows the second
half of the mission, which had been more quiet in terms of drastic changes than prior to about early 2019. 
The impact of these interventions can clearly be seen in Fig.~\ref{fig:Long_term_CRLB.fig}.
The two safe-mode events (see Table~\ref{tab:sign_events.tab}), which caused {\it Gaia} to be moved to a solar aspect angle of 0$^\circ$ instead of the usual 45$^\circ$
drastically altered the differential thermal exposure of the satellite. 
In the aftermath of these events, the CRLB was significantly larger than before, mostly caused by
a permanently degraded focus. Therefore the focus was adjusted shortly after thermal equilibrium had been reached. 
Planned interventions, such as refocusing and decontamination measures,
improved the images, thus lowering the CRLB values, which is also evident from the figure. 
In one case, a strong change in the CRLB was not caused by anything on board, but by a change
of the computation of the CRLB, indicated by a grey vertical line in Fig.~\ref{fig:Long_term_CRLB.fig}  -- thus the absolute values of the CRLB before and after this change, which was 
implemented at OBMT=3,300~rev, should not be compared. Figure~\ref{fig:Long_term_CRLB_event.fig} shows the evolution of the CRLB during both a safe-mode and a refocusing in detail.

A dependence of the CRLB on the source density is also evident in Fig.~\ref{fig:Long_term_CRLB.fig}, 
with the CRLB degrading at higher average source densities, i.e. during
Galactic Plane Scans. The reason for this is mainly the redder colour of objects near the Galactic plane,
 as the LSF of red stars are generally wider and less steep.
Another significant reason is the larger contamination of many stellar profiles by faint background stars, 
more unresolved multiple objects. A generally increased sky background, due to interstellar gas can also 
be a contributing factor.

Especially after early 2019, when the mission was more mature and fewer interventions were necessary, a seasonal pattern in the
behaviour of the CRLB became apparent. The CRLB was slightly but noticeably higher during the perihelion passage of Earth's orbit, 
i.e in early January of each year than at the aphelion, i.e.
July. This is due to the higher illumination by the Sun at perihelion, resulting in more residual stray light and a slightly higher overall payload temperature.        

Another effect seen especially in the undisturbed part of the mission (lower panel of Fig.~\ref{fig:Long_term_CRLB.fig}) is a more or less linear upward trend of the CRLB.
This is mostly due to the slow degradation of the focus. This defocusing effect seems to be slower since OBMT=8000~rev than during earlier parts of the mission. 
With the end of the {\it Gaia} operations in early 2025, the level of degradation was not deemed high enough to 
warrant a refocusing intervention in 2024.
However, given that the thermal equilibrium was significantly disturbed in May 2024 due to the failure of the
AF1\_3 detector control electronics (see Sect.~\ref{sect:FL:Trends:sign_events} and Table~\ref{tab:sign_events.tab},
and Fig.~\ref{fig:FPA.fig}), 
a final refocus was performed in June 2024. 

The CRLB was thus mainly affected by the seasonal distance from the Sun, the mean object colour, the average source density, 
the focus, and certain on board events, all of which can be seen in Fig.~\ref{fig:Long_term_CRLB.fig}. 
The significant dependence on the distance from the Sun mandates some further discussion. 
Potential influences on the data 
quality related to the variation of the intensity of the sunlight, {\it Gaia} was exposed to,
are stray-light and heat. Both increase with decreasing distance, i.e. compliant with the observed effect, 
i.e. larger values for the CRLB at lower distances. All CCD-detectors had an increase of dark current at higher
temperatures, so at first glance this is a likely origin. However, {\it Gaia}'s detectors were operated in a very cold environment where dark currents are miniscule. Furthermore,
 the annual temperature variation, shown in Fig.~\ref{fig:T_FPA.fig} is very small. However, when taking the whole payload into 
consideration, changes in the overall temperature, did slightly alter the optical system, due to the expansion and contraction of
its components. This would lead to a small seasonal variation of the focus, and as a consequence of the CRLB.
 There were also stray-light contaminations in varying degrees affecting the whole FPA (see \citet{2016A&A...595A...1G}). 
The slightly increased sunlight closer to perihelion did enhance the stray-light falling onto the FPA, 
thus increasing the background, leading to
a higher CRLB than at aphelion. Due to the degeneracy of the two influences, we cannot completely rule out either effect, and it is 
likely that both play a role.

While
the CRLB is mainly used to monitor the quality of the focus, the other quantities also factoring into the CRLB, 
leave their imprint. 
Therefore, for future missions, it might be prudent to compute multiple versions of the CRLB from several samples of stars, 
for example using only brighter objects for an overview of the focus, as these are less influenced by the background, and faint
stars for an estimation of the overall data quality. 

\subsubsection{Long-term ODAS results}\label{FL:Trends:ODAS}

\begin{figure*}[ht]
       \centering
        \includegraphics[width=\hsize]{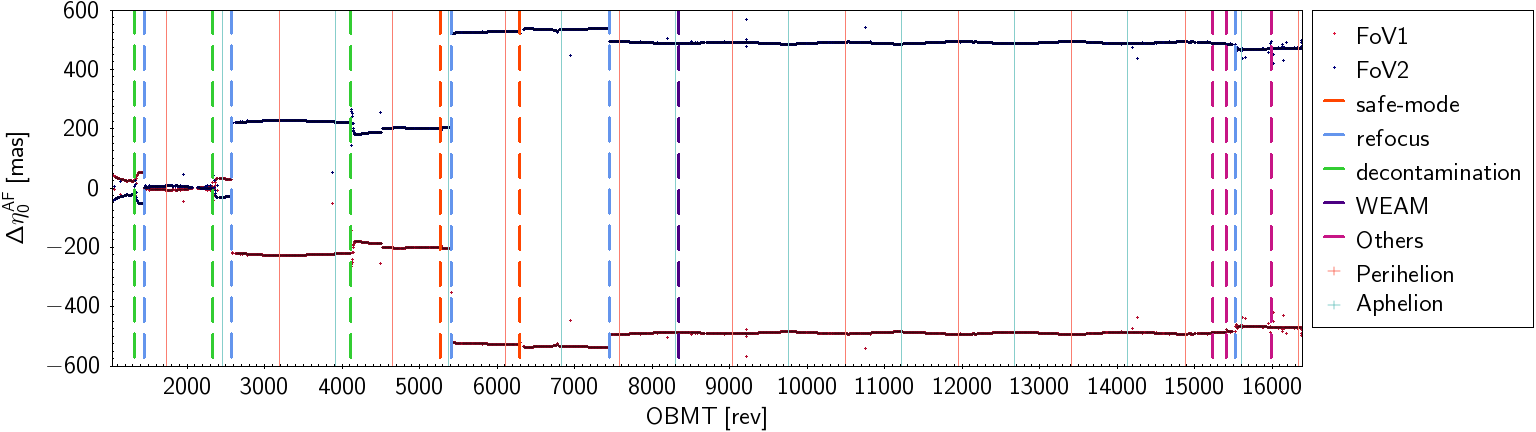}
        \includegraphics[width=\hsize]{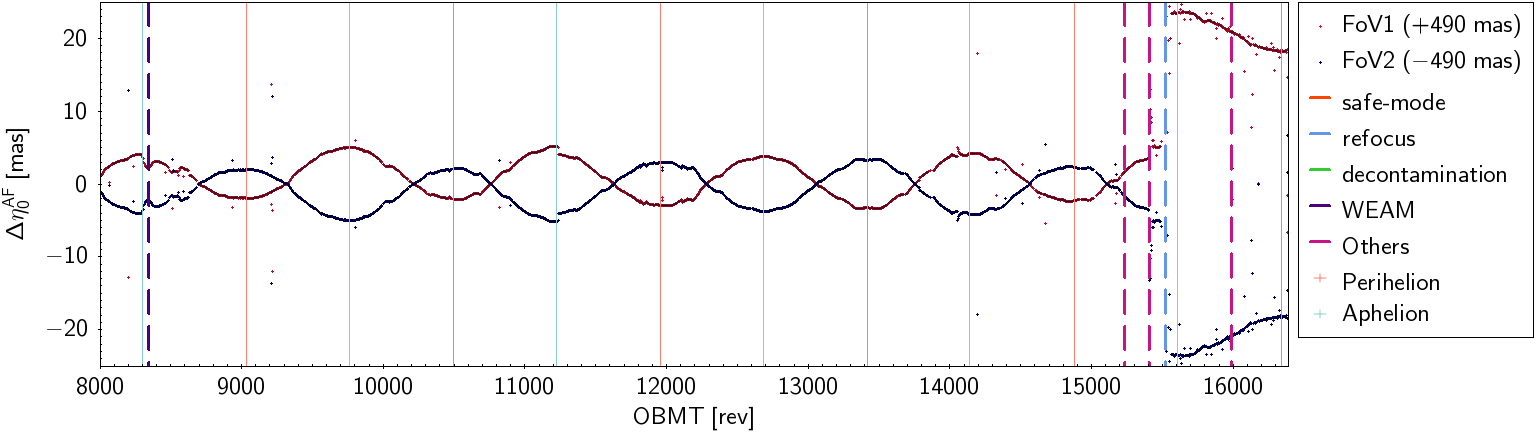}
        \caption{\small Zeroth order along-scan parameter of the astrometric large-scale 
        calibration. The upper panel shows the complete mission, while the lower plot depicts the quiet phase, 
        after OBMT=8200~rev. FoV1 is shown in red and FoV2 in blue. 
        As in the previous figures, the significant events listed in Table~\ref{tab:sign_events.tab} as well
        as the times of the extrema of the distance between Sun and Earth have also been indicated, as
        described in the plot legend.
        In the lower plot, the values have been shifted by $\pm490$~mas for better visibility.} 
       \label{fig:LSCP0_WC1.fig}
\end{figure*}

\begin{figure}[ht]
       \centering
        \includegraphics[width=\hsize]{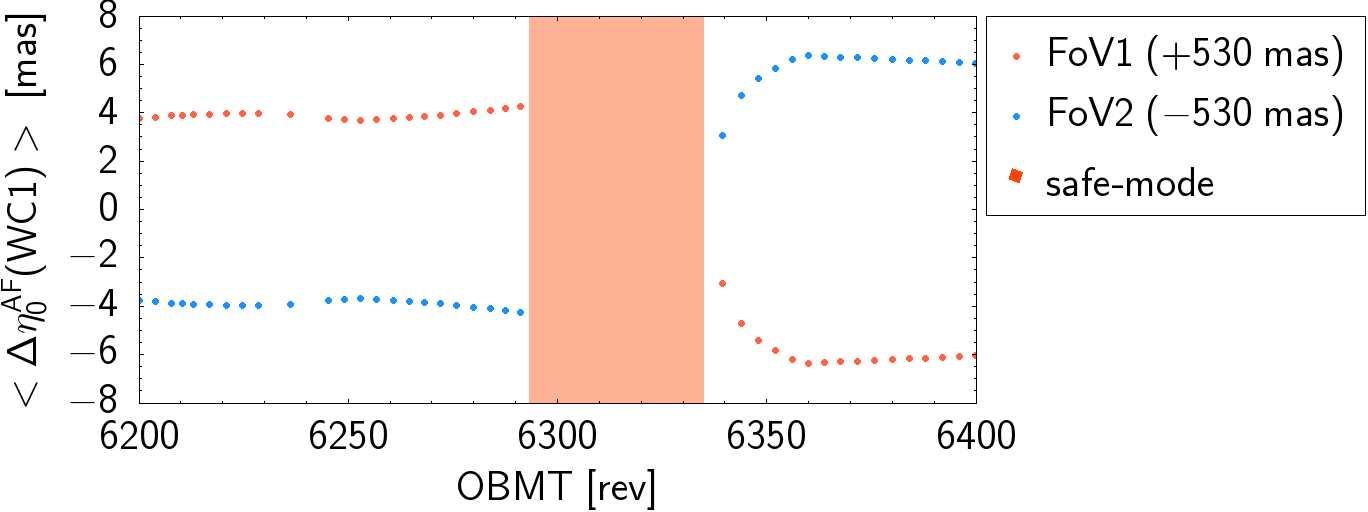}
        \includegraphics[width=\hsize]{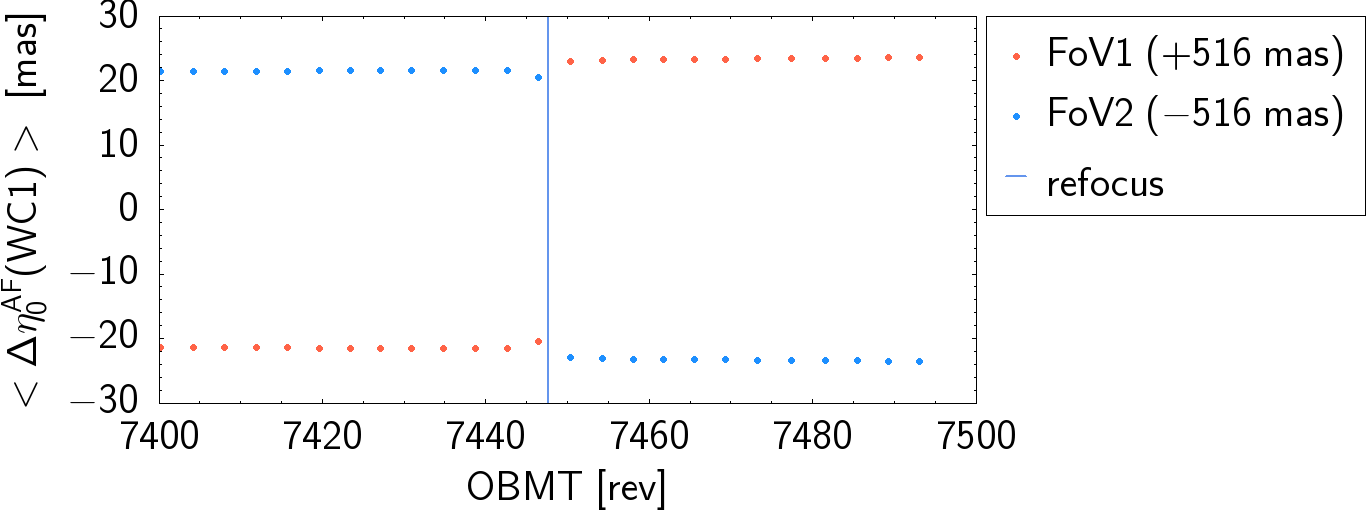}
        \caption{\small Zeroth-order along-scan parameter of the astrometric large-scale
        calibration  during a safe mode (upper panel) and a refocusing (lower panel). The time and duration 
        are indicated respectively by the shaded area and the vertical line.
        FoV1 is shown in red and FoV2 in blue.
        The values have been shifted by $\pm530$ mas (upper panel) and $\pm516$~mas (lower panel) for better visibility.
        }
       \label{fig:LSCP0_WC1_event.fig}
\end{figure}

\begin{figure*}[ht]
       \centering
        \includegraphics[width=\hsize]{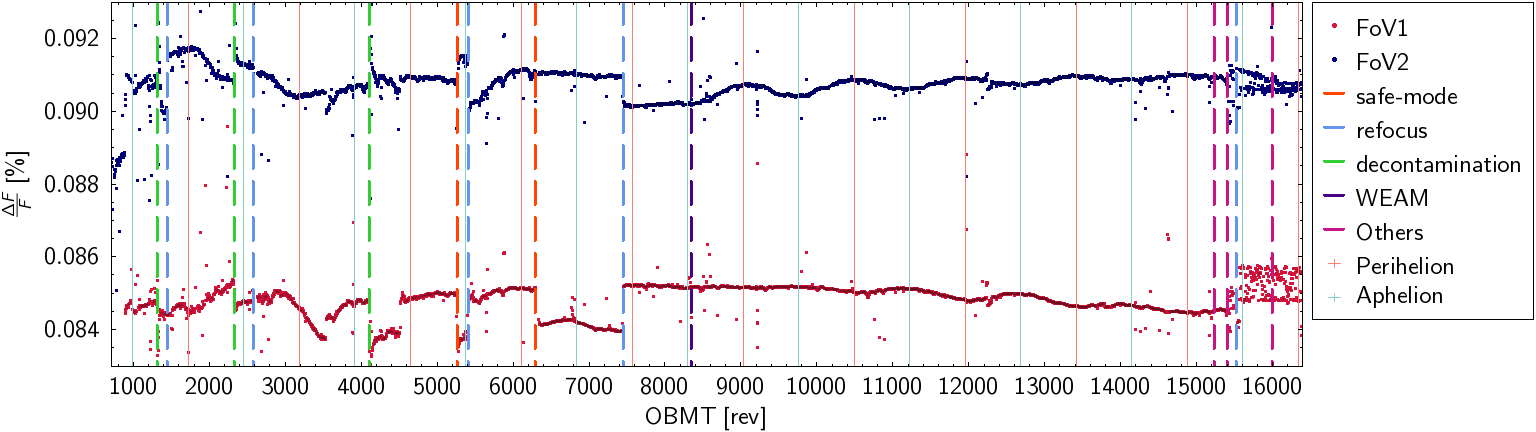}
        \caption{\small Evolution of the focal length of {\it Gaia} as determined by ODAS. Shown is the fractional change of the focal length with respect to the nominal value. FoV1 is represented by
        red data points, FoV2 by blue dots. As in the previous figures, the significant events listed in Table~\ref{tab:sign_events.tab} as well
        as the times of the extrema of the distance between Sun and Earth have also been indicated, as 
        described in the plot legend.
        In order to present the values in the same plot in a readable way,
         the FoV1 values have been shifted by +0.005\%.}
       \label{fig:Focal_length.fig}
\end{figure*}
\begin{figure}[ht]
       \centering
        \includegraphics[width=\hsize]{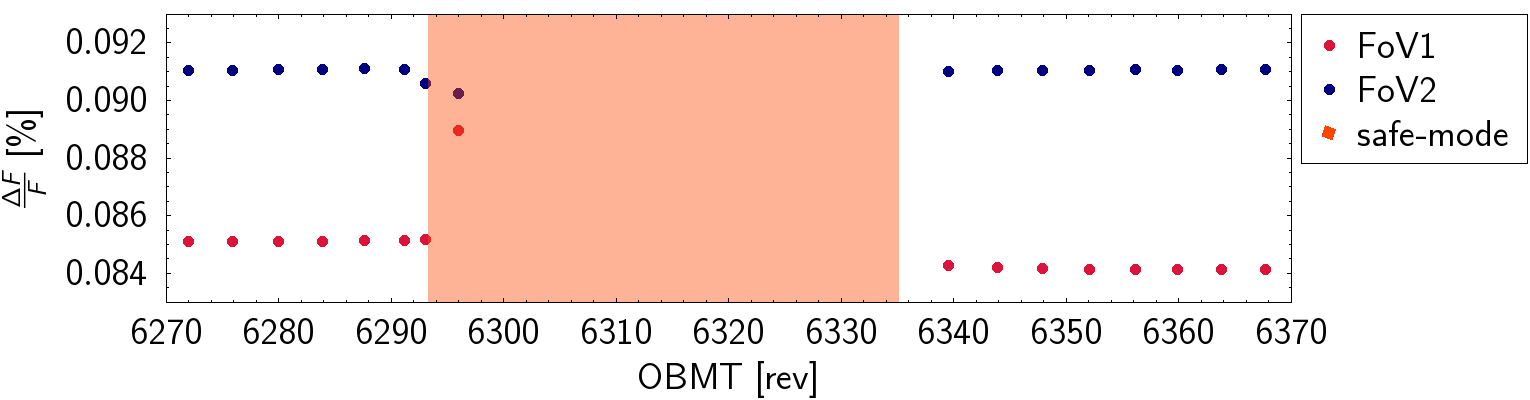}
        \includegraphics[width=\hsize]{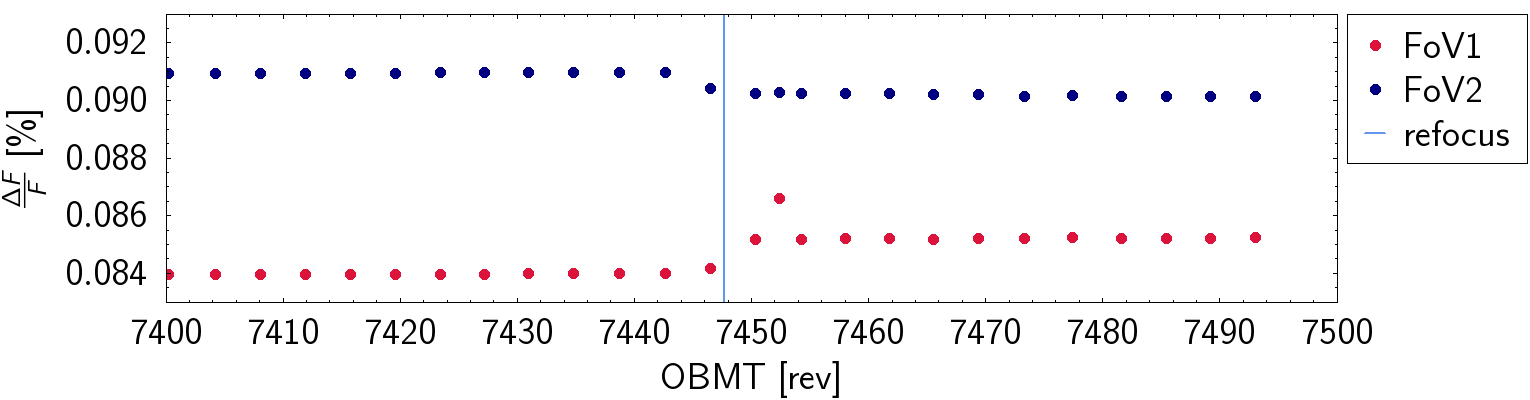}
        \caption{\small Behaviour of the focal length of {\it Gaia} as determined by ODAS during a safe mode (upper panel) and a refocusing (lower panel).
        Shown is the fractional change of the focal length with respect to the nominal value.  The time and duration 
        are indicated respectively by the shaded area and the vertical line.  FoV1 is represented by
        red data points, FoV2 by blue dots.}
       \label{fig:Focal_length_event.fig}
\end{figure}
\begin{figure*}[ht]
       \centering
        \includegraphics[width=\hsize]{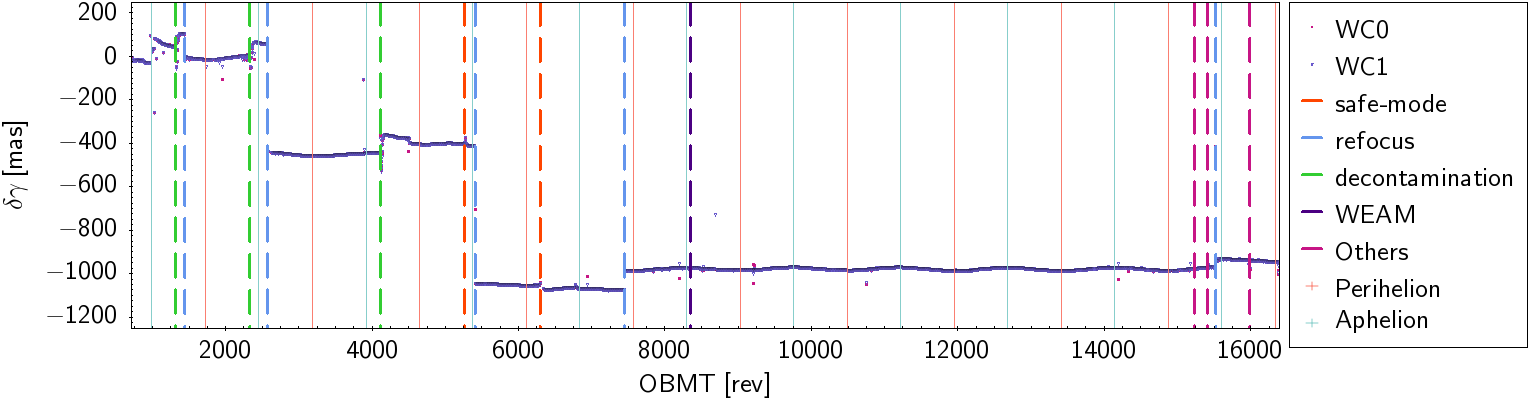}
        \includegraphics[width=\hsize]{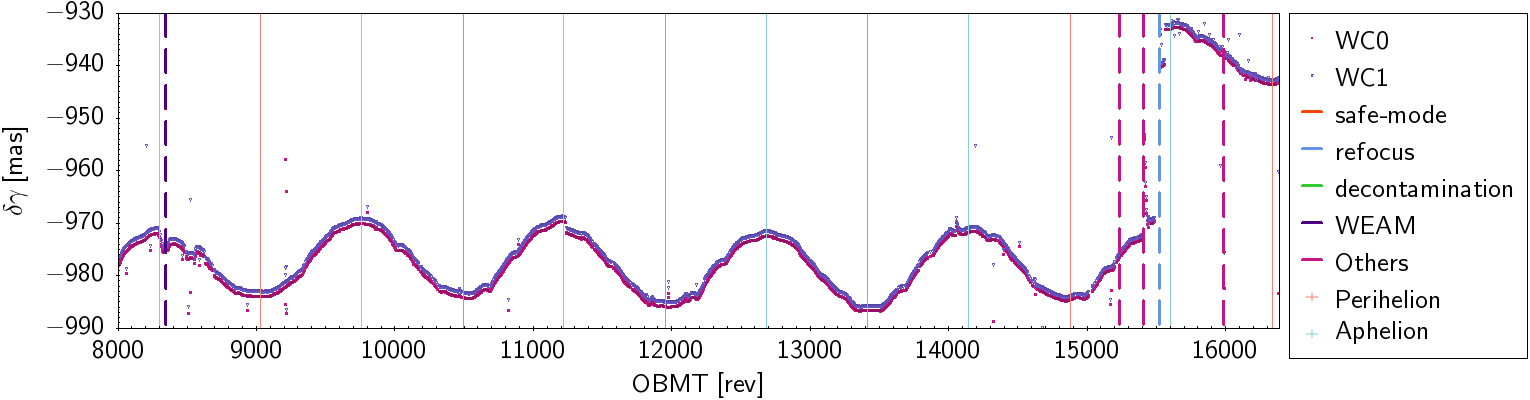}
        \caption{\small Long-term evolution of the {\it Gaia}
       Basic Angle as determined by ODAS. The values derived by the different window classes are represented by
       different colours.
         As in the previous figures, the significant events listed in Table~\ref{tab:sign_events.tab} as well
        as the times of the extrema of the distance between Sun and Earth have also been indicated, as
        described in the  plot legend.   The upper panel shows the evolution of the BA over the whole mission,
        while the lower panel zooms into the part (OBMT $\ge$ 8,000~rev) not affected by decontamination,
        refocusing measures, or safe mode events. It is to be noted that the 6-hour variations or oscillations
        of the BA are taken out by ODAS.
        }
       \label{fig:ODAS_BAM.fig}
\end{figure*}

\begin{figure*}[ht]
       \centering
        \includegraphics[width=\hsize]{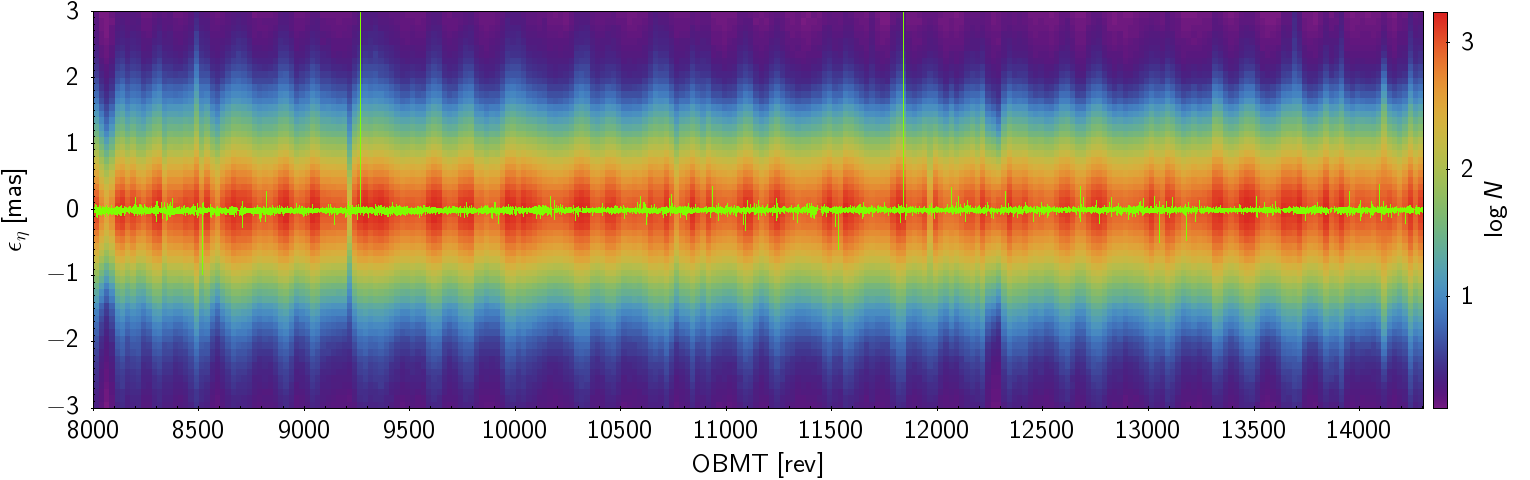}
        \includegraphics[width=\hsize]{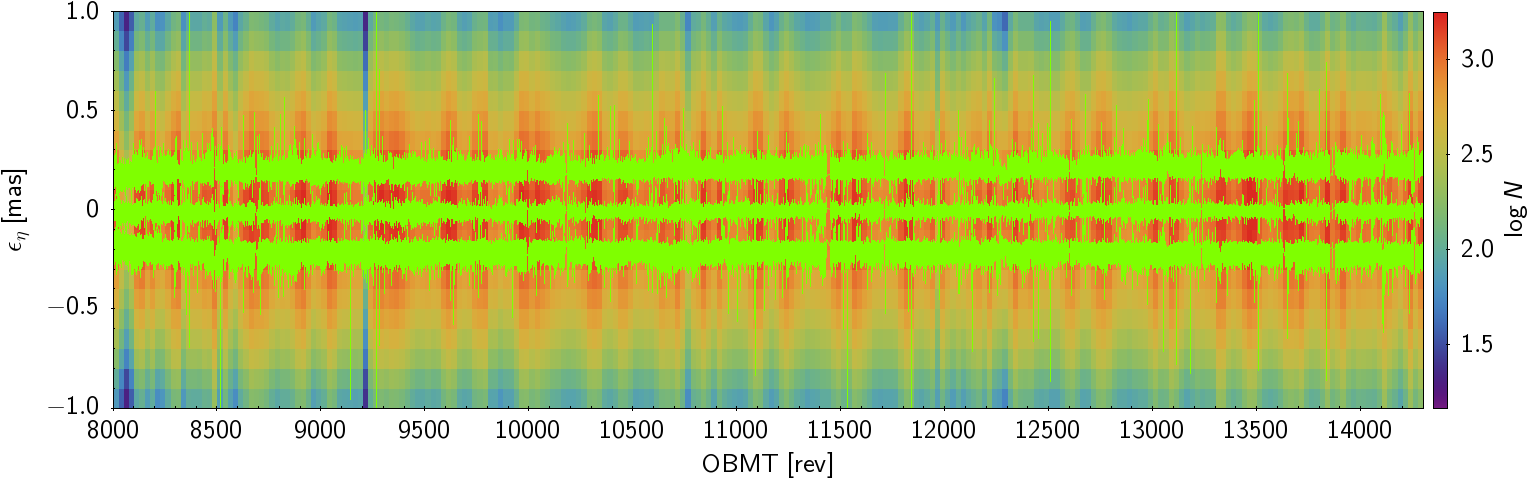}
        \caption{\small 2D histogram of the along-scan astrometric residuals of the ODAS primary sources between OBMT=8000~rev and
        14400~rev. Shown is FoV1.
        The upper panel shows a range of $\pm$3~mas, while the middle panel is a zoom-in on the $\epsilon_\eta$ (vertical) direction. 
        The median and $\pm 1-\sigma$ contour lines are shown in light green in the middle panel, while only the median contour is depicted in the
upper plot for reasons of clarity. We note that the colour map of the 2D histogram has an absolute scaling, i.e. the bin appears to be broader 
if there are more sources in a time bin in total.} 
       \label{fig:Long_term_ODAS.fig}
\end{figure*}
As illustrated in Sect.~\ref{sect:FL:ODAS}, one of the main components of the FL was the ODAS, which
was an astrometric solution with a limited scope, i.e. only being based on roughly one day of data, and therefore
limited sky coverage. While it was by no means comparable to the full astrometric data reduction, which covers 
years worth of data, it gave the basis for an analysis of the astrometry and a long-term monitoring of the data
quality. This way trends in the system could be followed, and potential issues 
identified long before the data was even touched in the global processing.

One important measure which gives information about the astrometric stability of the instrument is the astrometric
calibration. As described in \citet{ODASpaper}, the ODAS astrometric calibration model
utilises Legendre polynomials up to the
second order. Figure~\ref{fig:LSCP0_WC1.fig}, shows the mean zeroth-order along-scan calibration parameter $\Delta\eta_0^{\mathrm{}AF}$, 
averaged over all CCDs in the astrometric field, which determines the relative movements of the optical axes 
of the two fields of view on the focal plane.\footnote{The six-hour oscillations of the BA are eliminated before the ODAS adjustment of the geometric calibration.} As a joint angular AL shift of the two FoVs is mathematically degenerate with a shift of the AL spacecraft attitude, ODAS could not determine an absolute geometric calibration. 
This is why the values of the two FoV are exactly mirrored in Fig.~\ref{fig:LSCP0_WC1.fig}.
Changes in the relative position of the optical axes gave information on the causes of these. 
If the change was in AL only, they simply correspond to a change of the BA,
since this is the angle between the two apertures and thus only occurs in the AL direction,
 while changes in both the AL and AC direction were caused by a change of the focal length. 

Again the familiar seasonal variations are a rather prominent effect seen in the long-term
trends. Furthermore there are marks caused by short-term incidents, either planned, such as station-keeping manoeuvres,
or unplanned, such as stronger micro-meteorite impact events (some exemplary occurrences are
discussed in Sect.~\ref{sect:results:MM}), or even caused by propellant bubble movements. Notably, the two prominent dents in the curves
between OBMT=8,300 and 8,600~rev (see Fig.~\ref{fig:LSCP0_WC1.fig}, lower panel) are related to 
the WEAM (OBMT=8345.761~rev), especially to the heating and later cooling of the propellant (see also the development
of the fuel and oxidiser temperatures shown in Fig.~\ref{fig:Long_term_Temps.fig}). 

Fig.~\ref{fig:LSCP0_WC1_event.fig} highlights the behaviour of the astrometric calibration 
during a safe-mode event and a refocusing. The former event had a moderate effect on the zeroth-order astrometric calibration; however, it was 
 longer lasting, with the parameter returning back to normal after about 70~rev, and a lasting offset of about 1~mas.
 By contrast, the refocusing resulted in a much larger jump, with no event-related evolution afterwards. 
This reflects the nature of these very different types of events, with the safe-mode mainly being a thermal disruption,
 caused by the change in the spacecraft's orientation and the shutdown of a large part of the electronic systems. 
The refocusing on the other hand did not present a thermal influence of any significance, but presented a mandated manipulation of the 
optical system, with the desired part being the improvement of the instrument focus. However, subtle side-effects, such as tiny shifts 
in the optical alignment of the components of the optical assembly, shifted the zeroth-order LSCP values slightly, and that is what is
shown in the lower panel of Fig.~\ref{fig:LSCP0_WC1_event.fig}

Taking into account a longer timespan, the daily ODAS results aided in monitoring the overall stability of  
{\it Gaia}. As an example, Fig.~\ref{fig:Focal_length.fig} shows the evolution of the focal length $\frac{\Delta F}{F}$ 
of the {\it Gaia} instrument for both FoV. 
The refocusing campaigns show strong discontinuities, which were intended in this case. The decontamination and
safe mode events also caused step changes of the focal length, which were caused by the drastic change of the thermal stability
of the payload, due to the heating or change in orientation with respect to the Sun. As a secondary but periodic effect,
the annual temperature variation caused by the distance change of {\it Gaia} and the Sun manifests itself in an 
oscillation of the focal length. The linear drift in opposite directions for the two FoVs, seen mainly in the stable 
regime after OBMT $\simeq$ 8,000~rev, which had been untouched by major interventions,
is due to the slowly degrading focus and the resulting evolution of the focal length. Figure~\ref{fig:Focal_length_event.fig}
shows close-ups of the evolution of the focal length during a safe-mode and a refocusing. 
The safe-mode only had a small impact on the focal length, pushing the values for the
two FoV a bit apart, while the adaptation of the focus moves them closer together, as expected. 

Another product of ODAS is a daily average of the Basic Angle, which was computed by deriving the difference between 
the zeroth-order astrometric calibration in the AL direction for both FoV and each of the 62 AF detectors
and averaging over the individual values. Like in the case of the focal length,
we only have one value per FL day, meaning that ODAS was insensitive to changes on smaller time scales, such as the 
6-hour oscillations  in Fig.~\ref{fig:BAvar.fig}. However, we got access to the long-term evolution, which is
shown in Fig.~\ref{fig:ODAS_BAM.fig}. We note that this diagnostic parameter can only reflect those changes which occur
within the telescopes' optical path itself. Effects stemming from the inner workings of the BAM device would 
not be seen. 
Overall, the same effects are seen as in the previous cases.  The seasonal variations can again be clearly 
seen in the bottom panel of Fig.~\ref{fig:ODAS_BAM.fig}. In Sect.~\ref{FL:Trends:BAV} we   compare the 
ODAS-derived Basic Angle variations (BAV) with those measured with the BAM.

A final product of ODAS is an improved attitude, OGA2. As described in Sect.~\ref{sect:FL:ODAS}, ODAS
digests an attitude solution (OGA1) derived by IDT (Sect.~\ref{sect:FL:IDT}) and puts out an improved version called 
OGA2. The main difference between OGA1 and OGA2 is that the former is a simple Kalman filter reaching about 1~mas precision \citep{2016A&A...595A...3F}, while
the OGA2 was determined using the astrometric adjustment from ODAS.
 Given the importance of a precise knowledge of {\it Gaia}'s attitude, 
and to discuss the various internal and external effects impacting the {\it Gaia} attitude, 
a separate section, Sect.~\ref{sect:results:RE}, deals with the attitude-related results.

Finally, Fig.~\ref{fig:Long_term_ODAS.fig} displays the overall long-term stability of the ODAS along-scan astrometric residuals. 
The distribution and mean of the residuals provide a measure for the quality of the daily solution.  
In the longer-term they allow to monitor the stability of the instrument, as 
any change would manifest itself in changes in the distribution of the residuals. 
As the 2D-histograms
are very sensitive to the number of stars, the figure only shows the time interval after the last edition of the 
ODAS homogeniser was activated in on-ground operations. 
This filter ensures an as homogeneous distribution of stars in the ODAS solution as possible, taking into 
account the highly variable stellar density over the sky (see \citet{ODASpaper} or Sect.~\ref{sect:FL:ODAS}). 
Earlier versions of the ODAS suffered from no or insufficient homogenisation which left a far
larger residual inhomogeneity, making the comparison much more difficult. Therefore we limit
ourselves to the time after OBMT=8,000~rev. Figure~\ref{fig:Long_term_ODAS.fig} shows
the 2D-histogram of the ODAS along-scan residuals $\epsilon_\eta$, with the lower panel being a zoom-in. 
The median values are indicated in both of
these plots, with the lower plot also having the 1-$\sigma$ contour lines depicted. 
Given that
this plot shows a time range of about 4$\frac{1}{2}$ years, the remarkable stability of {\it Gaia} becomes evident.  

All these findings were made available to the rest of the data processing consortium, either via the daily data qualifications or 
by reports, which presented the results of investigations done by the FL team. 
\subsubsection{The Basic Angle}\label{FL:Trends:BAV}

\begin{figure*}[ht]
       \centering
        \includegraphics[width=\hsize]{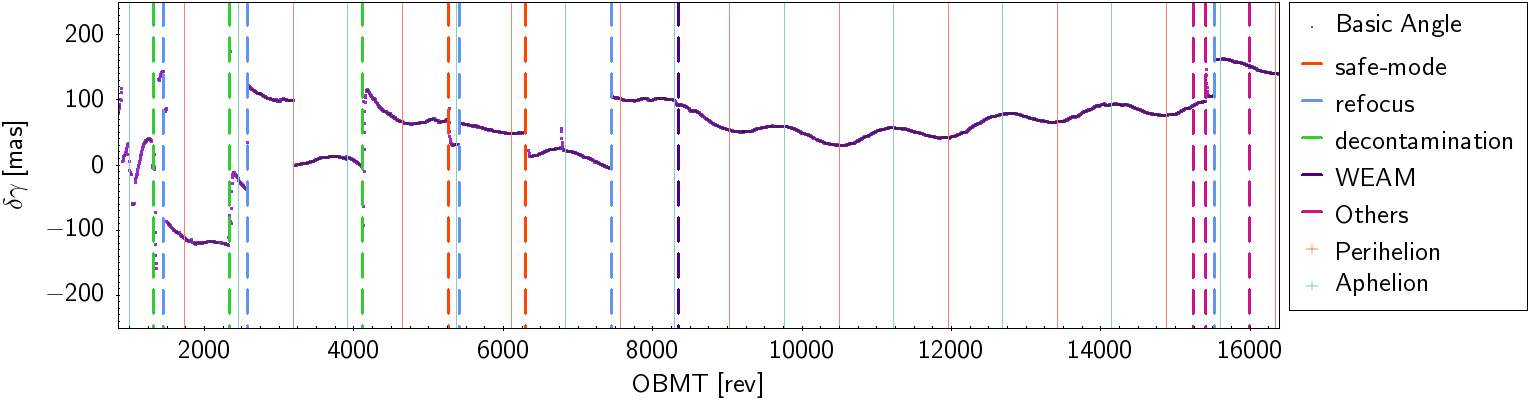}
        \includegraphics[width=\hsize]{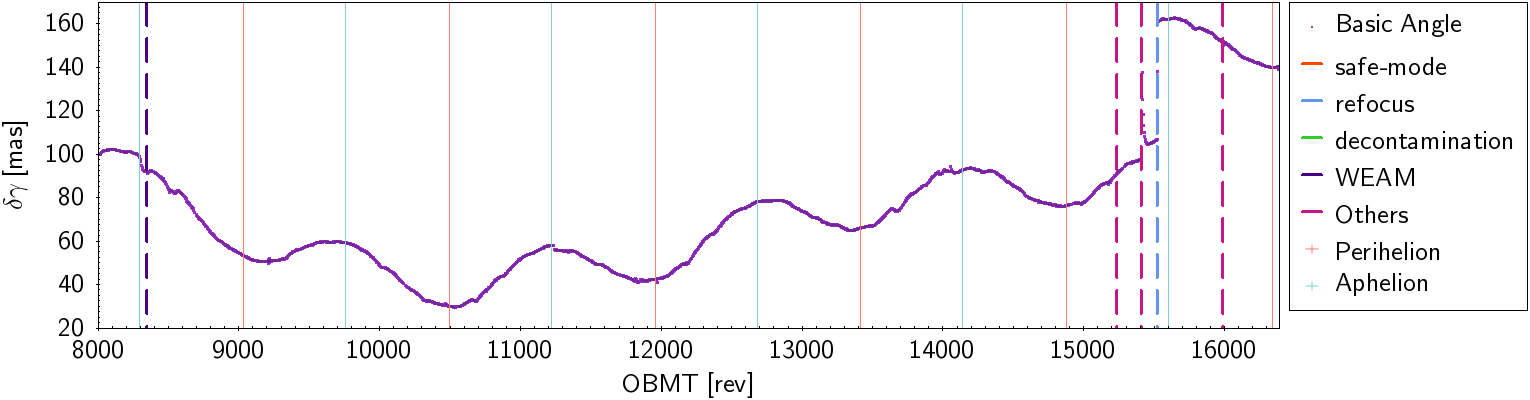}
        \caption{\small Long-term evolution of the {\it Gaia}
       Basic Angle as measured by the Basic Angle Monitor. The measurements shown in this plot have been
        averaged over one FL day.
        As in the previous figures, the significant events listed in Table~\ref{tab:sign_events.tab} as well
        as the times of the extrema of the distance between Sun and Earth have also been indicated, as
        described in the legend.
        The upper panel shows the evolution of the BA over the whole mission,
        while the lower panel zooms into the part (OBMT $\ge$ 8,000~rev), which had not been affected by decontamination,
        refocusing measures, or safe mode events.
        }
       \label{fig:BAM_BAV.fig}
\end{figure*}

\begin{figure}[ht]
       \centering
        \includegraphics[width=\hsize]{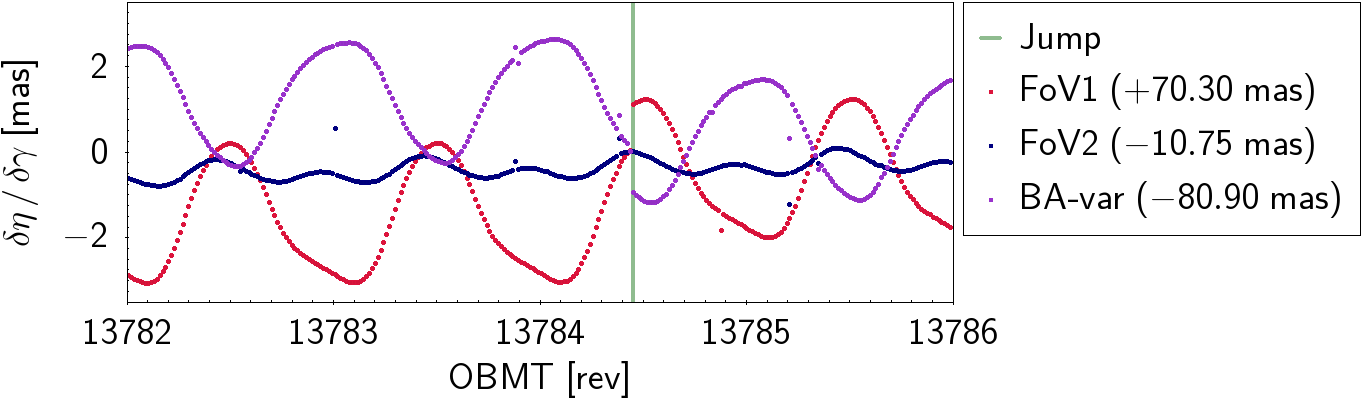}
        \includegraphics[width=\hsize]{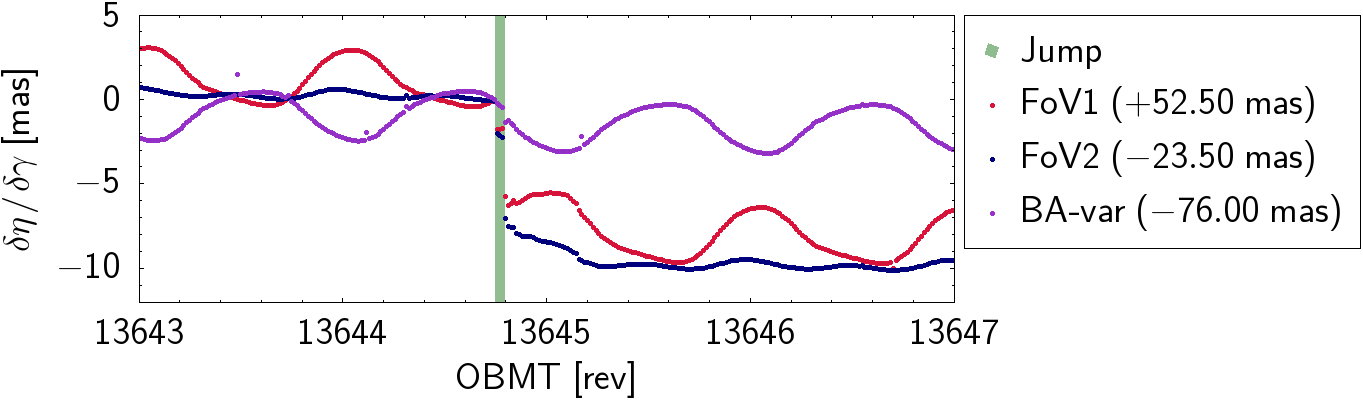}
        \caption{\small Examples of a unilateral jump event (upper panel) and bilateral jump event (lower panel).
The red points denote the AL fringe position of FoV1 and the blue points FoV2. The combined values, i.e. the  
BA variations are shown in violet.  We note that the absolute values are shifted by the amount indicated in the 
legends in order to fit them in the same plot.}
       \label{fig:BAvar_jumps.fig}
\end{figure}

\begin{figure}[ht]
       \centering
        \includegraphics[width=\hsize]{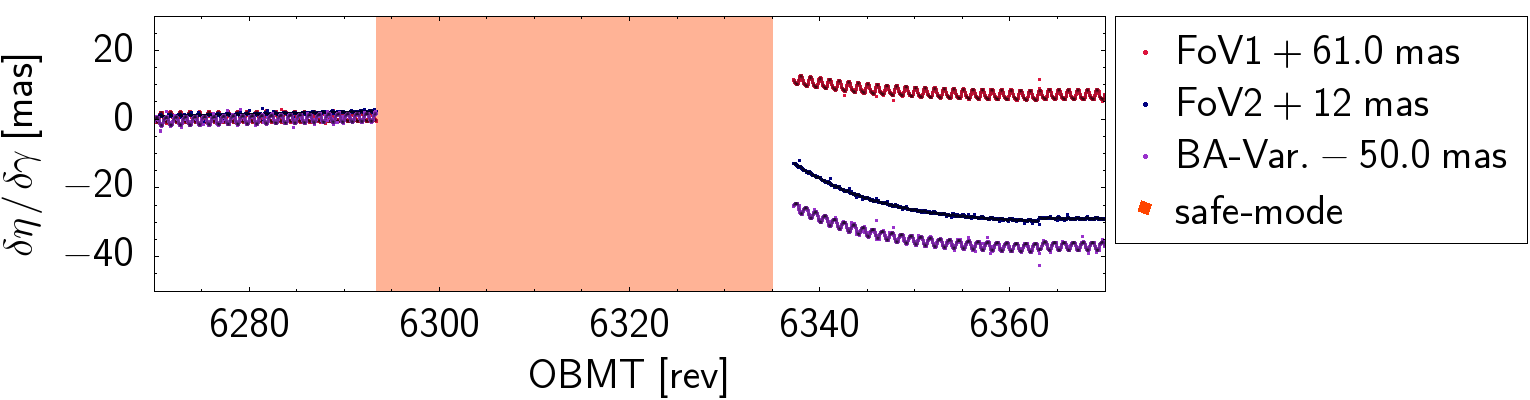}
        \includegraphics[width=\hsize]{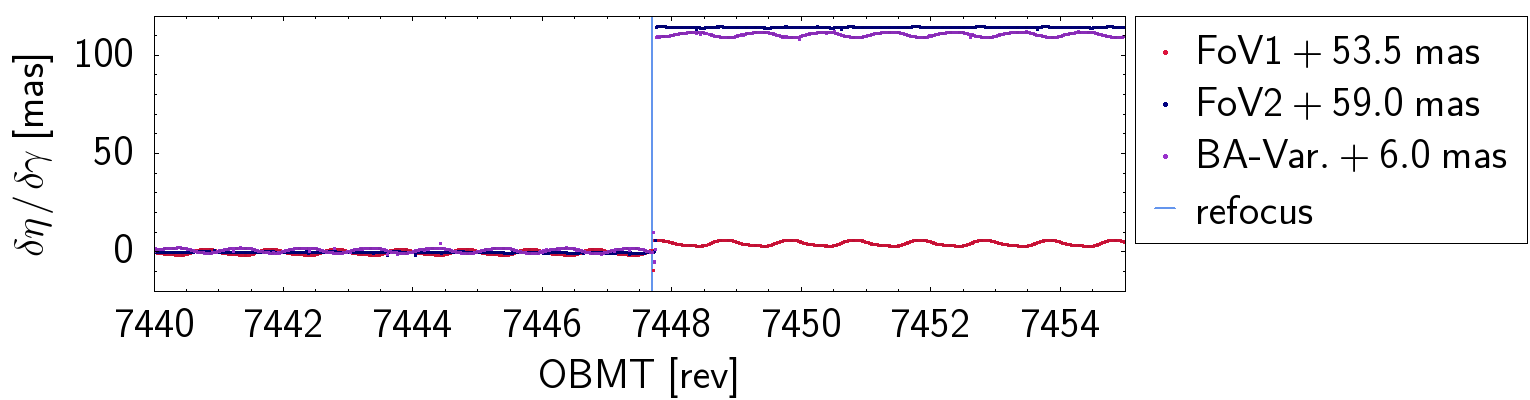}
        \caption{\small Behaviour of the Basic Angle of {\it Gaia} as determined by the BAM during a safe mode (upper panel) and a refocusing (lower panel).
        The red points denote the AL fringe position of FoV1 and the blue points FoV2.  The combined values, i.e. the
BA variations are shown in violet. The time and duration 
        are indicated respectively by the shaded area and the vertical line.  
We note that the absolute values are shifted by the amount indicated in the
legends so that all three datasets coincide prior to the event.}
       \label{fig:BAvar_events.fig}
\end{figure}

With the knowledge of the Basic Angle being such as critical quantity for the {\it Gaia} astrometric data reduction, there was
an instrument exclusively devoted to measuring the BA, the BAM, which is described in Sect.~\ref{sect:FL:BAM}.  Soon after the launch of the 
spacecraft, it became evident that the Basic Angle was not behaving as expected, i.e. being a constant in the short-term, and just having 
the scatter of the measurements as BAV. 
Instead, it followed a heliotropic oscillation (of about 6 hours, i.e.\ about 1~rev).
This oscillation pattern turned out to be mostly stable, and 
could thus be largely compensated by fitting (see \citet{2016A&A...595A...1G}). 

As with other key parameters, the BA was not only affected by these oscillations but also by long-term regular
 variations. These were, just as previously
shown, mostly caused by the seasonal temperature variation. Like in the case of the astrometric calibration (see Sect.~\ref{FL:Trends:ODAS}), changes of the ambient
temperature caused the whole optical assembly of {\it Gaia} to slightly expand or contract, impacting the BA. This effect can be seen in Fig.~\ref{fig:BAM_BAV.fig}.
This figure also shows the impact of significant events (see Sect.~\ref{sect:FL:Trends:sign_events}) on the payload, 
such as refocusing, safe modes, or decontaminations, which obviously 
have had strong influences on the structure, both thermal and 
mechano-optical.\footnote{Under mechano-optical influences, we here understand non-thermally induced changes in the optical path (e.g. by adapting the focus)} In contrast to the ODAS BA, a longer-term trend can be seen, namely a curve, at first going to lower values, culminating at around OBMT=10,500~rev, then increasing until the end of the measurements. Before OBMT=8000~rev, this effect was not seen, mainly due to the large step-changes caused by some of the significant events (see Sect.~\ref{sect:FL:Trends:sign_events}),
which dominate this more subtle change (see Fig.~\ref{fig:BAM_BAV.fig}). The cause of this multi-year trend is at current not known, prime suspects are drifts in the BAM
electronics, and other gradual technical effects. In a future paper dedicated to the BA, 
this will be explored in greater detail. It is also to be noted that the BAM was by design a short-term 
instrument, thus the results of measurements are only coherent over timespans of several days.

Comparison of the BAM-derived long-term BAV with the long-term BA derived from the ODAS astrometric calibration (see Sect.~\ref{FL:Trends:ODAS}), 
shown in Fig.~\ref{fig:ODAS_BAM.fig} shows a lot of similarities but also differences. 
Some of these differences can be explained by the fact that the light path involved was slightly different from that of the BAM for technical reasons. The effects
of the seasons, the significant events (see Sect.~\ref{sect:FL:Trends:sign_events}), and smaller disturbances are 
clearly seen in both figures. However, the step changes during the major disruptions are significantly different with the BAM ones being larger. 
This can be attributed to changes in the optical structure affecting the telescopes' geometry
 (on which the ODAS-derived BAV is based) and the BAM 
in different ways. Looking closer, the seasonal amplitudes differ from each other, with the BAM amplitudes again being the larger ones. This can be explained by slight
differences in the thermal expansion of the two set-ups. We note that the values shown here are differential, and differ only by a few mas. The multi-year
curve seen in the BAM-derived BAV is absent in the ODAS-derived BAV. This means that this phenomenon is indeed intrinsically related to the BAM.   

A final phenomenon seen in the BAM data are sudden jumps (see Fig.~\ref{fig:BAvar_jumps.fig}). We distinguish between unilateral jumps, which affect only one branch of 
the BAM, thus the AL fringe position in one FoV, and bilateral jumps, which affect both branches. 
In the former case, this always leads to a jump in the BA, as it is the 
difference of the FoV1 and FoV2 AL fringe position variations or line of sight variations measured by the BAM. 
In the bilateral case, the resulting BA is mostly not or only slightly affected, as in most cases the jumps in each of the AL fringe positions are the same or at least similar
as our example shows. This hints at a common singular disturbance as a cause for the bilateral BA jumps. 
However, in some cases there is also a noticeable jump in the Basic angle itself, which means that the individual
jumps in the two lines of sight do not completely cancel out.
 In other cases of bilateral jumps, the BA jump is even larger than those in the lines of sight. 
Obviously this depends on the direction and the magnitude of each of the two AL fringe position jumps. 
Interpretation of such events is more complicated. Given that the AL fringe position jumps occur at the same time, they should be caused by a common event, which affects each branch of the BAM in a different way.  

The reason for either type of BA-jump event cannot always be
determined; however,  it is often related to the thermal relaxation after a larger intervention into the payload, 
or a micro-meteoroid impact. In many cases, the imprints
of these jumps are seen both in the BAM- and ODAS-derived BAV. This means that the origin of the jump lies inside 
the common part of the light path. This could be
a shift of one of the mirrors, the FPA, or another component of the optical set-up. In cases, where the ODAS value does 
not show the jumps, the origin must lie in the
BAM itself, which can suffer slight displacements on rare occasions. The jumps displayed 
in Fig.~\ref{fig:BAvar_jumps.fig} present an ideal example, 
how diverse the origins of such jumps are. The unilateral jump, which occurred on April 5, 2023, coincided 
with an apparent micro-meteoroid impact, which left its marks
on many parts of {\it Gaia} telemetry. This event is discussed in more detail in Sect.~\ref{sect:results:MM}. 
The bilateral event, which is the first of two similar events, both occurring in March 2023, has no convincing 
explanation, and was not seen in any other diagnostic to which the FL had access. 
Therefore the cause can only be a sudden change within the BAM itself. This is further
evidenced by co-temporal jumps occurring in other derivatives of the BAM measurements.   

Some of the significant events (see Sect.~\ref{sect:FL:Trends:sign_events}) also had an influence on the BA. Figure~\ref{fig:BAvar_events.fig} shows the impact from a safe-mode event and a refocusing. During the safe-mode, 
which had a rather long duration, a relatively small jump with respect to the situation prior to the event could be seen, 
relaxing to a new normal situation, while the refocusing resulted in a large bi-lateral
jump, and no longer-term evolution of both the lines of sight or the BA itself, apart from the seasonal trend. 
This demonstrates how different the consequences of such events could be. 

While in the scope of this article, we can only scratch the surface of the Basic Angle measurements and their intricacies, 
a further study will be entirely devoted to this topic.
\subsubsection{Discussion of the long-term evolution of {\it Gaia}}\label{FL:Trends:discussion}

The Sun-Earth L2-region is known for its great overall stable conditions,
 and had for this very reason been chosen for the {\it Gaia} mission. However, even in this region of space, there is no
complete stability and tranquillity. 
As shown in the previous parts of this section, there were a number of factors influencing the stability of {\it Gaia}.

Some cannot be avoided, as the overall temperature variation caused by the ellipticity of Earth's orbit and the Lissajous type orbit of {\it Gaia}
around the L2-point, and these had to be accounted for in the data reduction. This also
applied to the source density. It could vary by more than a factor of 500 in one revolution, as shown in 
Fig.~\ref{fig:ASD.fig}.

Another class of disturbances are operational interventions, such as refocusing and decontamination campaigns, station-keeping and other manoeuvres, 
and automated responses of the spacecraft to potentially catastrophic situations, i.e.
safe modes. While these also could not be avoided entirely, 
their effects on the payload were minimised, so that the impact
of these was kept as low as possible. However, the more a space probe is left untouched, the more stable it is, as had been the case
for {\it Gaia} since 2019. Luckily, even the large step changes described in this section do not imply a significant detrimental effect on the overall data quality 
and usability. In fact some, such as the better CRLB after refocusing, 
are the desired outcome of the corresponding intervention. 
These drastic changes did
mean, however, that the whole optical system had changed slightly, 
which led to the need to recalibrate the astrometric model.
This way, we could ensure the best possible calibration of the astrometric model of {\it Gaia}.

Individual events can be divided into those with a significant thermal disruption, and those which do not feature
large abrupt changes in temperature. The decontamination measures, which involved a significant heating of the
optical system, and the two safe modes, which have led the satellite to be slewed to a solar aspect angle of $0^\circ$
(see Table~\ref{tab:sign_events.tab}), so
that the sun-shield directly pointed towards the Sun, had caused strong changes in the temperature distribution throughout the payload. 
This resulted in the need of a longer stabilisation period after the intervention, 
during which {\it Gaia}'s temperature distribution relaxed to a 
stable condition. In general the data obtained during these periods of (thermal) instability is more difficult to
calibrate. Therefore it was paramount to 
minimise the duration of such time intervals. Nonetheless the ground breaking {\it Gaia} releases DR1-3 
(see e.g. \citet{2016A&A...595A...1G}, \citet{2018A&A...616A...1G}, and \citet{2023A&A...674A...1G}) are all based on at least the parts of the data, 
which features these step changes, highlighting the science data quality which is achievable, even with these disruptions. 
 
Other interventions, for example refocusings, do not involve a significant change of the temperature distribution; thus while 
the jump in the astrometric calibration, or the Basic Angle look impressive, thermal stability was
reached soon afterwards. 

The station-keeping manoeuvres (SKMs),
required in regular intervals to maintain {\it Gaia}'s orbit, had noticeable influences onto the payload, 
which were by no 
means as dramatic as the impacts from the events, described previously. 
Station-keeping manoeuvres produced subtle changes of the velocity, without much change in the optics apart from some thermal influence caused by 
the heating of the propellant. The FL monitored the aftermath of every one of these 62
instances, and  their influence onto the payload is indeed modest. It had happened, however, that individual SKMs have caused or ended series of 
periodic rate excursions in the attitude, which is discussed in Sect.~\ref{sect:results:RE}.

Finally other non-anticipated events, such as temporary component switch-offs or impacts from micro-meteoroids
have left their temporary --- or in some cases lasting --- impacts on the {\it Gaia} instrument. Some of these are discussed in Sect.~\ref{sect:results:MM}.  
\subsection{The OGA2 and attitude rate excursions}\label{sect:results:RE}
\begin{figure}[ht]
       \centering
        \includegraphics[width=1.00\hsize]{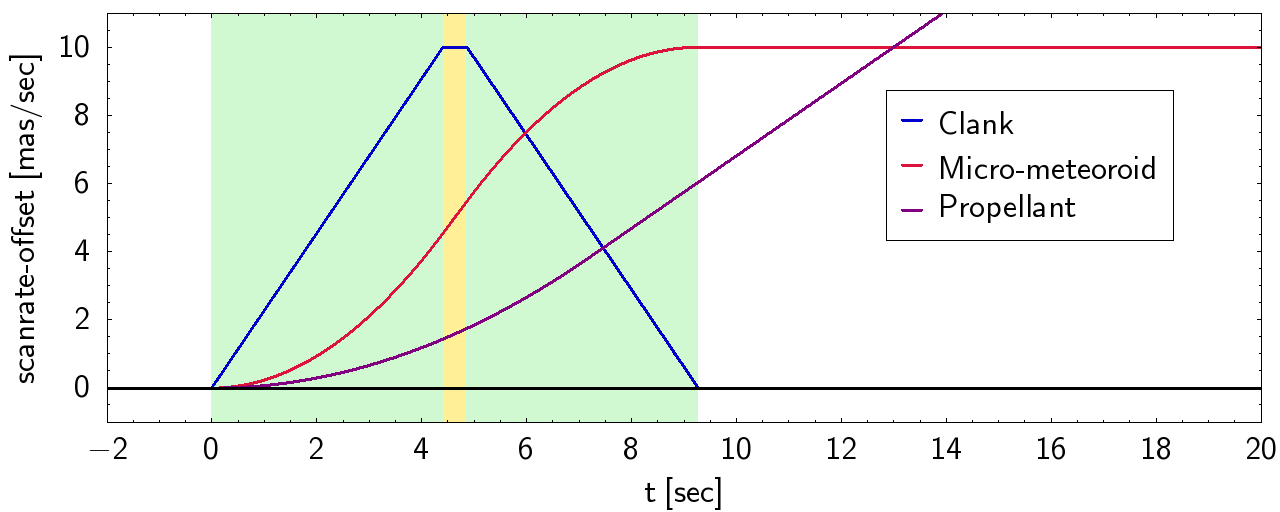}
        \caption{\small Schematic representations of the three main
        types of rate excursions seen in the transition time differences. The green shaded areas
         indicate the times required to traverse one detector, the yellow areas the transition time for the gap between 
         two consecutive detectors. All schematic curves
         have a magnitude of 10~mas/sec; the duration time 
        of the propellant movement is 7.2 sec.}
       \label{fig:PR_scheme.fig}
\end{figure}
\begin{figure}[ht]
       \centering
        \includegraphics[width=1.00\hsize]{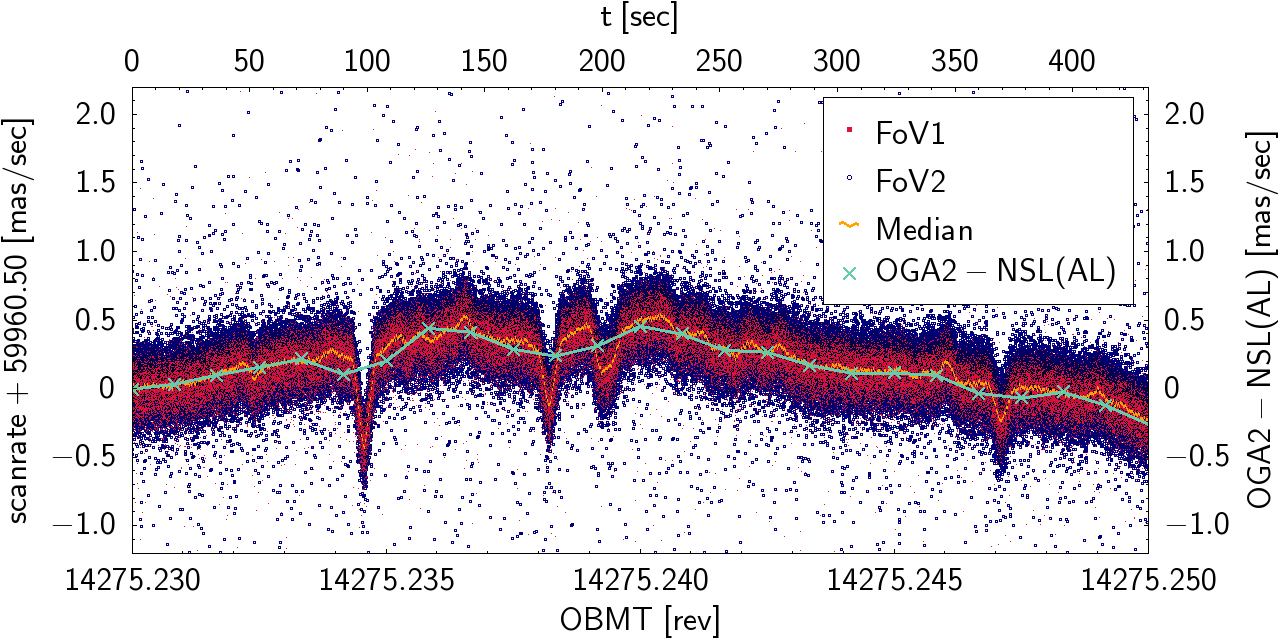}
        \caption{\small Comparison of OGA2--NSL and pair rates for an undisturbed stretch of attitude data. Shown are the pair rate 
         analysis derived scanrates for Fov1 (red symbols) and FoV2 (blue symbols), as well as the 50th percentile for both FoV in orange. 
         The data points of the OGA2-NSL attitude approximation are shown as green crosses connected by a green solid line. 
         A number of micro-clanks of various amplitudes can be seen.}
       \label{fig:OGA2vsPR.fig}
\end{figure}
\begin{figure}
       \centering
       \includegraphics[width=\hsize]{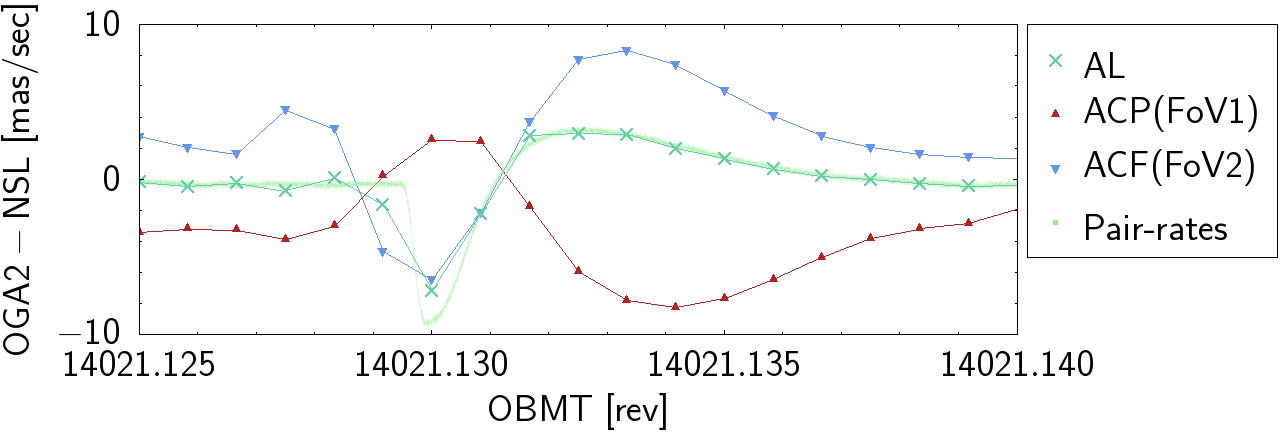}
       \caption{\small Comparison of the OGA2--NSL and pair rates for a micro-meteoroid impact. The same event, occurring at OBMT=14021.1300~rev, is shown in more detail in Fig.~\ref{fig:RE.fig}.
       The AL component of the OGA2--NSL is shown as green crosses, while the pair rate
       data as small green dots. The AC component of the OGA2--NSL is shown as
       red-brown (ACP=FoV1) and blue (ACF=FoV2) triangles.}
       \label{fig:RE_OGA2vsPR.fig}
\end{figure}       
\begin{figure}
       \centering
        \includegraphics[width=\hsize]{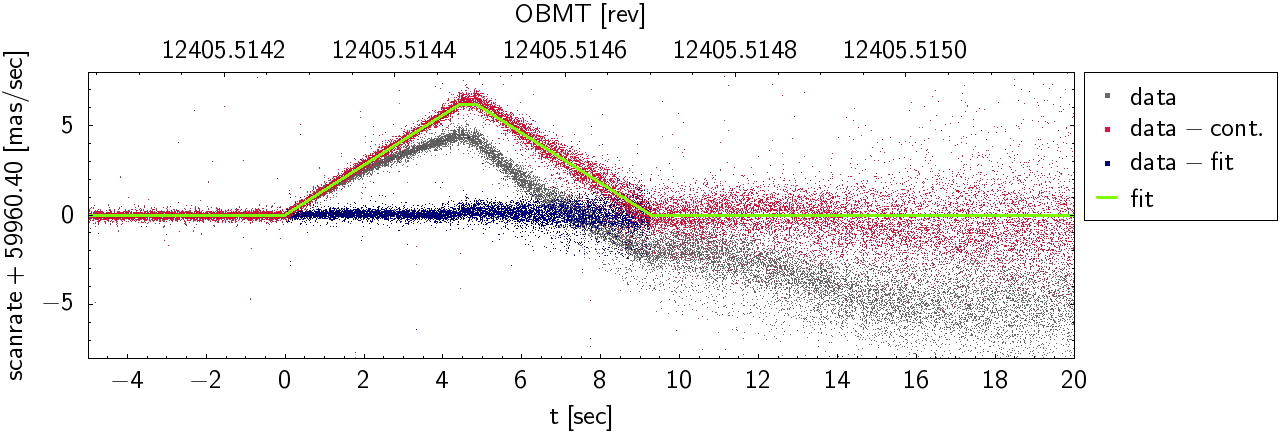}\\
        \includegraphics[width=\hsize]{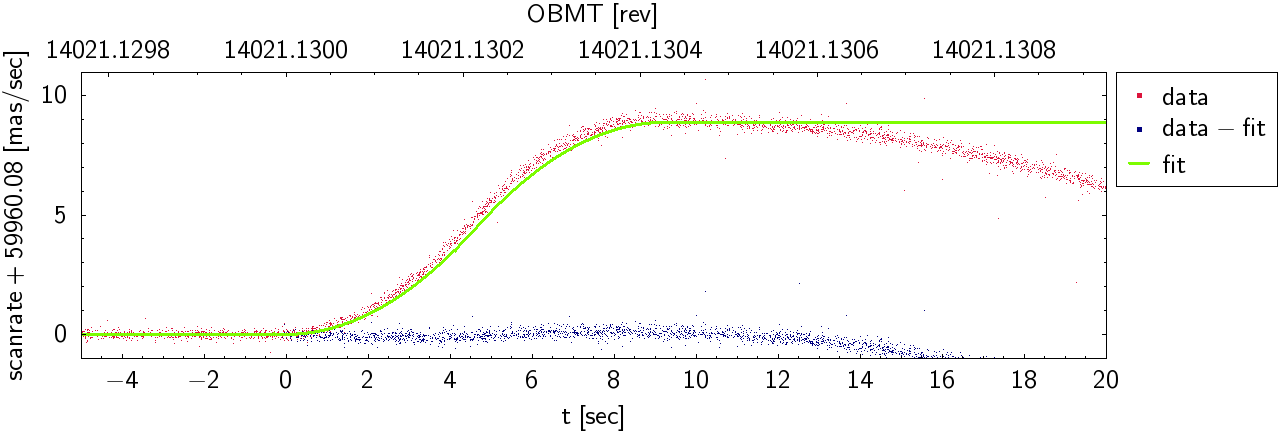}\\
        \includegraphics[width=\hsize]{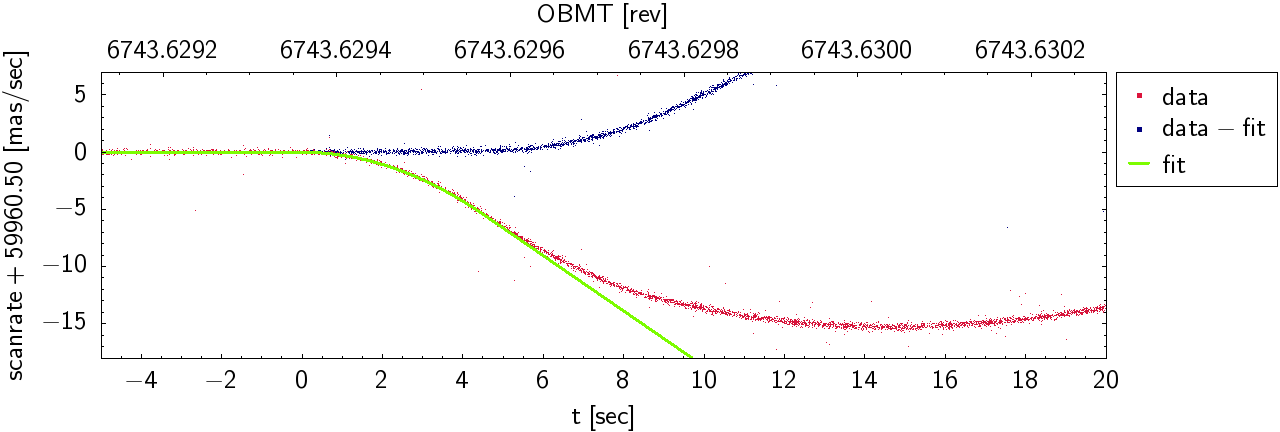}\\
        \caption{\small Pair rates for typical examples of 
         the three main types of rate excursion. The upper row
        shows a clank event that occurred at OBMT=12405.5142~rev, the second row a micro-meteoroid impact that occurred
        at OBMT=14021.1300~rev, 
        and the bottom row a propellant movement (oxidiser) that happened at OBMT=6743.6293~rev. 
        The plots show the unsubtracted data in red and the subtracted data in dark blue. 
        Since the clank, which was triggered by a micro-meteoroid event, 
        contained other rate excursions in the original data, these contaminations had to be removed first. Thus, in the
        upper panel, the original data is shown in grey, while the data with the contaminating features removed
         is shown in red. The deviations of the fits from the data, starting at 6 to 8 seconds after the events, 
        are due to the corrective actions of the on board AOCS.}   
       \label{fig:RE.fig}
\end{figure}

{\it Gaia} recorded the sky using the drift-scan method, i.e. continuously scanning the sky by rotating once around itself every 6~hours, 
and following a defined scan-law, called the Nominal Scanning Law (NSL), 
which ensured an optimal coverage of the complete sky. 
Thus, the astrometric measurements of {\it Gaia} are a combination of temporal 
and positional measurements. While in the ideal commanded world of the scanning-law, all motions are smooth, this is not the case in reality. Although great effort was
taken to ensure everything to be as smooth as possible, there were small residual deviations from the ideal.
The system, which ensured an optimised attitude was called the Attitude and Orbit Control System (AOCS). 
This subsystem relied on the current astrometric measurements of
the astrometric field, supplemented by input from the {\it Gaia} Star Tracker. It controlled and stabilised the attitude with the help of a micro-propulsion
system, which consisted of two sets of six cold gas thrusters (see \citet{2013A&A...551A..19R}), with the second set having been the redundancy. 

Apart from the attitude noise, occasionally larger deviations occurred, which are the main focus
of this section. These could  be due to the impact of small particles, i.e. micro-meteoroids; to internal processes, 
such as the spontaneous relaxation of a component
within the payload; or to propellant movement. Moreover, the AOCS could itself cause such deviations, 
as it relied, in addition to the actual measurements of a large subset of objects, on input from the star tracker. 
This device was far less precise, and its input catalogue contains some entries with degraded astrometric quality. 
The lack of proper
motions in this catalogue resulted in the coordinates of its entries degrading with time, especially for those objects with large 
proper motions. 
Furthermore, the AOCS was also responsible for the attitude oscillations after micro-clanks (see Sect.~\ref{sect:results:RE:clanks}), 
as initially it could not distinguish
between real disturbances, for example those caused by micro-meteoroid impacts, and the perceived ones, thus causing the micro-clanks.
In any of these instances, the AOCS counteracted all the attitude disturbances that could be detected by applying correctional torques via the thrusters. 
If the deviation of the attitude grew too large, a loss of convergence (LoC) was declared, 
and the data obtained during this situation is deemed compromised.
In more extreme cases, the data acquisition itself was disrupted, 
and the AOCS then commanded consecutively less precise modes of detector operation, 
or even determined the spacecraft orientation
from scratch, using the gyroscopes or sun sensors.

In order to have been able to take these deviations into account and to analyse them, the actual attitude needed
 to be derived. 
This was done in several steps (see Sect.~\ref{sect:FL:ODAS}). 
Here we  look at the most accurate attitude to which the FL had access to, the OGA2, produced by the ODAS. 
As described in Sect.~\ref{sect:FL:ODAS}
and \citet{ODASpaper}, this is one of the outputs of the ODAS, and is essentially a spline fit with a knot interval of $\sim$30 seconds. 

The aforementioned small attitude rate excursions and the resulting response of the AOCS usually only lasted
 a couple of seconds, 
which is unfortunately too short for the
temporal resolution of the OGA2 to accurately fit them. Therefore, another approach was derived at least for the AL direction. 
It relies on the actual astrometric data, and which allows for
a much more detailed analysis than relying on the difference between the OGA2 and the NSL. This is a prime example, where the data routinely used for the 
FL monitoring did not suffice, and further data had to be procured from the {\it Gaia} data base (see Sect.~\ref{sect:FL:Other_data}).

This method utilises the very nature of the {\it Gaia} measurements, i.e. being a drift scan over an array of several strips of detectors. 
When unperturbed, the passage time of an object over a given detector was constant, likewise the passage over the gap between two consecutive detectors. 
In {\it Gaia}'s case, an object traversed a CCD in 4.41~sec, and the detector gap in 0.45~sec. 
The nominal angular velocity of {\it Gaia} was $-$59.96''/sec.\footnote{The minus sign is a mere convention, caused by the definition of the field coordinates inside the FoV.} 
Any disturbance in the motion of {\it Gaia} in the along-scan direction showed up as a
deviation from this nominal value. As both the transition time across the light-sensitive surfaces of the detectors, and the light-insensitive gaps are finite,
the change of transition time caused by the perturbation incident gets smeared out (see \citet{2005A&A...438..745B}).
Therefore such a transition time difference analysis (also known as pair rate analysis) 
does not depict the true
shape of the event, but features a distinct integrated signature depending on the type of disruption. Figs.~\ref{fig:OGA2vsPR.fig} and \ref{fig:RE_OGA2vsPR.fig} show a short time interval of the
transition time differences, compared with the difference of the ODAS-derived OGA2 and the NSL, the OGA2$-$NSL. The overall similarities but also differences between the two
methods are evident, also highlighting the limitations of both. Given that the more precise method relies on the transition times over the FPA, it is only feasible in
the along-scan- and not the across-scan direction. 

From a purely mathematical point of view, one can distinguish between four basic types of disruptions, 
namely (i) an instantaneous change in the along-scan angular orientation, 
without a change in angular rate, (ii) an instantaneous change in the along-scan angular rate,
 without a jump in the pointing position, and (iii) and (iv) the same as (i) and (ii),
just with the events themselves having a finite duration. Reverting back to the actual, physical world, we do find real events for all four of the previously mentioned
event types, of which the important three are discussed here.
 Here, instantaneous means that the total duration of the event is much shorter than the integration time across the detector, 
i.e. $t_{\rm event}\ll4.41$~sec. In principle, the origin of attitude perturbations could lie both inside the spacecraft, or act from the outside on it. 
This origin had strong implications for the signature seen in the attitude. An intrinsic origin, for example a sudden minute movement of some component within the payload,
changes the distribution of the angular momentum within the payload, but not the total momentum of {\it Gaia} itself, 
which then, due to the conservation of angular momentum, leads to a change in
the overall spatial orientation, without changing the rotation velocity. Conversely, an external mass acting on the spacecraft will transfer momentum, leading to
a true change in the angular momentum and thus the rotation speed. The corresponding signatures on the attitude 
differed accordingly. In order to classify and
quantify individual events, we have derived a set of parametrisations, which fit the different signatures of the transition times (see Fig.~\ref{fig:PR_scheme.fig} for a schematic representation). As the equations themselves are rather specialised, 
these are kept out of the main text, but 
given in Appendix~\ref{sect:Appendix_B}. 

These rate excursion events degraded the attitude, if they were large enough. 
Therefore the AOCS would intervene, and attempt to correct for the perturbation. 
Since the AOCS started its intervention
after a couple of seconds, the actual shapes for events larger than a certain magnitude do not look like in the schematic representation shown in 
Fig.~\ref{fig:PR_scheme.fig}, but start to deviate from the model after about 6-8 seconds, as seen in Fig.~\ref{fig:RE.fig}.  

In the following, we  briefly explain the most important types of attitude perturbations.
\subsubsection{Micro-clanks}\label{sect:results:RE:clanks}
The micro-clank is an example of an instantaneous internal perturbation event. While the exact origin of clanks is unknown, they are most likely related to flexing of the 
Deployable Sunshield Assembly (DSA). Micro-clanks have also been recorded in {\it Gaia}'s predecessor mission, 
Hipparcos (see \citet{2007ASSL..350.....V}). 
The seemingly undisturbed stretch of attitude data shown in Fig.~\ref{fig:OGA2vsPR.fig} contains several such
small features, the larger ones being quite prominent.
But upon closer inspection more micro-clanks with very small amplitude can be identified, some overlapping each other.
While small micro-clanks are ubiquitous in the data, larger micro-clanks occurred in the aftermath of larger velocity 
changes, such as station-keeping manoeuvres, when whole series of large amplitude micro-clanks were observed.
 They could also be triggered by micro-meteoroid impacts (see also 
Sect.~\ref{sect:results:RE:MM}), which is the case for the specimen shown in the top panel of Fig.~\ref{fig:RE.fig},
where the triggering micro-meteoroid impact occurred almost simultaneously. 
The scan-rate data shows the characteristic trapezium shape, indicating an instantaneous change in the orientation of the satellite without a change in the rotation velocity. 
Given the perpetual appearance of smaller micro-clanks this type of perturbation is the most frequent.
\subsubsection{Micro-meteoroid impacts}\label{sect:results:RE:MM}
One of the hazards concerning any spacecraft is the impact of a small body, due to the usually high velocity differences between the two. 
In the Sun-Earth L2 region these micro-meteoroids are particles of interplanetary dust. The signature of such an event 
is shown in Fig.~\ref{fig:RE_OGA2vsPR.fig}, comparing the OGA2$-$NSL rates and the pair rates,
 and the middle panel of Fig.~\ref{fig:RE.fig}. In contrast to the clank, which shows a trapezium shaped signature
in the scan-rate data, a double-parabolic trend with an intervening linear segment is seen, due to the change in angular
velocity caused by the impactor. The physical reaction by the AOCS 
to a disturbing impulse was typically delayed by several seconds due to the time required to measure the 
star transits in the focal plane and the communication protocols between the on board systems.
This can be seen in Fig.~\ref{fig:RE.fig} 
in the deviation of the actual measurements from the model starting after about 11~sec (middle panel) and 6~sec (bottom panel) in the cases shown. 
It is to be noted that a micro-meteoroid impact only appears in the along-scan attitude 
if it results in a change of the angular momentum, i.e. the rotation speed. Should the impact be close to the axis of rotation 
or in perfect radial direction from or towards the axis of rotation, 
no perturbation of the along-scan attitude may be seen, even if the impact was substantial. 
In Sect.~\ref{sect:results:MM}, in which we discuss some of the most
noteworthy and momentous impacts, several of these did not cause any noticeable attitude deviation.  

\subsubsection{Propellant movement}\label{sect:results:RE:PM} 

The final type of perturbation is also the longest lasting, with the duration of the disturbance generally being larger than the integration time of 4.41~sec.
It is supposed to be caused by moving bubbles of propellant liquid. 
The parametrisation of this type of event is very difficult, as the  exact nature of the motion of the liquid in question is generally not known. An example is depicted in
the bottom panel of Fig.~\ref{fig:RE.fig}. 
 
\subsubsection{Quasi-periodic rate excursions}\label{sect:results:RE:QPRE}
\begin{figure}[ht]
       \centering
        \includegraphics[width=1.00\hsize]{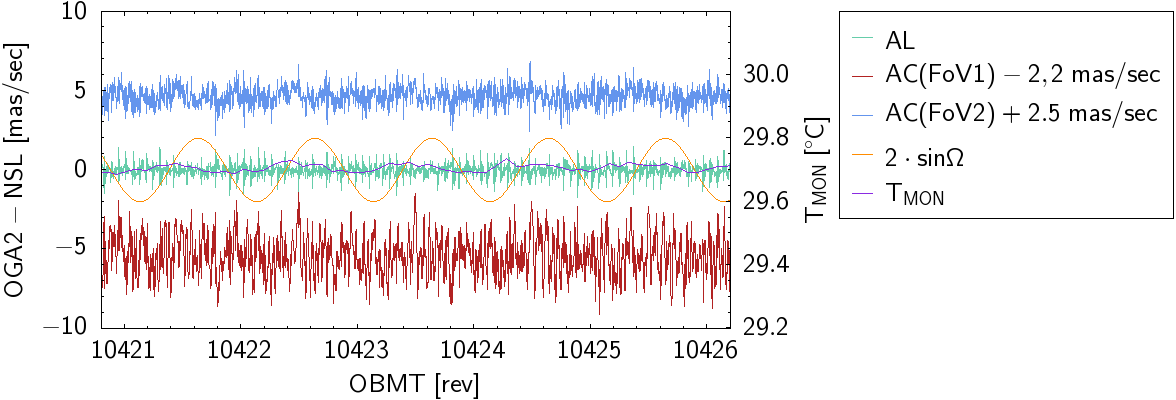}
        \includegraphics[width=1.00\hsize]{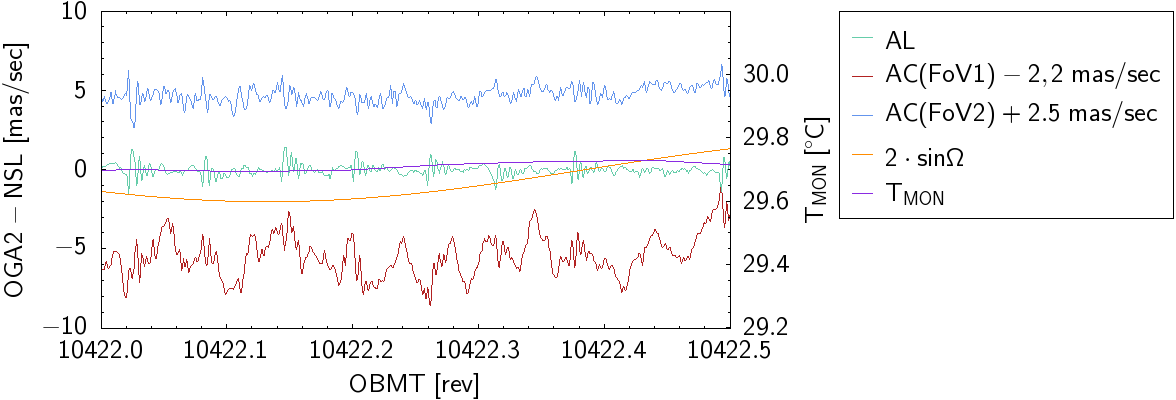}
        \includegraphics[width=1.00\hsize]{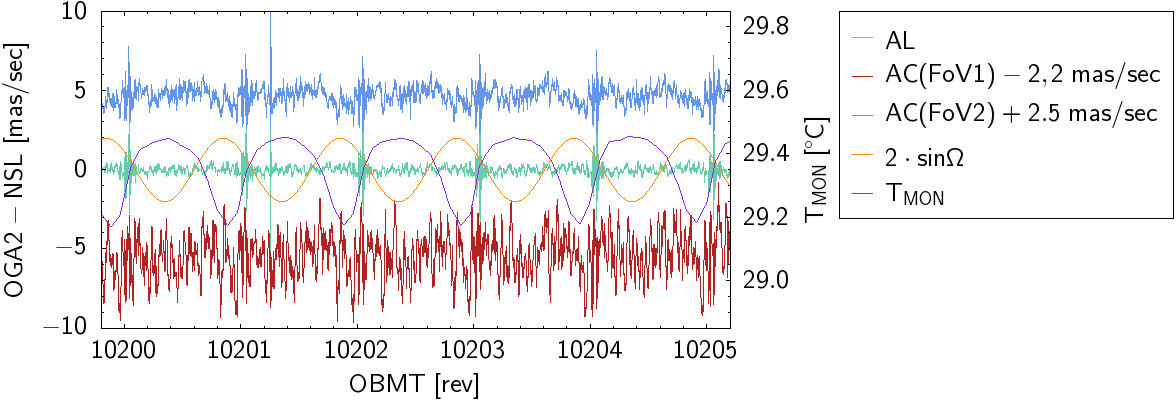}
        \caption{\small The two types of quasi-periodic rate excursions (QPREs). The upper panel shows the high-cadence QPREs, 
        which generally occur at time intervals between
        15 and 25~minutes (0.04 and 0.07~rev). The centre panel is a zoomed-in image  highlighting the individual QPRE signatures (especially in the
AL direction, green curve). The lower panel shows the higher-amplitude QPREs with a cadence of one heliotropic revolution. The two across-scan curves have been shifted
by the amount indicated in the  legend, so that the overlap between the three curves is reduced. Additionally, we have plotted 
the sine of the heliotropic angle ($\Omega$)
and the temperature curve of one of the oxidiser temperatures.}
       \label{fig:OGA2NSL_QPRE.fig}
\end{figure}
While in general the rate excursions described in the previous sections occurred irregularly and unpredictably, 
apart from series of large clanks after a SKM, there were time intervals in which rate excursions occurred regularly and persistently. 
These are known as quasi-periodic rate excursions (QPREs). Two distinct types of QPREs have been
identified, samples of which are shown in Fig.~\ref{fig:OGA2NSL_QPRE.fig}.

The first, and more common type had a typical cadence of between about 15 and 25 minutes, i.e. 0.04 to 0.07~rev. 
Observed cadences have been 16-17 minutes and 21-22 minutes. The magnitudes of the individual events are small, 
usually in the vicinity of 1~mas/sec or less in the AL direction. An example of this type is shown in the upper and centre panel
 of Fig.~\ref{fig:OGA2NSL_QPRE.fig}. 
In the majority of cases, the start of a series of high-cadence QPREs is preceded by an initial larger rate excursion event.
As these events were remarkably similar, it is presumed that the same physical origin is the cause of these high-cadence QPREs. 
Additionally, temperature changes in the fuel tanks are associated with the occurrence of these QPREs.
Often the series have been stopped by a routine intervention into the {\it Gaia} spacecraft, such as a SKM. 
Conversely, some series also started in the aftermath of a SKM. 
However, not every SKM stopped an on-going QPRE series, and much less initiated one. This also means that these 
series often lasted over months. 
Given their slow evolution and low magnitude, below the thresholds of LoCs, the scientific quality of the data derived during these timespan 
is not affected in a significant way, as they can be well modelled in the data processing.

The second type are the 1-rev-cadence QPREs. We note that the exact cadence is slightly less than one event per revolution since they follow the heliotropic
angle, accounting for the apparent motion of {\it Gaia} around the Sun. 
These QPREs, having a much larger magnitude, are fortunately far less common. Here, the magnitude is usually in the region of 3~mas/sec, but there have been
times, where it could go up to 7 or 8~mas/sec, causing losses of convergence here and there. There is a strong relation between the magnitude of these 
1-rev-cadence QPREs and the oscillation amplitude of the temperatures in the oxidiser tanks (see Fig.~\ref{fig:Long_term_Temps.fig}). In the lower panel of this figure,
two time intervals with a high-amplitude temperature oscillation can be seen in two of the temperature curves. The first one lasted from about 
OBMT=10200~rev to 10350~rev, and the second one from 10600 to 10800~rev. This amplitude has a direct correlation to the magnitude of the resulting QPREs, with a lower 
threshold for the appearance of these disturbances near 0.1~K. During most of the mission these two oxidiser temperatures have shown 1~rev oscillations
with amplitudes between 0.02 to 0.05~K, for which no corresponding QPREs were seen in the attitude. 

It is interesting to note that both types of QPRE series appeared only between July 2019 and April 2023, 
with the lower-cadence higher-magnitude events being even more
constrained in time. Furthermore, these series started more often in northern hemisphere autumn than in spring. 
The reason for this remains unknown, possibly it is related to the filling level of the propellant tanks.      
\subsection{Effects of micro-meteoroids on the payload}\label{sect:results:MM}

\begin{figure}[ht]
       \centering
       \includegraphics[width=0.48\hsize]{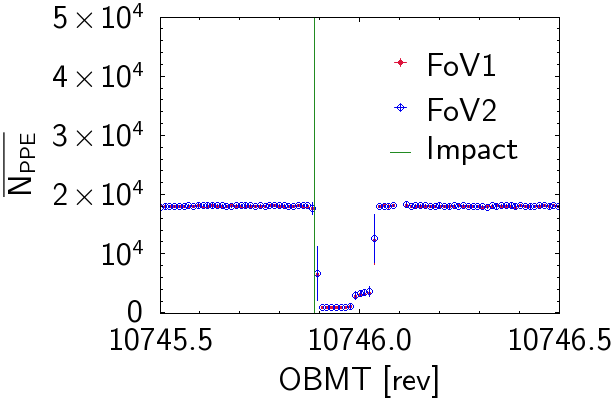}
       \includegraphics[width=0.48\hsize]{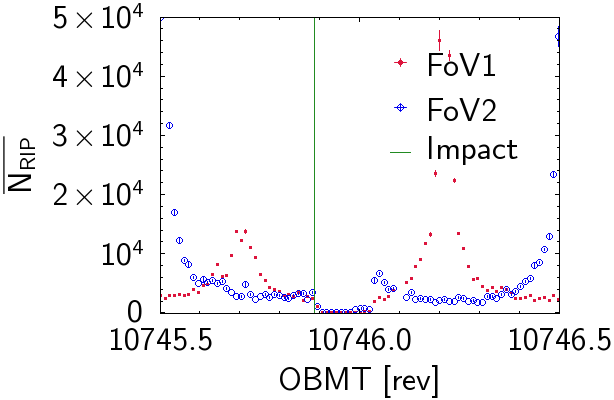}\\
       \includegraphics[width=0.48\hsize]{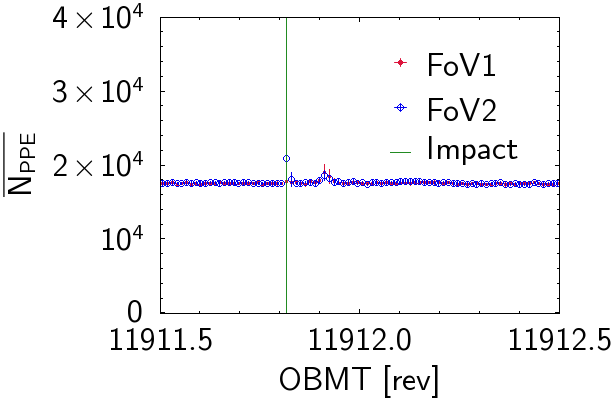}
       \includegraphics[width=0.48\hsize]{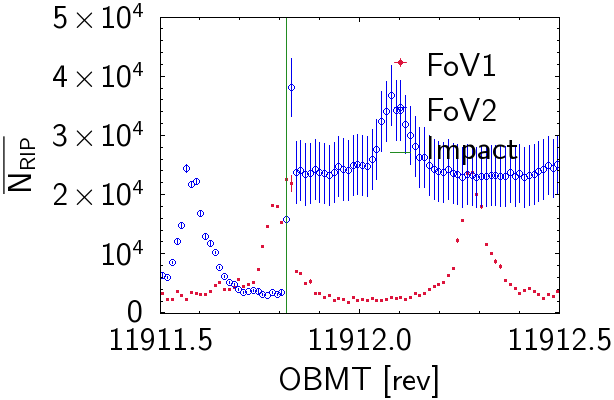}\\
       \includegraphics[width=0.48\hsize]{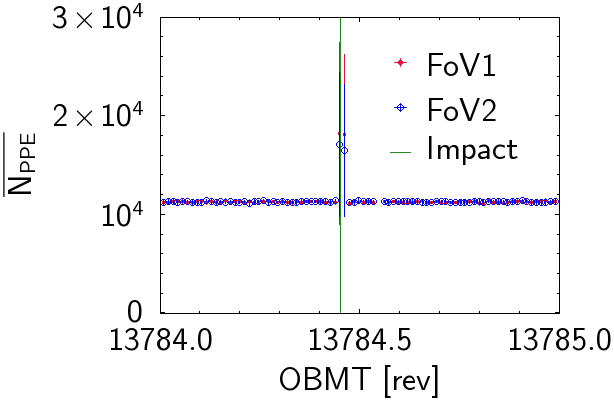}
       \includegraphics[width=0.48\hsize]{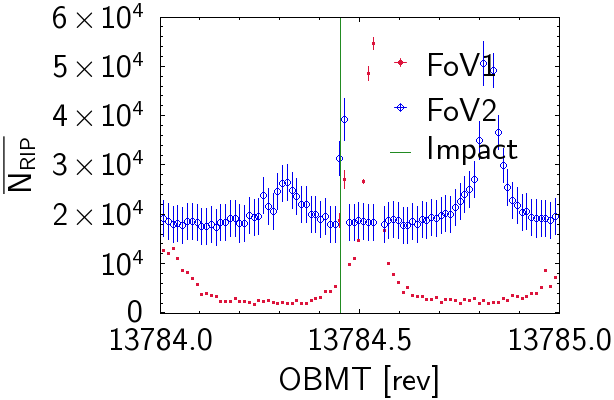}\\
       \includegraphics[width=0.48\hsize]{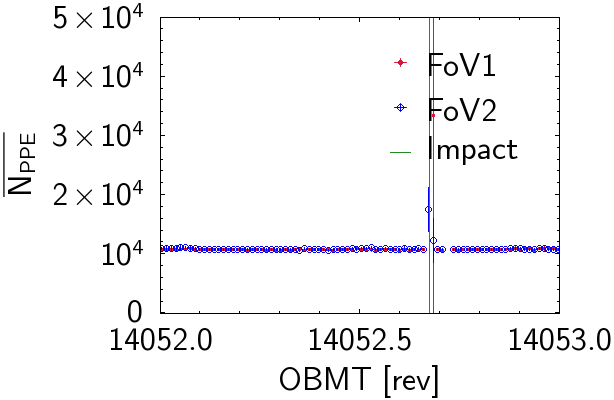}
       \includegraphics[width=0.48\hsize]{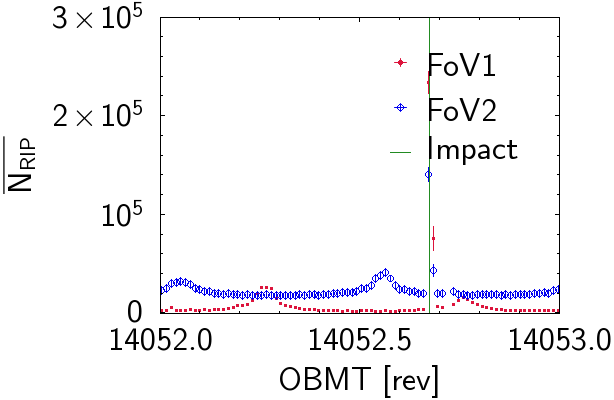}\\
       \includegraphics[width=0.48\hsize]{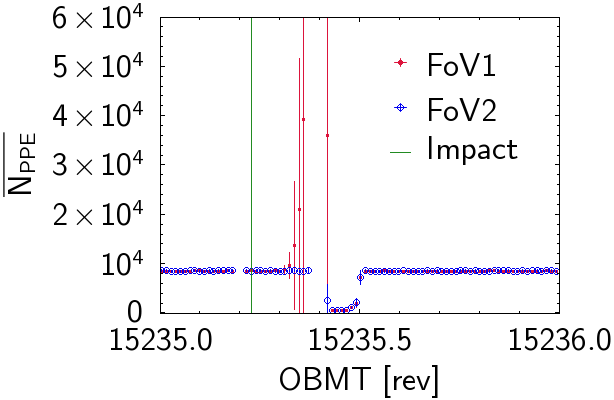}
       \includegraphics[width=0.48\hsize]{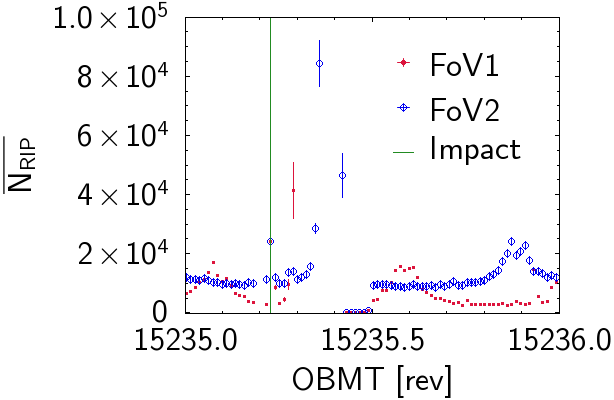}
        \caption{\small PPE and RIP count rates around each of the five micro-meteoroid impact events discussed in this section. Shown are for each FoV the mean values over all seven CCD rows, and their 
       standard deviation (error bars), which for the RIP is scaled by $\frac{1}{10}$ for better visibility. The left column depicts the PPE rates and the right column the rejected ripples (RIP).}
       \label{fig:MM_ASD4.fig}
\end{figure}

\begin{figure}[ht]
       \centering
       \includegraphics[width=0.48\hsize]{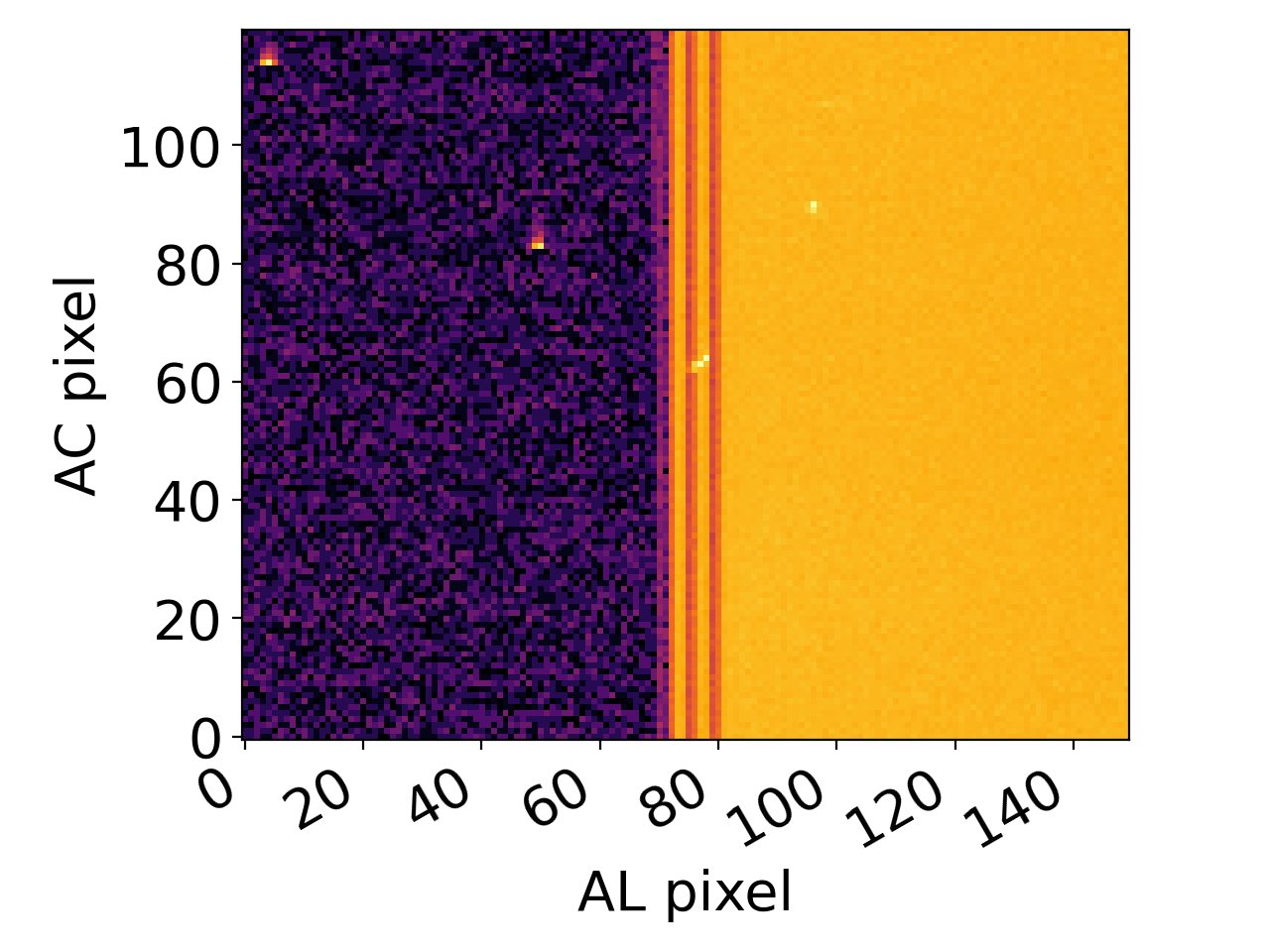}
       \includegraphics[width=0.48\hsize]{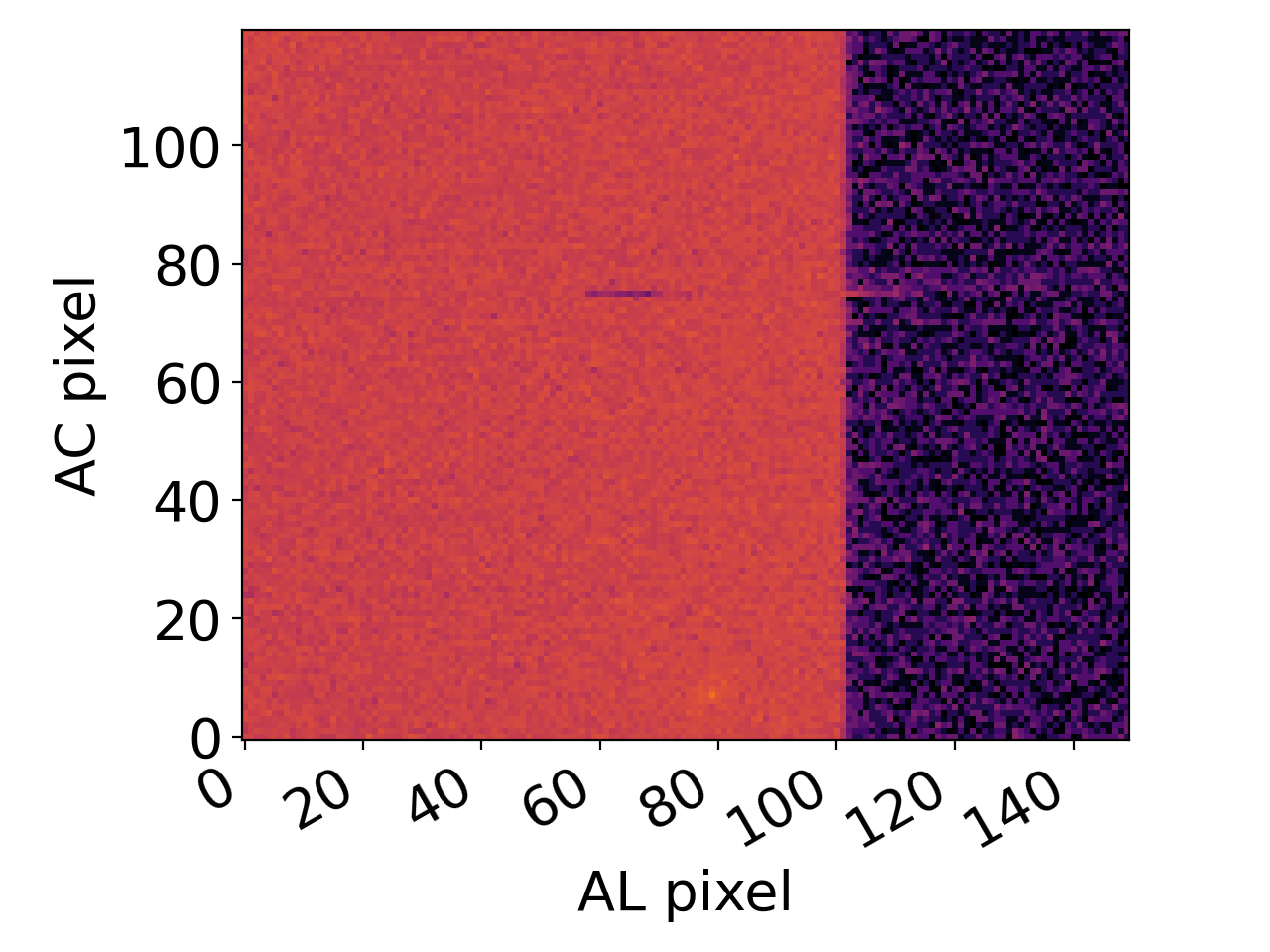}
       \includegraphics[width=1.00\hsize]{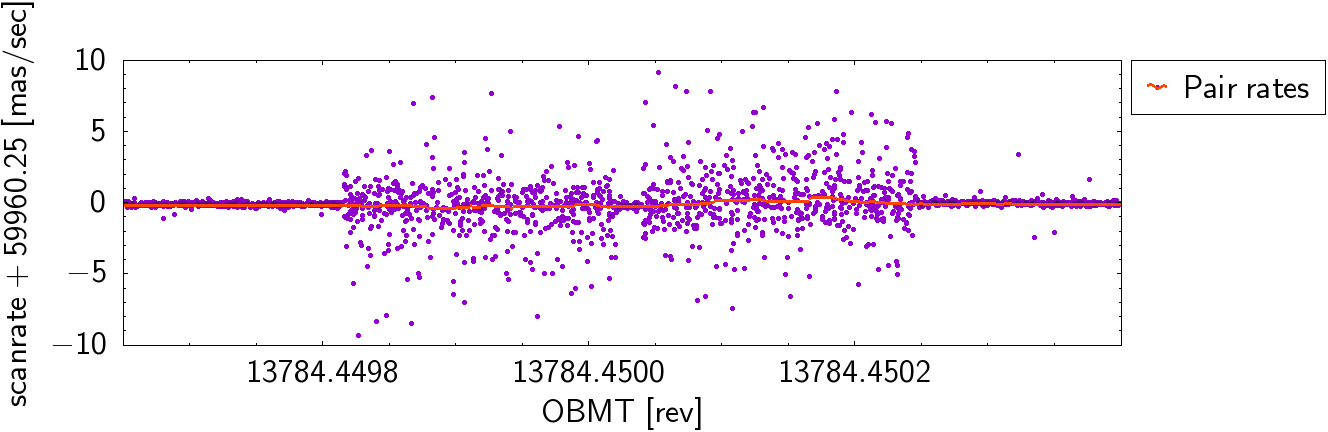}
 \caption{\small Micro-meteoroid impact event of April 5, 2023. 
The upper panels show the Wave Front Sensor  images at the start and the end of the optical flash. 
The lighter colours show the illuminated part. We note that the WFS CCDs are read out in a similar way as the other CCDs of the {\it Gaia} FPA. 
The lower panel depicts 
the pair rates. The red line shows a continuous median.}
       \label{fig:MM_PR_13785.fig}
\end{figure}
\begin{figure}[ht]
       \centering
       \includegraphics[width=1.00\hsize]{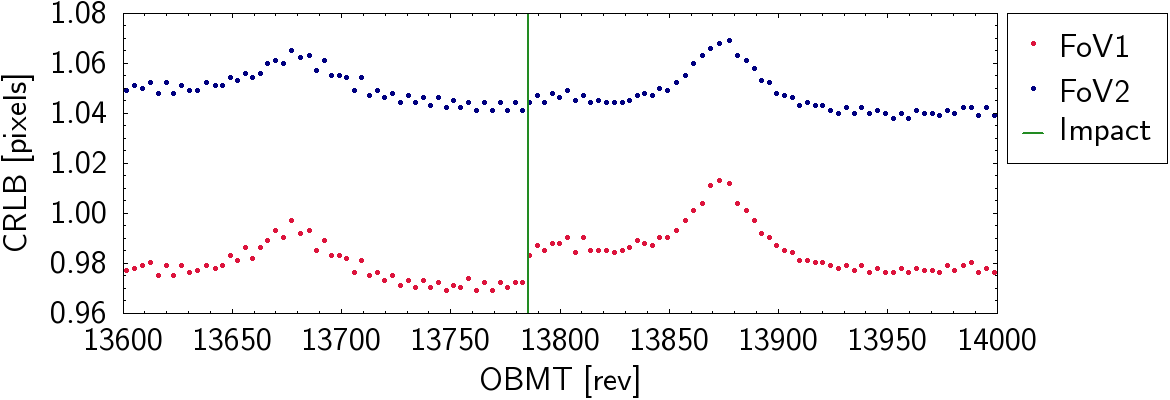}
\caption{\small Cram\'er--Rao lower bound at the time of the impact of April 5, 2023. FoV1 is shown in red, FoV2 in blue, and the incident is indicated as a green vertical line.}
       \label{fig:MM_CRLB_13785.fig}
\end{figure}

\begin{figure}[ht]
       \centering
       \includegraphics[width=1.00\hsize]{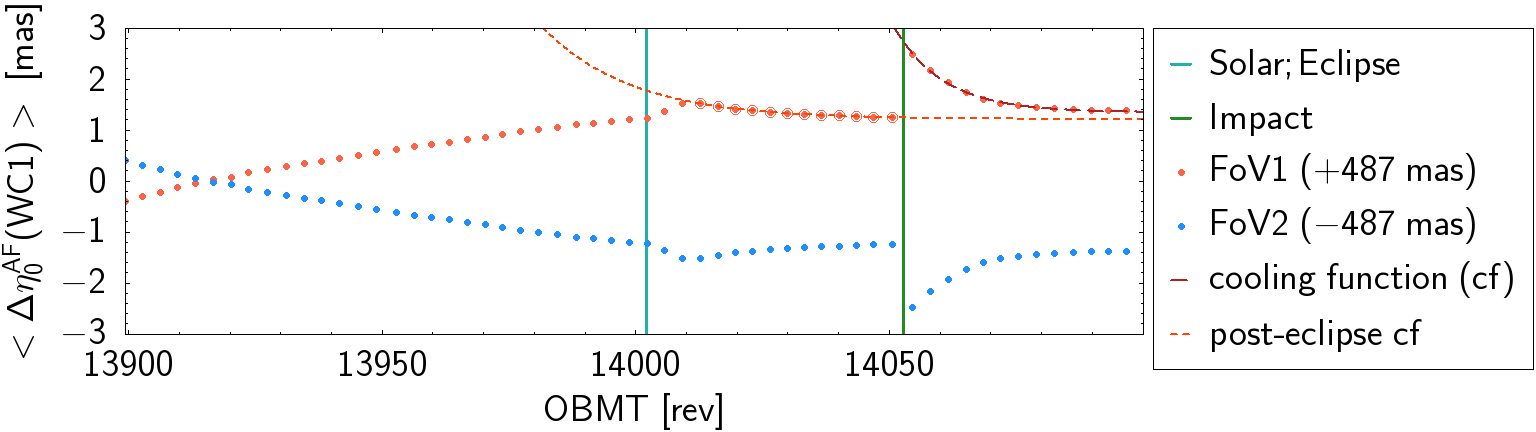}
\caption{\small Zeroth-order astrometric calibration averaged over all AF detectors at the time of the solar eclipse by the moon 
and micro-meteoroid impact in May--June 2023. 
FoV1 is depicted in  red, and FoV2 in  blue. 
Exponential functions, parametrising the relaxation process after both events have been fitted to the data and are shown here for the FoV1 data. 
For the eclipse the data points used for its
computation are indicated by larger open circles around the data points (for the impact, all data points after the event are used for the computation).
We note that for reasons of visibility the data has been shifted by $\pm487$ mas.}
       \label{fig:MM_LSCP0_14052.fig}
\end{figure}

\begin{figure}[ht]
       \centering
       \includegraphics[width=1.00\hsize]{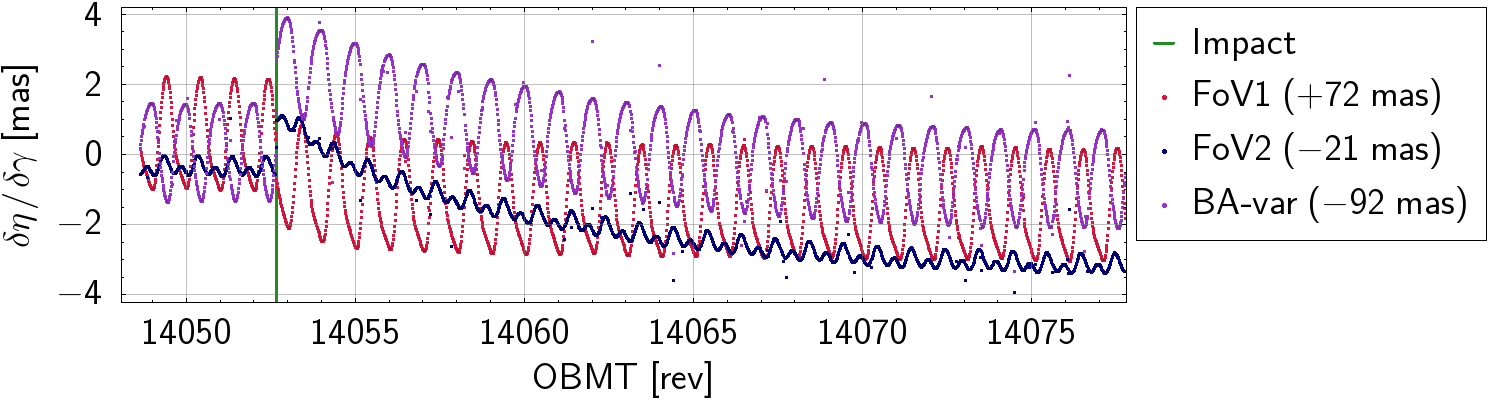}
       \includegraphics[width=1.00\hsize]{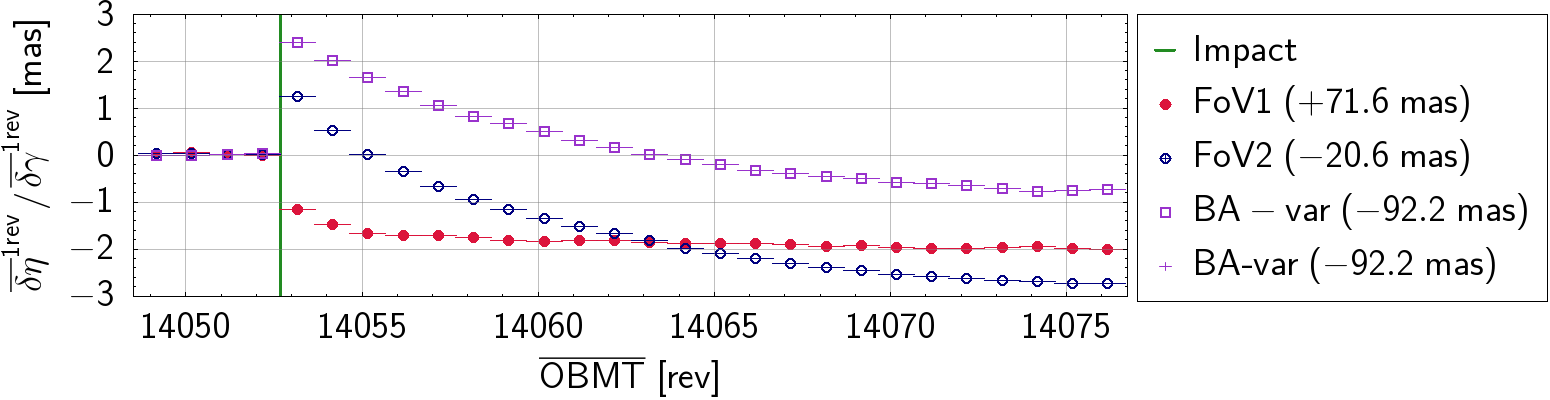}
\caption{\small Basic angle during the time of the impact of June 12, 2023. The upper panel shows the BA with the same binning as Figs.~\ref{fig:BAvar.fig} 
and \ref{fig:BAvar_jumps.fig}, while the lower panel is averaged over 6 hours (one revolution) per point, thus removing the 6-hour oscillation.
We note that for reasons of visibility the data has been shifted by the amounts stated in the legend.}
       \label{fig:MM_BA_14052.fig}
\end{figure}

\begin{figure}[ht]
       \centering
       \includegraphics[width=1.00\hsize]{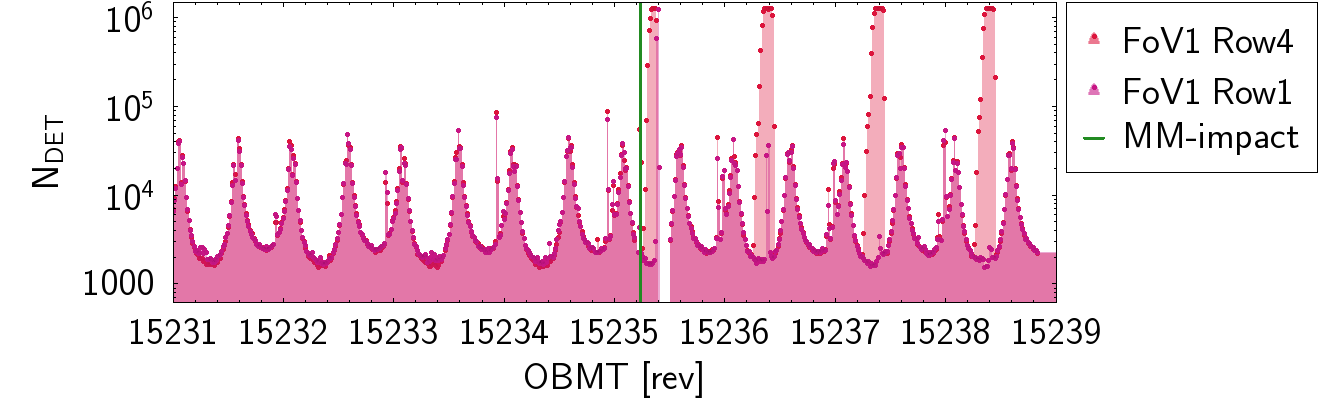}
       \includegraphics[width=1.00\hsize]{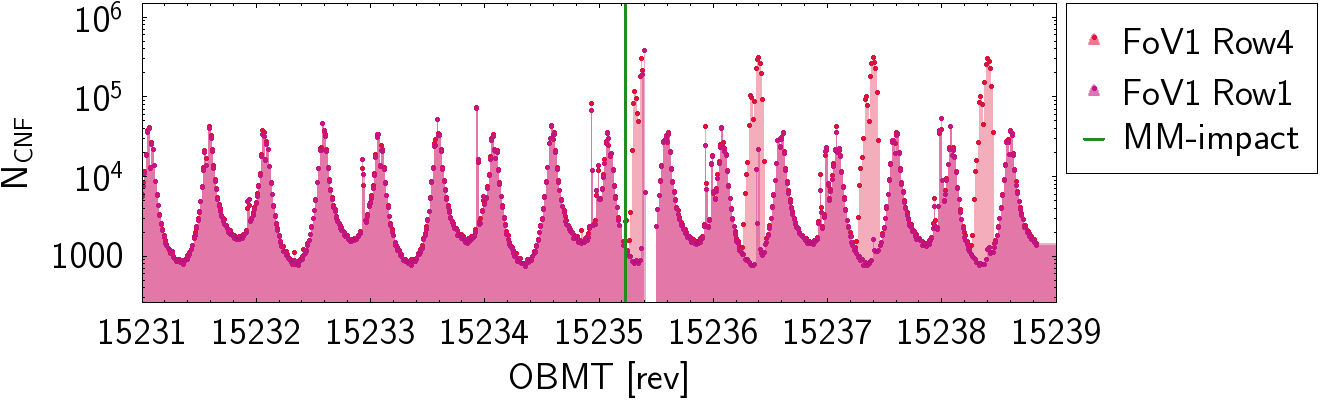}
       \caption{\small ASD4 count-rates for FoV1 during the micro-meteoroid impact event of April 2, 2024.
                Shown is the most affected Row~4 in red and for comparison Row~1, which is only minimally affected (in dark red). 
                The upper panel shows the counts of the detected objects and the lower panel those of the objects
                confirmed by the on board  procedure. The initial micro-meteorite impact is indicated by a green vertical line.}
       \label{fig:MM_NWC2_15235.fig}
\end{figure}

\begin{figure}[ht]
       \centering
       \includegraphics[width=1.00\hsize]{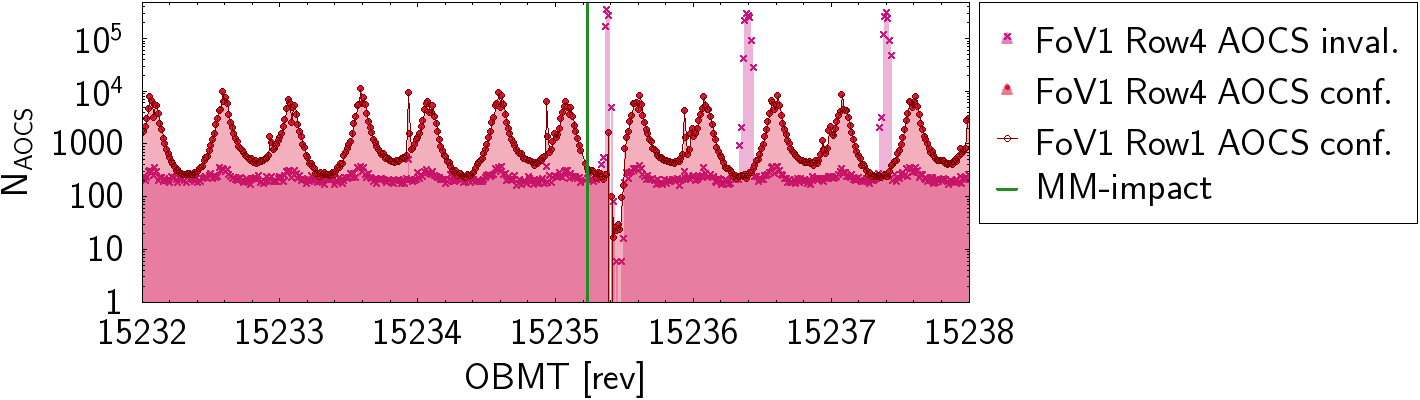}
       \includegraphics[width=1.00\hsize]{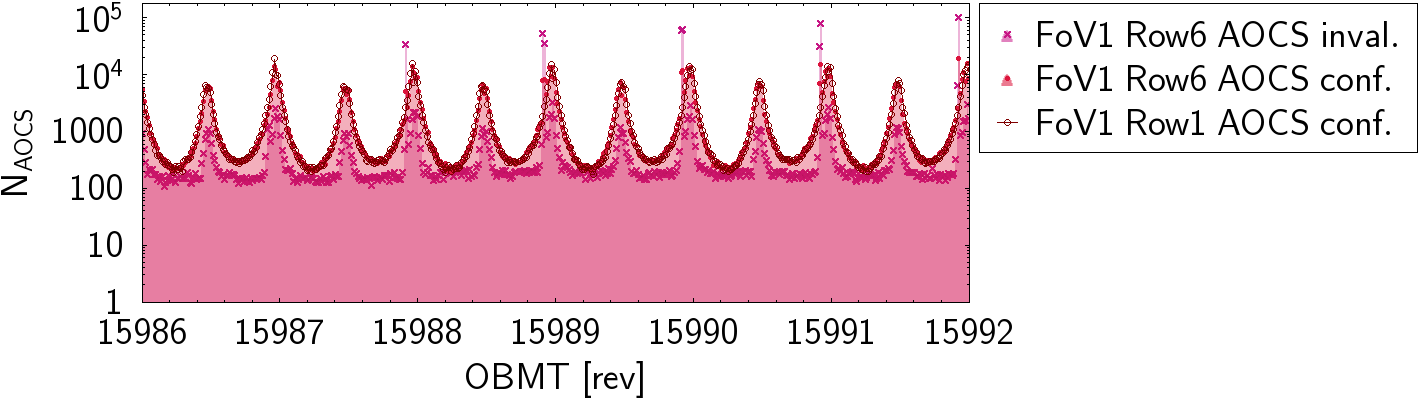}
       \caption{\small Efficiency of the AOCS object confirmation after the initial event of April 2, 2024 (upper panel), 
       and after the stray-light enhancement in October 2024 (lower panel). The upper panel shows FoV1, Row~4, 
       the most affected row;  the lower panel shows FoV1, Row~6 as the Row~4 data was inhibited for  
       usage for the AOCS during the stray-light peaks. The ASD4 counts for the objects invalidated for the AOCS use 
       are shown in dark red, while the confirmed objects are depicted in red. In both panels, FoV1, Row~1 is shown for comparison. 
       The initial MM impact is shown as a green line in the upper panel.}
       \label{fig:MM_NAOCS_SC106.fig} 
\end{figure}

After having discussed the long-term evolution of {\it Gaia} in Sect.~\ref{sect:FL:Trends} and 
presented one type of regular occurrences, i.e. attitude rate excursions in Sect.~\ref{sect:results:RE}, 
we now turn to 
unplanned disruptive events, using as an example larger micro-meteoroid impacts. 
Most of such impact events, as described in Sect.~\ref{sect:results:RE} have passed without serious consequences, only showing an attitude rate 
excursion, at worst being labelled as an LoC, which is then quickly corrected by the AOCS. 
However, some specific impacts did have damaging consequences in the near term or in some cases even longer-term or permanently. 
Every small body hitting a satellite will cause some amount of damage, as its kinetic energy is deposited onto the spacecraft,
 and that within a very small cross-section. In most cases this will be a scratch, dent, or even hole in a non-critical area. 

In order to present the diverse consequences in the aftermath of micro-meteoroid impacts, we have selected five noteworthy examples. These and their impact on the payload are summarised in Table~\ref{tab:MMevents}.
This sample includes the largest-magnitude impact recorded during the mission, as well as four others, which led to some change in the payload environment. 
All of these events have been intensively analysed by
the FL team in collaborations with other groups within DPAC. 

Before we start off, we need to briefly mention one instrument on board  {\it Gaia},
 which was not normally used for FL activities, and to which the FL usually did not have access to. 
This is the wavefront sensor (WFS), which
was primarily used in the very early phases of the mission during the initial set-up and calibrations after the launch. 
The functionality of this instrument is beyond the scope of this study, and it is relevant here only
because it was sensitive to light of the night sky coming through the apertures, and it made full-frame images, 
rather than the tiny cutout windows
 usually made by the science detectors. Other than that fact, the WFS had
the same scanning properties as the science detectors. Therefore flashes entering the optical assembly, 
for example due to an impact of a small body, would show up in the WFS data. This was also seen in the ASD4 count rates
(see Sect.~\ref{sect:FL:ASD}), especially the PPE and ripples counters. 
Unfortunately when an impact occurred on the far side of the apertures, nothing was seen, neither in the WFS nor in the ASD4 counters.
As the WFS was continuously read out in TDI mode, it is possible to directly measure the time and duration of the recorded illumination. An example is shown in the upper panels of Fig.~\ref{fig:MM_PR_13785.fig}.

The first incident listed in Table~\ref{tab:MMevents}, which took place on March 7, 2021, turned out to be the micro-meteoroid impact with the highest transfer of energy. 
This led to the nominal data taking ceasing immediately, and the AOCS cycled into its lowest-precision mode in order to keep control. 
Therefore there is no OGA2 available for this impact as shown in Table~\ref{tab:MMevents}. Likewise the ASD4 counters,
which supply the statistics of observations in regular mode, stopped counting at the time of impact, 
as evident from the top panel in Fig.~\ref{fig:MM_ASD4.fig}, where the PPE and ripples counts went to zero.
 Additionally, lacking actual astrometric measurements, no pair rate analysis (see Sect.~\ref{sect:results:RE}) could be made.
Therefore all information on the magnitude of the impact comes from other sources, such as the star tracker unit and optical
 gyroscope data. In the along-scan direction, the velocity deviation
was more than 1000~mas/sec, in the across-scan direction, it was more than 6300~mas/sec. 
The Sun Aspect Angle (SAA) of 45$^\circ$ suffered an instant deviation of more than 3'. 
While it took more than one hour
to mitigate the situation, so that nominal operations could continue, there have been no obvious detrimental consequences for the payload. The only noticeable change was a persistent temperature increase in one of the 
sun-shield arrays, presumably where the impactor hit. This shows that it is not necessarily the most energetic impacts causing issues in the payload.  

A second noteworthy event occurred on Dec. 23, 2021 (OBMT=11911.94~rev). 
In this case, the rate excursion resulting from the impact was rather small; however,  the impact did  show up in the WFS data. No impression of this event was seen in the CRLB,
the BA, the temperatures, or the astrometric calibration. However, in the ASD4 ripples counts of the SM2\_7 detector a dramatic increase in the counts from the usual few thousand per data bin to about 150,000 counts per
bin was observed. This could be attributed to a damage in a couple of pixel columns in this particular detector, possibly caused by debris originating from the impact on one of the primary mirrors. The signal in the SM2\_7 ripples count rates can be
clearly seen in the corresponding panel in Fig.~\ref{fig:MM_ASD4.fig} as a step change occurring at the time of the impact. The persistence of this change is also seen in all later ripples count plots in 
Fig.~\ref{fig:MM_ASD4.fig}, by the higher level of the mean value, and the significantly increased standard deviation. 
While this is a near-permanent damage, the consequences for the scientific performance of the spacecraft
have been limited to a slightly higher number of objects classified on board as ripples. 
As such detections are rejected and not sent down to Earth along with the valid data, this is a negligible degradation.

Another remarkable incident took place on April 5, 2023 (OBMT=13784.45~rev). This one was first detected by the bright WFS flashes, 
shown in Fig.~\ref{fig:MM_PR_13785.fig} (upper panels). In this case, no sign of a significant
rate excursion was seen in the OGA2$-$NSL data. The pair rate data shown in the lower panel of Fig.~\ref{fig:MM_PR_13785.fig} did not show any 
systematic deviation in the scan rates. What it did show, however, is that the 
scatter was significantly enhanced in two 4.41~sec time intervals separated by a 0.45~sec interval. This is due to the illumination caused by the flash of the impact, i.e. the light also recorded by the WFS. This caused
the background to be strongly enhanced during the duration of the flash, 
which can be assumed to have been instantaneous as verified by the WFS data. 
This in turn led to the quality of the position measurements to be severely degraded, resulting in increased scatter for all objects being observed at the time of the impact. As the pair rate analysis compares the positions
on consecutive detectors, the second timespan with large scatter is caused by the comparison of good data obtained after the flash with the degraded data during the event. This impact also resulted in
strong peaks in both the PPE and ripples ASD4 counters (see Fig.~\ref{fig:MM_ASD4.fig}. More detailed analysis had shown that there was a minor persistent increase in the ripples counts for the SM2\_4 detector, albeit
much smaller than the strong increase in the SM2\_7 ripples counts after the December 2021 event. Both the FoV1 AL fringe position branch 
of the BAM (see Fig.~\ref{fig:BAvar_jumps.fig}, 
upper panel) and the astrometric calibration exhibited a jump. More 
severe was the increase in the FoV1 CRLB of about 0.015 pixels, shown in Fig.~\ref{fig:MM_CRLB_13785.fig}. While in the other aperture the effect was negligible and only temporary, it was quite pronounced in FoV1, also 
being visible in Fig.~\ref{fig:Long_term_CRLB.fig}, which shows the long-term evolution on the CRLB. Very likely the micro-meteoroid had dislodged or slightly damaged something in the optical path of FoV1. This is also 
supported by the fact that the AL fringe position branch had shown a jump in FoV1. The question remains, how can the lack of a significant rate excursion as seen in both the OGA2$-$NSL and astro-elementary data comply with 
the significant effects on various components of the payload? Most likely the impacting body struck the satellite very close 
to the latter's centre of gravity, so that no discernible changes in the rotational behaviour was seen, despite the impact being of substantial energy.  

The fourth event listed in Table~\ref{tab:MMevents}, occurred on June 12, 2023 (OBMT=14052.67~rev). A strong rate excursion, qualifying as a LOC, was observed, 
most dominantly in the across-scan direction. In this case there was also a bright flash, as indicated by the PPE and ripple counter data, as well as the 
WFS response. The CRLB did not show any response to this impact. 
The thermal changes were striking, as they were detected in all measurements by the FPA detectors and many of the devices on the mirrors as well. 
By the time of this impact,  the payload was still relaxing from a partial obscuration of the Sun by the Moon,
which took place on May 30, 2023 (OBMT$\simeq$14002~rev). 
These longer-term changes in the temperatures left a persisting imprint
on both the astrometric calibration (see Fig.~\ref{fig:MM_LSCP0_14052.fig}) and the BA (see Fig.~\ref{fig:MM_BA_14052.fig}). 
The former suffered
a 1.5~mas jump in both FoV, resulting in a total offset of about 3 mas, which was about five times 
the size of the jump observed in the event of April 5, 2023. 
After the event, as the temperatures started to rebound, the astrometric calibration followed in a similar fashion, 
relaxing in an exponential way, as shown in Fig.~\ref{fig:MM_LSCP0_14052.fig}. 
This behaviour is also very alike to the aftermath of the partial obscuration of the Sun, 
which is also indicated in
Fig.~\ref{fig:MM_LSCP0_14052.fig}. The same holds true for the Basic Angle. In contrast to the previous impact, a jump was seen 
in both branches of the BAM measurements (see Table~\ref{tab:MMevents}), albeit in opposite direction, so that unlike in most bilateral jumps 
(see Sect.~\ref{FL:Trends:BAV}) the resulting jump in the 
Basic angle was larger than in each of the branches. After the jump, the mean values
started to relax in an exponential manner (see Fig.~\ref{fig:MM_BA_14052.fig}). Eventually, the BA settled asymptotically on a value about 
0.8~mas lower than before the impact. It thus seems that there was a combination effect at play.  
The mechanical disturbance was 
due to the micro-meteoroid impact on one of the primary mirrors, causing the jumps, while the
longer-term thermal change, which caused the BA and the astrometric calibration to evolve to new average values, was caused by the thermal radiation of the
evaporating impactor. While these effects were noticeable,
they are subtle enough to not cause significant adverse effects on the payload.

The final impact discussed in this paper had shown itself to be the most momentous for the remainder of the mission. 
On April 2, 2024, {\it Gaia} was impacted by a micro-meteoroid. The impact itself was large but rather unremarkable. 
While it was clearly seen in the attitude data, 
in the OGA2$-$NSL and the scan rates, and was large enough to lead to a loss of AOCS convergence, albeit not an extreme one, 
it did not show up in the WFS  or in the PPE or ripples counters. However,
about 0.1~rev later, there appeared a huge spike in just about all ASD4 counters, not only the PPE and ripples counters, which usually 
record incoming 
light flashes. Most affected was CCD Row~4 in FoV1, closely followed by Row 5. One revolution later similar signals were seen, albeit with smaller
magnitudes, and more concentrated in Rows~4 and 5, with Rows~1 and 2 and most of FoV2 not showing anything, or only small peaks. 
This is shown in Fig.~\ref{fig:MM_NWC2_15235.fig}, using the FoV1 WC2 ASD4 counts as an example. 
This figure shows that while the on board object confirmation could cull the vast majority of bogus detections caused by the event,
the fraction which remained was large enough to overwhelm the on board storage capacities and resources. 
This by itself rather unremarkable impact event must have created a stray-light path straight 
onto the FPA. {\it Gaia} had been built, so that a singular puncture cannot by itself produce a stray-light path. Therefore, this impact must have created a 
hole, which is aligned to a previously existing damage, so that stray-light can be directed onto the FPA. 
Therefore every time the Sun aligned with this path, excess light caused the large number of spurious detections. 
The final entry point
of this path had later been identified as a pre-existing gap in the MLA cover between the base plate and the thermal tent very close to the direction
opposite to the centre between the pointing directions of both apertures.

As it soon became clear that these stray-light features would be permanent, 
and given that the spurious objects use up a significant amount of the of the on board and downlink resources, 
measures had to be promptly implemented in order to minimise the adverse effects caused 
by the stray-light contamination. Thus, unlike the four
previous events discussed in this section, the aftermath of this micro-meteoroid impact had significant long-lasting consequences, 
which is why it is also included in Table~\ref{tab:sign_events.tab}. 

Things were further complicated because of the failure, on May 15, 2024, 
 of the  Proximity Electronics Module (PEM) controlling unit for the first astrometric detector (AF) in CCD Row~3 (AF1\_3 detector; see Fig.~\ref{fig:FPA.fig}). This failure was completely unrelated to the April 2, 2024, micro-meteoroid impact, with the presumed origin being
 a late effect of the Sun's coronal mass ejection of May 10, 2024. 
At first the whole Row~3 was inoperable. However, eventually it could be activated again, without the AF1\_3 detector, 
which could not be recovered. As the AF1 strip sets itself apart from the other astrometric strips, i.e. AF2-9, 
in that it performed the object confirmation, the loss of this particular unit 
had significant consequences for the operation of Row~3. With the loss of the detection confirmation for WC1 and WC2 
objects, and the fall-back to direct confirmation, bad-object culling was 
compromised, and nearly every detected object was passed on to further data taking.\footnote{PPE and ripples are 
still identified and sorted out}
 The same applied for the validation 
process for the AOCS stars. While Row~3 was not the row most effected by the impact of April 2, 2024, 
it now passed on a large quantity of bogus detections, and further mitigation issues had to be implemented 
for this row.

While a satisfactory situation was reached by June 2024, and maintained throughout the following months, the stray-light peaks 
suddenly increased in magnitude\footnote{A plausible possibility of what might have happened is the following: An already existing loose flap of the sunshield or thermal tent (caused by the impact of April 2) was charged up   by the ions of the solar plasma cloud, and then flipped over due to electrostatic forces.} on October 7, 2024, possibly by 
the solar coronal mass ejection happening around that date (see Sect.~\ref{sect:FL:Trends:sign_events} and Table~\ref{tab:sign_events.tab}).
 This resulted in the appearance of larger attitude rate 
excursions during the stray-light peaks, which were not seen before this enhancement. The reason for this change in the 
behaviour of the AOCS is shown in Fig.~\ref{fig:MM_NAOCS_SC106.fig}. The upper panel shows the numbers of the invalidated and confirmed AOCS stars on April 2, 2024, i.e. before, during, and after the initial impact, while the lower panel
depicts the situation after October 7, 2024. It can be clearly seen that despite the huge number of bogus AOCS candidates, virtually all of those were weeded out by the confirmation process, in the time between the initial MM-impact and the 
stray-light enhancement in October. Afterwards, a number of bogus objects passed the confirmation and were used by the AOCS
to control the attitude, which obviously led to confusion, as these sources were not real but spurious objects. This in turn led to 
the AOCS trying to correct deviations from the scan law which were not really there, thus causing its own rate excursions. Thus, this aggravation
of the stray-light peaks, from October 7, 2024 on, 
had a much stronger detrimental effect on the science performance than the initial impact event, of April 2, 2024.
Fortunately, this last set of events happened in the last year of operations, and while it had certainly compromised a certain fraction of the data 
taken, the overall impact was rather limited.

These five examples give an impression of the wide spectrum of effects caused by events of apparently
 very similar nature. This also means that it was not straightforward
to classify an observed peculiarity on board as being caused by a certain phenomenon, in our case here, a micro-meteoroid impact. While the most obvious diagnostic
sign that such an impact had occurred are the attitude disturbances, followed by spikes in the PPE or ripples counters (or the WFS), by far not all impacts cause these signals.
Actually, as demonstrated above, some of the more momentous events did not show one or both of these telltale signs. 
The reason is that optical signatures only appear when
they fall onto the FPA, i.e.  through the official optical pathways or through stray pathways, 
caused by holes in the optical tent, reflections off surfaces within, or sunlight scattering off an expanding debris cloud, among other factors. 
Likewise, disturbances of the attitude, which are caused by the deposit of momentum onto the payload by the impacting body, 
and their strength not only depend on the mass of the impactor but also on the location of the impact and the impact 
vector. This means that the analysis of
the OGA2$-$NSL or the even more thorough pair rate analysis, as described in Sect.~\ref{sect:results:RE:MM} can provide valuable insight into the frequency of micro-meteoroid
impacts, and give also a rough estimate of the energy function, i.e. the statistics of momentum transfer. However, there will always be a selection bias, due to the aforementioned limitations in the detectability. Actually an unknown but substantial fraction of impacts may go undetected at all. 
Therefore it seems to be paramount to monitor such events, not only for one mission -- in our case, {\it Gaia} -- but for all missions in the L2 region. This is all the more important 
as L2 missions are becoming more and more popular given the decisive advantages this type of orbit has.
\section{Summary and discussion}\label{sect:discussion}

In the previous sections we have given an overview of some of the findings of the {\it Gaia} First Look. 
We note that in this article we could only present a few highlights, representing a small part of the full scope of the FL's activities.
The FL had no goal in itself; it served a purpose within
 the greater scheme of {\it Gaia} operations. Primarily, it was a diagnostic tool that 
focused both on longer-term trends and discrete events, which merited the attention of either the operators of 
{\it Gaia} or the data processing. Therefore, all significant findings by the FL have been
 distributed to the broader {\it Gaia} community, 
or rather the relevant parts thereof. Partly, 
this was done by the daily qualifying of ten crucial parameters,\footnote{These were the following results from ODAS (see Sect.~\ref{sect:FL:ODAS}), i.e. 
the Astrometric calibration, the residuals of the source updates of the primary and secondary ODAS, from CODC (see Sect.~\ref{sect:FL:CODC})
 the Bias non-uniformity calibration library integrity check results, the CCD Health, the Charge injection, and release integrity check results, 
and from LODC (see Sect.~\ref{sect:FL:LODC}) the Empirical LSF (ELFS) library integrity check results, the optical ELSF integrity check results,
and the Mean LSF check results} thus regularly
conveying information about the quality of a given set of data to down-stream
consumers of these data. This gave long-term access to a standardised assessment of
 the quality of the data in a quick and concise way, even years after the data in question 
had actually been taken. While this was certainly one aspect,
 the main task of the FL focused on monitoring the data quality during {\it Gaia}'s operations. 
The diagnostic tools
 employed during the actual data reduction campaigns (i.e. those resulting in the Gaia data releases),
 surpass by far what the FL had available. 

In the case of unforeseen events, 
such as those discussed in the previous sections, 
the FL alerted other parties of anomalous situations, and 
presented input for an analysis of the impact and consequences of those events, thus 
aiding to find a solution, if needed. 
This was done in close interactions with the various counterparts,  within DPAC, the Mission Operations Centre (MOC), or the Science Operations Centre (SOC), and with the manufacturer of {\it Gaia}, 
Airbus Defense and Space. Here not only the initial findings and analyses were presented, 
but also follow-up investigations defined, as discussed with the partners, in order to achieve a more precise bearing
on the problem. This then provided part of the foundation for necessary decisions about how to tackle the problem at hand.
In the same way, FL monitored a number of characteristic parameters over the longer term (see for example Sect.~\ref{sect:FL:Trends}), 
and raised alerts when the evolution of 
these indicated that something on board was approaching the permissible limits.
A prime example for this was the focus, which needed to be adjusted from time to time 
once a defined level of focal degradation had been reached.

First Look's input was especially crucial in the commissioning phase of the mission, prior to
the start of nominal operations. During this phase, `teething' problems had to be identified and remedied, 
as well as the real-life characteristics of the satellite determined with respect to 
the pre-launch specifications. One example for such an issue discovered during the 
commissioning was the six-hour oscillation of the BA (see Sect.~\ref{sect:FL:BAM}).

Overall, the FL has served the mission well. In collaboration with the relevant counterparts within the {\it Gaia} project, 
it had contributed to the conduct of the operational part of the mission that was  as smooth as possible 
between {\it Gaia}'s launch on December 19, 2013, to
its final decommission on March 27, 2025. This also means that the basic set-up of the {\it Gaia} DPAC FL was efficient and remained so for all of the 11 years 
of operations. As with all such enterprises there are some aspects that imposed some limitations. 
There are some obvious drawbacks, such as the limited amount of data per day, and the fact that the data had to be pre-processed to a certain degree.
These could not easily be remedied without compromising the overall timeliness of the FL results. However, the very long daily 
processing phase of the data delayed the
access to parameters, such as the ASD countrates or the housekeeping data, which did not need to go through significant pre-processing. These are often the
parameters that indicate problems the soonest, yet they only became available to the FLSs, along with the other data  
that had gone through  pre-processing. Therefore, these data was usually only inspected many hours after it was downloaded from the 
spacecraft. This means that
the first hint of issues with the payload often came through other channels, mostly MOC, who had quicker access to the most recently downloaded data.
It was, however,  possible to trigger a daily FL report prematurely, which then contained everything that was processed by the time
of the triggering. Generally, such an early report triggering was only done when the FLSs were aware of problems.

While this issue was not as significant as it seemed at first, for the implementation of something akin to the {\it Gaia} FL in the frame of future missions,
it would be beneficial to implement a more flexible diagnostic output 
than the FL daily reports, in which all information regarding one FL day was condensed. 
Moreover, many investigations involve data stretching over more than one day. Therefore a database system in addition to the 
daily reports would have been highly beneficial.

Despite these minor shortcomings, we have proven that the FL has fulfilled its mandate over a long time. Looking  
ahead, our experiences with such an undertaking can serve as a template for future scientific missions. 
Obviously, the nature of such new missions
will lead to overall different architectures, catering to the need of the scientific content to be obtained by that project. With the end of {\it Gaia} operations
on March 27, 2025, the purpose of the {\it Gaia} First Look had been fulfilled, but its legacy will continue to aid the data processing in the years to come.

\begin{acknowledgements}
This work was financially supported
by the European Space Agency (ESA) in the framework of the {\it Gaia} project, and the 
Ministry of Economy and Technology of the Federal Republic of Germany
through the German Space Agency DLR (Deutsches Zentrum f\"ur Luft- und Raumfahrt) 
via grants 50QG0501, 50QG1401, and 50QG2101, 
and the United Kingdom Space Agency (UKSA) through the following grants
to the University of Edinburgh and the University of Leicester, ST/S000976/1,
ST/S001123/1. 
The authors would like to thank Jordi Portell i de Mora for
his thorough reading of the article, and the Gaia MOC
as well as the Gaia Payload Expert group for the long-term close collaboration.
We gratefully acknowledge the support of our colleagues at DPCE (Data Processing Centre ESAC) in the operation of the First Look software and in supporting the activities of the FLSs in various ways. Their contribution was essential to the successful conduct of FL operations throughout Gaia’s operational phase.
\end{acknowledgements}

\bibliographystyle{aa}
\bibliography{aa59204-26}

\begin{appendix}
\section{The focal plane array}\label{sect:Appendix_B:FPA}

\begin{figure}[ht!]                                             
  \centering  
      \includegraphics[width=1.00\hsize]{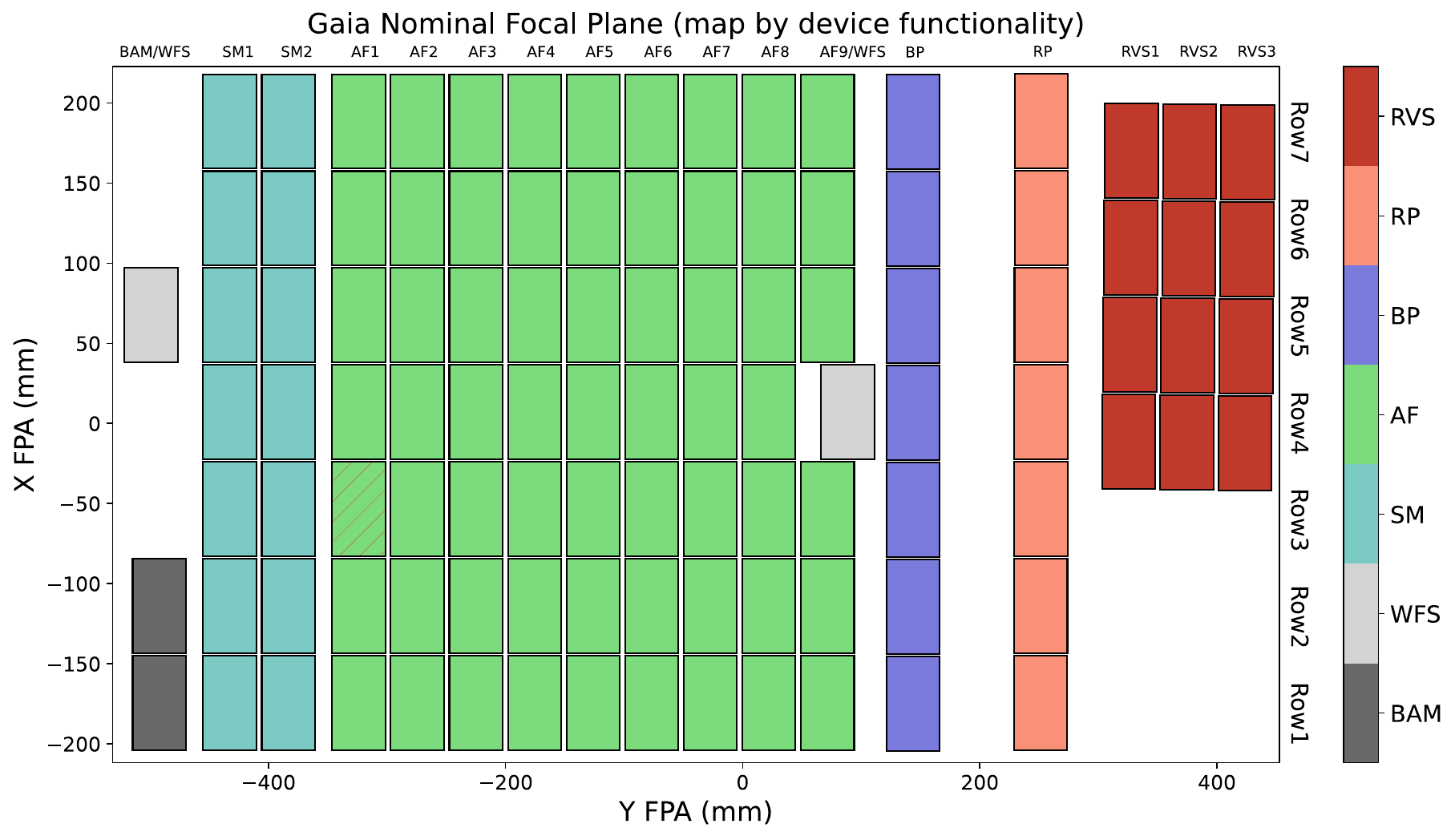}
\caption{\small Layout of Gaia's focal plane array. The Sky Mapper (SM) is shown in blue, the Astrometric field (AF) 
in green, the Blue Photometer (BP) in purple, the Red Photometer (RP) in light red. The Radial Velocity
Spectrometer (RVS) is depicted in dark red, while the detectors of the Basic Angle Monitor (BAM)
 are coded in dark grey, and those of the wavefront sensor (WFS) in light green. The scales on the x- and y-axes
indicate the dimensions of the FPA. We note that the slight internal shifts in alignment of the RVS reflects
their actual positions. The AF detector, which also shows a red hatching, is the A1\_3 unit, 
which failed on May 15, 2024 (see Sect.~\ref{sect:FL:Trends:sign_events} and Table~\ref{tab:sign_events.tab}).
This figure is adapted from Fig.~1 in \citet{2016A&A...595A...6C}}
\label{fig:FPA.fig}
\end{figure}

Since the Gaia Focal Plane Array (FPA) is of utmost relevance throughout this article, we present it here for reference.
\newpage

\section{Parametrisation functions of the main attitude rate excursion types}\label{sect:Appendix_B}

Shown here are the parametrisation functions of the pair rate analysis of the three main disturbance
effects of the OGA2, as described in Sect.~\ref{sect:results:RE}. Here, $t$ is the independent variable, i.e.
the time since the start of the perturbation in seconds,
 $h$ is the magnitude in mas/sec, $\tau$ is the finite length of the event, also given in seconds.
For the micro-clank and micro-meteoroids the duration is zero.

\subsection{Parametrisation of clanks}\label{sect:Appendix_B:clank}

\begin{align}
    r_\mathrm{clank}(t) & =
    \begin{cases}
       0,                               &        t \leq 0,    \\
       \frac{h}{4.41} t,                & 0    < t \leq 4.41, \\
       h,                               & 4.41 < t \leq 4.86, \\
       h -\frac{h}{4.41} (t-4.86),      & 4.86 < t \leq 9.27, \\
       0,                               & 9.27 < t
    \end{cases}
    \label{eqn:wl model micro clank}
\end{align}
\subsection{Parametrisation of micro-meteoroids}\label{sect:Appendix_B:MM}

\begin{align}
	r_\mathrm{micro-meteoroid}(t) & =
	\begin{cases}
		0,                                                           &        t \leq 0,    \\
		\frac{h}{8.82\cdot4.86} t^2,                                 & 0    < t \leq 4.41, \\
		\frac{h}{4.86}t - \frac{4.41h}{2\cdot4.86},                   & 4.41 < t \leq 4.86, \\
		h - \frac{h}{8.82\cdot4.86}(t-9.27)^2,                       & 4.86 < t \leq 9.27, \\   
		h,                                                           & 9.27 < t
	\end{cases}
	\label{eqn:wl model micro meteoroid}
\end{align}

\subsection{Parametrisation of propellant movement events}\label{sect:Appendix_B:propellant}

\begin{align}
	r_\mathrm{propellant}(t) & =
	\begin{cases}
		0,                                                     &        t \leq 0,    \\
		\frac{h}{2\tau^2 + 4.41\tau} t^2,                      & 0    < t \leq \tau, \\
		\frac{h}{\tau+4.41/2} t - \frac{h\tau}{2\tau + 4.41},  & \tau < t \\
	\end{cases}
	\label{eqn:wl model propellant}
\end{align}

\section{List of noteworthy micro-meteoroid impacts and their effects on the payload}\label{sect:app:mm}

In Sect.~\ref{sect:results:MM} we have analysed a number of micro-meteoroid impacts of significance to a certain degrees, and their
different effects on various parts of the {\it Gaia} instrument. For reference these are summarised in a concise form and presented in 
Table~\ref{tab:MMevents}. We note that this list comprises only those events discussed in this publication and not many others, so this is not meant to be exhaustive.  
\begin{sidewaystable*}
\caption{\small \label{tab:MMevents}Summary of the noteworthy micro-meteoroid impact events} 
\begin{center}
\begin{tabular}{lrccp{2cm}rp{1.5cm}rrcp{3.5cm}}
\hline
UTC                       & OBMT & OGA2$-$NSL        & ASD4        & BA  & Astr. Cal. & CRLB   & $T_{\rm FPA}$  & $T_{\rm Mirror}$  & WFS & Remarks \\
                          &      & AL/ACP/ACF        &             &     & WC1, $0^{th}$ order           &        &                &                   &     &        \\
yyyy-mm-dd{\it T}hh:mm:ss & rev  & mas/sec           &             & mas & mas        & pixels & K      & K         &     &       \\
\hline
\hline
2021-03-07T07:32:01 &  10745.885 &        ?          &   ?         & --- &  ---       &  ---    &      ---     &     ---      &  n  & largest magnitude impact event during the {\it Gaia} mission (from the AOCS; see text)        \\
2021-12-23T19:50:32 &  11911.817 &   $-$1/$-$3/+4    & PPE/RIP     & --- &  ---       &  ---    &     ---        &   ---            &  y  & caused permanent damage to a part of the SM2\_7 detector        \\
2023-04-05T22:55:17 &  13784.450 &   ---             & PPE/RIP     & +0.9(FoV1)   & $\pm0.3$   & +0.015 (FoV1)    &     ---        &    ---         &  y  & strong optical signal, damage to optics in FoV1, slight damage to SM2\_4 detector        \\
2023-06-12T00:10:06 &  14052.674 &   +54/+783/$-60$  & PPE/RIP     & +1.7(FoV1) $-0.9$(FoV2) +2.7(total)   & $\pm1.5$   &  ---    &  $\sim0.05$       &   $\sim0.06$             &  y  & strong optical signal, persistent thermal changes \\
2024-04-02T15:35:02 &  15235.227 & $-$5/+20/$-$123   & All\tablefootmark{*} & ---   & ---   &  ---   &  ---  &  ---  &  n  & caused stray-light path affecting the FPA, especially 
FoV1 CCD Rows~4 and 5, most momentous impact event in mission history        \\
\hline
\end{tabular}
\end{center}
\tablefoot{
A long dash (---) indicates that no signal was seen for this parameter, while  N/A means that the data was not available for this parameter at 
the time of the event, in some cases as a direct consequence of the impact event.
For the column OGA2$-$NSL, the first entry, labelled AL, refers to the along-scan direction,
ACP to the across-scan direction in the preceding field of view (FoV1), and ACF to the following field of view (FoV2; see Sect.~\ref{FL:Trends:ODAS}, and \ref{sect:results:RE}). 
In the column ASD4, RIP means the ripples counter and PPE means the prompt particle event counter (see Sect.~\ref{sect:FL:ASD}). 
The $\pm$ signs shown in the astrometric calibration column indicates the direction of the jump in the 
zeroth-order calibration (see Sect.~\ref{sect:FL:ODAS}). This indicates that the FoV1 value jumped
to a higher value, pushing the two values further apart. The total magnitude of the jump is twice the value indicated.\\
Remarks:\\
\tablefoottext{$*$}{Due to the permanent damage that occurred periodically to the spikes in most ASD4 counters, for up to 40 minutes during every 6 hour revolution. We note that the peaks did not coincide with the initial impact.}
}
\end{sidewaystable*}
\newpage
\newpage
\section{Acronyms}
\begin{table}[h!]
\caption{\small \label{tab:acronymns} List of the acronyms used throughout  this article}
\begin{center}
\begin{tabular}{lp{0.37\textwidth}}
\hline
\text{Acronym} & \text{Description}  \\
\hline
AC&Across Scan \\
AF&Astrometric Field (in Astro) \\
AGIS&Astrometric Global Iterative Solution \\
AL&Along Scan \\
AOCS&Attitude and Orbit Control Sub-system \\
ASD&Auxiliary Science Data \\
BA&Basic Angle \\
BAM&Basic-Angle Monitor \\
BAV&Basic-Angle Variation \\
BP&Blue Photometer \\
CCD&Charge-Coupled Device \\
CFS&Calibration Faint Star \\
CODC&CCD One-Day Calibration (FL) \\
CRLB&Cram\'er-Rao Lower Bound \\
CTI&Charge Transfer Inefficiency \\
DLR&Deutsches Zentrum f\"ur Luft und Raumfahrt \\
DPAC&Data Processing and Analysis Consortium \\
DR1&Gaia Data Release 1 \\
DR2&Gaia Data Release 2 \\
DSA&Deployable Sunshield Assembly \\
ESA&European Space Agency \\
ELSF&Empirical Line Spread Function\\
FL&First Look \\
FLS&First-Look Scientist \\
FPA&Focal Plane Array \\
FoV&Field of View (also denoted FOV) \\
GC&Galactic Centre \\
GP&Galactic Plane \\
GPS&Galactic Plane Scan \\
HK&Housekeeping (also denoted H/K) \\
IDT&Initial Data Treatment \\
IDU&Intermediate Data Update \\
IOGA&Initial On-Ground Attitude \\
LOC&Loss of Convergence \\
LODC&LSF/PSF One-Day Calibration (FL) \\
LSCP&Large-Scale Calibration Parameter \\
LSF&Line Spread Function \\
LTE&Local Thermal Equilibrium \\
LoC&Loss of Convergence \\
MIT&MOC Interface Task \\
MLA&Multi-Lateral Agreement \\
MOC&Mission Operations Centre \\
NSL&Nominal Scanning Law \\
OBMT&On-Board Mission Timeline \\
ODAS&One-Day Astrometric Solution \\
ODC&One-Day Calibration \\
OGA1&First On-Ground Attitude determination (in IDT) \\
OGA2&Second (and improved) On-Ground Attitude determination (in ODAS/FL) \\
PDHU&Payload Data Handling Unit \\
PEM&Proximity Electronics Module \\
PPE&Prompt Particle Event \\
PSF&Point Spread Function \\
QPRE&Quasi-Periodic Rate-Excursion (Event) \\
RODC&RVS One-Day Calibration (FL) \\
\hline
\end{tabular}
\end{center}
\end{table}

\addtocounter{table}{-1}

\begin{table}[h!]
\caption{\small continued}
\begin{center}
\begin{tabular}{lp{0.37\textwidth}}
\hline
\text{Acronym} & \text{Description}  \\
\hline
RP&Red Photometer \\
RVS&Radial Velocity Spectrometer \\
SAA&Solar Aspect Angle \\
SM&Sky Mapper \\
SMO&Suspect Moving Object \\
SOC&Science Operations Centre \\
SP&Star Packet \\
TDI&Time-Delayed Integration (CCD) \\
UTC&Coordinated Universal Time \\
VPU&Video Processing Unit \\
WEAM&Whitehead Eclipse Avoidance Manoeuvre \\
WFS&WaveFront Sensor \\
XP&Shortcut for BP and/or RP (generic name for Blue or Red Photometer) \\
\hline
\end{tabular}
\end{center}
\end{table}

\end{appendix}
\end{document}